\documentclass[10pt, twocolumn]{article}
\usepackage{amsmath,epsfig,subfigure,geometry,ccaption,nature,naturefem,cite,citesupernumber}

\renewcommand{\thesection}{\hspace*{-6mm}}
\renewcommand{\thesubsection}{\hspace*{-5mm}}

\begin{document}

\title{Extraclassical receptive field phenomena \& short-range connectivity in V1}

\author{{\large Jim Wielaard and Paul Sajda}\\
{\em Laboratory for Intelligent Imaging and Neural Computing}\\
Department of Biomedical Engineering \\ Columbia University, New York,  USA}

\maketitle
\section{Abstract}
{\bf Neural mechanisms of extraclassical receptive field
phenomena in  V1 are commonly assumed to result from long-range lateral 
connections and/or extrastriate feedback. We address two such phenomena: 
surround suppression and contrast dependent receptive field size. We present
rigorous computational support for the hypothesis that the phenomena largely 
result from local short-range ($< 0.5$ mm) cortical connections and LGN input. 
Surround suppression in our simulations results from (A) direct cortical inhibition 
or (B) suppression of recurrent cortical excitation, or (C) action of both these mechanisms 
simultaneously. Mechanisms B and C are substantially more prevalent than A. We observe 
an average growth in the range of spatial summation of excitatory and inhibitory synaptic 
inputs for low contrast. However, we find this is neither sufficient nor necessary to explain 
contrast dependent receptive field size, which usually involves additional changes 
in the relative gain of these inputs.}
 \\ \\ 
{\bf keywords:} visual cortex, spatial summation, surround suppression,
receptive field, model, simulation
\vspace{1cm}
\section{Introduction}
In mammals, the very first stage of cortical visual processing takes
place in the striate cortex (area V1). Already at this level spatial summation of visual
input displays a considerable complexity. This is manifest from the fact that single cells in V1
display surround suppression (suppression for increasing stimulus size, ``size tuning'')
of their responses and have receptive field sizes (preferred sizes) that depend on contrast and 
context. Such behavior is seen throughout striate cortex, including all cell types
in all layers and at all eccentricities
\cite{dow81,sch76,sil95,sce99,kap99,sce01,and01,cav02,oze04}.  The suppression
seen is substantial, 30-40\% on average in macaque V1\cite{cav02}.
Similarly profound is the growth of receptive field size at low contrasts.  Typical
is a doubling in receptive field size for stimulus contrasts
decreasing by a factor of 2-3 on the linear part of the
contrast response function \cite{sce99}.  Apparently, neurons in V1
sacrifice spatial sensitivity in return for a gain in contrast
sensitivity at low contrasts \cite{sce99}. Neural mechanisms
responsible for these two so called extraclassical receptive field
phenomena are very poorly understood. Understanding these mechanisms
is potentially important for developing a theoretical model of early signal integration and 
neural encoding of visual features in V1.

Popular working hypotheses are that extraclassical receptive field
phenomena in V1 are a product of long-range horizontal connections
within V1 \cite{dea94,dra00,hup01,ste02} and/or feedback from
extrastriate areas \cite{sce01,cav02,ang02,bai03}. Arguments in
support of these hypotheses are based on the observed surround sizes and the cortical magnification
factor, and claim that short-range ($< 0.5$ mm) and even long-range
horizontal ($< 5 $ mm) connections in V1 do not have sufficient
spatial extent to be responsible for surround suppression or receptive field growth
\cite{sce01,cav02}. Further support along this line was presented using 
anterograde and retrograde tracer injections\cite{ang02} and timing
experiments \cite{bai03}. So far, however, all support for these hypotheses is entirely 
based on indirect experimental observations, while they also lack any rigorous 
computational support.  

The hypothesis that the phenomena result from local short-range ($< 0.5$ mm) cortical connections 
and LGN input is largely ignored or dismissed. However, support for it can be found 
in the experimental data. For instance, surround suppression and contrast dependent receptive field size are equally profound throughout V1\cite{sce99,sce01}, including in layers that do not receive extrastriate feedback 
and do not have long-range horizontal connections. Both phenomena have been
observed in the lateral geniculate nucleus (LGN), and are likely to be partially inherited by V1 cells via
feedforward connections from LGN\cite{sol02,oze04}. Finally, there is experimental evidence for contextual modulations mediated by local short-range connections in cat\cite{das99}.

In this paper we show that local short-range connections in V1 and LGN input are, in principle, sufficient 
to explain these two extraclassical receptive field phenomena in layers $4C\alpha$ and $4C\beta$ of macaque. 
We do this by means of a large-scale computational model which is constructed, as much 
as possible, from basic, established experimental data. We suggest neural mechanisms for the phenomena
by analyzing the synaptic inputs that generate them. An illustration of the model's architecture
is given in Figure 1. A brief summary of the model is given in Methods.

\section{Methods}
\subsection{The Model}
Our model consists of 8 ocular dominance columns and 64
orientation hypercolumns (i.e. pinwheels), representing a 16 ${mm}^{2}$
area of a macaque V1 input layer $4C\alpha$ or $4C\beta$. 
The model contains approximately 65,000 cortical cells and the corresponding 
appropriate number of LGN cells. Our cortical cells are modeled as conductance 
based integrate-and-fire point neurons, 75\% are excitatory cells and
25\% are inhibitory cells. Our LGN cells are rectified spatio-temporal linear filters.
The model is constructed with isotropic short-range cortical connections ($<500\mu m$), 
realistic LGN receptive field sizes and densities, realistic sizes of LGN axons in
V1, and cortical magnification factors and receptive field scatter that are
in agreement with experimental observations. We will only give a very brief description
of the model here, it is explained in detail in Supplementary Materials. 
Some background information can also be found in previous work\cite{mcl00,wie01} by one of the authors (JW).

Dynamic variables of a cortical model-cell $i$ are its membrane potential $v_{i}(t)$ 
and its spike train ${\cal S}_{i}(t)=\sum_{k}\delta(t-t_{i,k})$, where $t$ is time 
and $t_{i,k}$ is its $k$th spike time.  Membrane potential and spike train of each
cell obey a set of $N$ equations of the form
\[
C_{i}\frac{dv_{i}}{dt} = -g_{L,i}(v_{i}-v_{L})
-g_{E,i}(t,[{\cal S}]_{E},\eta_{E})(v_{i}-v_{E})
\]
\begin{equation}
\label{eq:mem}
-g_{I,i}(t,[{\cal S}]_{I},\eta_{I})(v_{i}-v_{I})
\; , \;i=1,\dots ,N \: .
\end{equation}
These equations are integrated numerically using a second order
Runge-Kutta method with time step 0.1 ms.  Whenever the membrane
potential reaches a fixed threshold level $v_{T}$ it is reset to a
fixed reset level $v_{R}$ and a spike is registered. 
The equation can be rescaled so that $v_{i}(t)$ is dimensionless and $C_{i}=1$, $v_{L}=0$, 
$v_{E}=14/3$, $v_{I}=-2/3$, $v_{T}=1$, $v_{R}=0$, and conductances (and currents)
have dimension of inverse time.

The quantities $g_{E,i}(t,[{\cal S}],\eta_{E})$ and $g_{I,i}(t,[{\cal S}],\eta_{I})$
are the excitatory and inhibitory conductances of neuron $i$. They are 
defined by interactions with the other cells in the network, external 
noise $\eta_{E(I)}$, and, in the case of $g_{E,i}$ possibly by LGN input. 
The notation $[{\cal S}]_{E(I)}$ stands for the spike trains of all 
excitatory (inhibitory) cells connected to cell $i$.  Both, the excitatory 
and inhibitory populations consist of two subpopulations ${\cal P}_{k}(E)$ 
and ${\cal P}_{k}(I)$, $k=0,1$, a population that receives LGN input ($k=1$) and one 
that does not ($k=0$). 
In the model presented here 30\% of both the excitatory and inhibitory cell populations 
receive LGN input.  
We assume noise, cortical interactions and LGN input
act additively in contributing to the total conductance of a cell,
\[
g_{E,i}(t,[{\cal S}]_{E},\eta_{E})   = 
\eta_{E,i}(t) +g_{E,i}^{cor}(t,[{\cal S}]_{E})+\delta_{i} g^{LGN}_{i}(t)
\]
\begin{equation}
g_{I,i}(t,[{\cal S}]_{I},\eta_{I})  = \eta_{I,i}(t)
+g_{I,i}^{cor}(t,[{\cal S}]_{I}) \; ,
\end{equation}
where $\delta_{i}=\ell$ for $i\in \{ {\cal P}_{\ell}(E), {\cal P}_{\ell}(I)\}$, $\ell=0,1$.
The terms $g_{\mu,i}^{cor}(t,[{\cal S}]_{\mu})$
are the contributions from the cortical excitatory ($\mu=E$) and inhibitory ($\mu=I$) neurons and
include only isotropic connections,
\[
g_{\mu,i}^{cor}(t,[{\cal S}]_{\mu})=
\]
\begin{equation}
\int_{-\infty}^{+\infty}ds\sum _{k=0}^{1} \; \sum_{j\in
{\cal P}_{k}(\mu)}{\cal C}_{\mu^{\prime},\mu}^{k^{\prime},k}
(||\vec{x}_{i}-\vec{x}_{j}||)G_{\mu,j}(t-s){\cal S}_{j}(s) \; ,
\end{equation}
where $i\in {\cal P}_{k^{\prime}}(\mu^{\prime})$
Here $\vec{x}_{i}$ is the spatial position  (in cortex) of neuron $i$,
the functions $G_{\mu,j}(\tau)$ describe the synaptic dynamics of
cortical synapses and the functions
${\cal C}_{\mu^{\prime},\mu}^{k^{\prime},k}(r)$ describe the cortical
spatial couplings (cortical connections). The length scale or excitatory
and inhibitory connections is about $200\mu$m and $100 \mu$m
respectively. 

An important class of parameters are the geometric parameters, which define and relate 
the model's geometry in visual space and cortical space. Geometric properties are different 
for the two input layers $4C\alpha, \beta$ and are different at different eccentricities.
As said, the two extraclassical phenomena we seek to explain are observed to be insensitive to those differences \cite{sce99,kap99,sce01,cav02}. 
In order to verify that our explanations are consistent with this observation, we have performed numerical simulations for four different sets of parameters, corresponding to the $4C\alpha, \beta$ layers 
at para-foveal eccentricities $<5^{\circ}$ and at eccentricities around $10^{\circ}$. 
These different model configurations are referred to as M0, M10, and P0, P10 in the text.
Reported results are qualitatively similar for all four configurations unless otherwise noted.

In agreement with experimental findings (see references in \cite{mcl00}), the LGN 
neurons are modeled as rectified center-surround linear spatiotemporal filters.  A
cortical cell, $j\in{\cal P}_{1}(\mu)$ is connected to a set
$N^{LGN}_{L,j}$ of left eye LGN cells, or to a set $N^{LGN}_{R,j}$ of
right eye LGN cells,
\[
g^{LGN}_{j}(t) = \sum_{\ell \in N^{LGN}_{Q,j}} [ g^{0}_{\ell}
+g^{V}_{\ell}
\]
\begin{equation}
\label{eq:lgncon}
\int_{-\infty}^{\infty} ds
\int d^{2}y\;
G^{LGN}_{\ell}(t-s)\;  {\cal L}_{\ell}\, (\,||\vec{y}_{\ell}- \vec{y}||)\; I({\vec{y}},s) ]_+
\:,
\end{equation}
where $Q=L$ or $R$.
Here $[x]_{+}=x$ if $x\geq 0$ and $[x]_{+}=0$ if $x\leq 0$, ${\cal L}_{\ell}(r)$ and
$G^{LGN}_{\ell}(\tau)$ are the spatial and temporal LGN kernels respectively,
${\vec{y}}_{\ell}$ is the receptive field center of the $\ell$th left or right eye LGN cell, which is
connected to the $j$th cortical cell, $I({\vec{y}},s)$ is the visual stimulus.
The parameters	$g^{0}_{\ell}$ represent the maintained activity of LGN cells and
the parameters $g^{V}_{\ell}$ measure their responsiveness to visual stimuli.
Their numerical values are taken to be identical for all LGN cells in the model,
$g^{0}_{\ell}= 2 s^{-1}$ and $g^{V}_{\ell}= 25\; cd^{-1}m^{2}s^{-2}$.
The LGN kernels are of the form \cite{ben99} 
\begin{equation}
\label{eq:tlgn}
G^{LGN}_{\ell}(\tau) = \left\{ \begin{array}{ll}
0 & \mbox{if $\tau \leq \tau^{0}_{\ell}$} \\
k \:\tau^{5}\left( e^{-\tau/\tau_{1}}-c\: e^{-\tau/\tau_{2}}\right) & \mbox{if
$\tau > \tau^{0}_{\ell}$}
\end{array} \right.
\end{equation}
and
\begin{equation}
\label{eq:xlgn}
{\cal L}_{\ell}\, (r)=\pm (1-K_{\ell})^{-1}\left\{ \frac{1}{\pi \sigma^{2}_{c,\ell}}
e^{-(r/\sigma_{c,\ell})^{2}} - \frac{K_{\ell}}{\pi \sigma^{2}_{s,\ell}}
e^{-(r/\sigma_{s,\ell})^{2}} \right\} \; ,
\end{equation}
where $k$ is a normalization constant, $\sigma_{c,\ell}$ and
$\sigma_{s,\ell}$ are the center and surround sizes respectively, and
$K_{\ell}$ is the integrated surround-center sensitivity.  The
temporal kernels are normalized in Fourier space,
$\int_{-\infty}^{\infty} | \widehat{G}_{\ell}^{LGN} (\omega) | d\omega
= 1$,
$\widehat{G}_{\ell}^{LGN}(\omega)=(2\pi)^{-1}\int_{-\infty}^{\infty}
G_{\ell}^{LGN}(t) e^{-i\omega t} dt$.  For the magno cases (M0, M10)
the time constants $\tau_{1}=2.5$ ms and $\tau_{2}=7.5$ ms and
$c=(\tau_{1}/\tau_{2})^{6}$ so that $\widehat{G}_{\ell}^{LGN}(0) =0$,
in agreement with experiment \cite{ben99}.  For the parvo cases (P0,
P10) the time constants $\tau_{1}=8$ ms and $\tau_{2}=9$ ms and
$c=0.7(\tau_{1}/\tau_{2})^{5}$. The delay times $\tau^{0}_{\ell}$ are
taken from a uniform distribution between 20 ms and 30 ms, for all
cases.  Sizes for center and surround were taken from experimental
data \cite{hic83,der84,spe84,sha90,cro95} and were
$\sigma_{c,\ell}=\sigma_{c}=0.1^{\circ}$, $0.2^{\circ}$,
$0.04^{\circ}$, $0.0875^{\circ}$ (centers) and
$\sigma_{s,\ell}=\sigma_{s}=0.72^{\circ}$, $1.4^{\circ}$,
$0.32^{\circ}$, $0.7^{\circ}$ (surrounds), for M0, M10, P0 and P10
respectively.  The integrated surround-center sensitivity was in all
cases $K_{\ell}=0.55$ \cite{cro95}.  By design, no diversity has been
introduced in the center and surround sizes in order to demonstrate
the level of diversity resulting purely from the cortical interactions
and the connection specificity between LGN cells and cortical cells
(i.e. the sets $N^{LGN}_{Q,j}$, see specifications below).  Further, no 
distinction was made between ON-center and OFF-center LGN cells other than 
the sign reversal of their receptive fields ($\pm$ sign in Eq. \ref{eq:xlgn}). 
The LGN RF centers $\vec{y}_{\ell}$ were organized on a square lattice with
lattice constants $\sigma_{c}/2$, $\sigma_{c}$, $\sigma_{c}/2$, and
$\sigma_{c}/2$ for M0, M10, P0 and P10 respectively.  These lattice
spacings and consequent LGN receptive field densities imply LGN cellular
magnification factors that are in the range of the experimental data
available for macaque\cite{con84,mal96}.
The connection structure between LGN cells
and cortical cells, given by the sets $N^{LGN}_{Q,j}$, is made so as to establish
ocular dominance bands and a slight orientation preference which is
organized in pinwheels\cite{bla92}. It is further constructed
under the constraint that the LGN axonal arbor sizes in V1 do not
exceed the anatomically established values of 1.2 mm for magno and 0.6
mm for parvo cells\cite{bla83,fre89}.  
A sketch of the model is given in Figure 1. Further details are given in Supplementary Information.

\begin{figure}[here]
\label{ss}
\centering
\begin{minipage}{0.35\columnwidth}
\centering
\vspace{0.3cm}
\includegraphics[height=2.75cm,width=2.75cm]{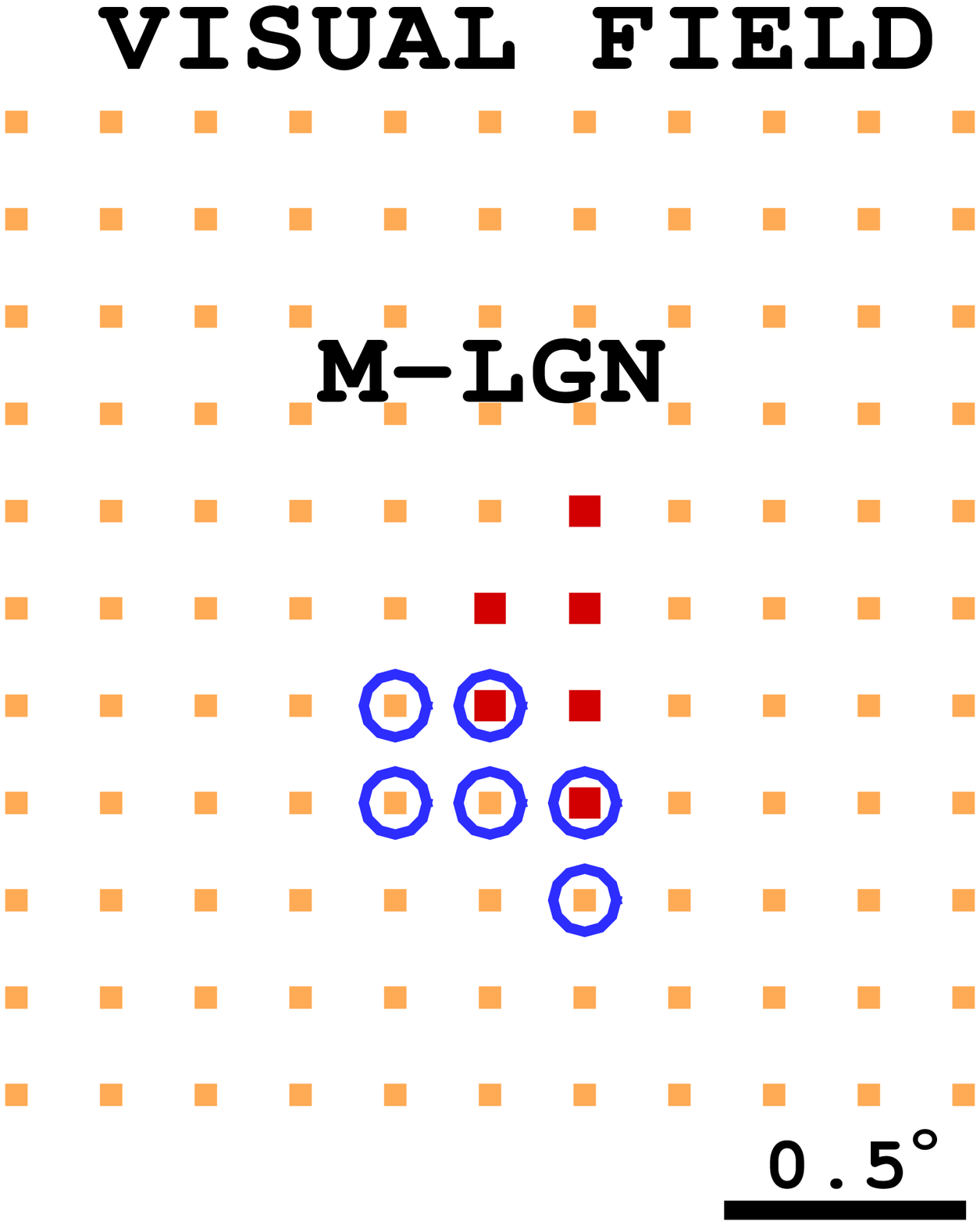}
\includegraphics[height=2.75cm,width=2.75cm]{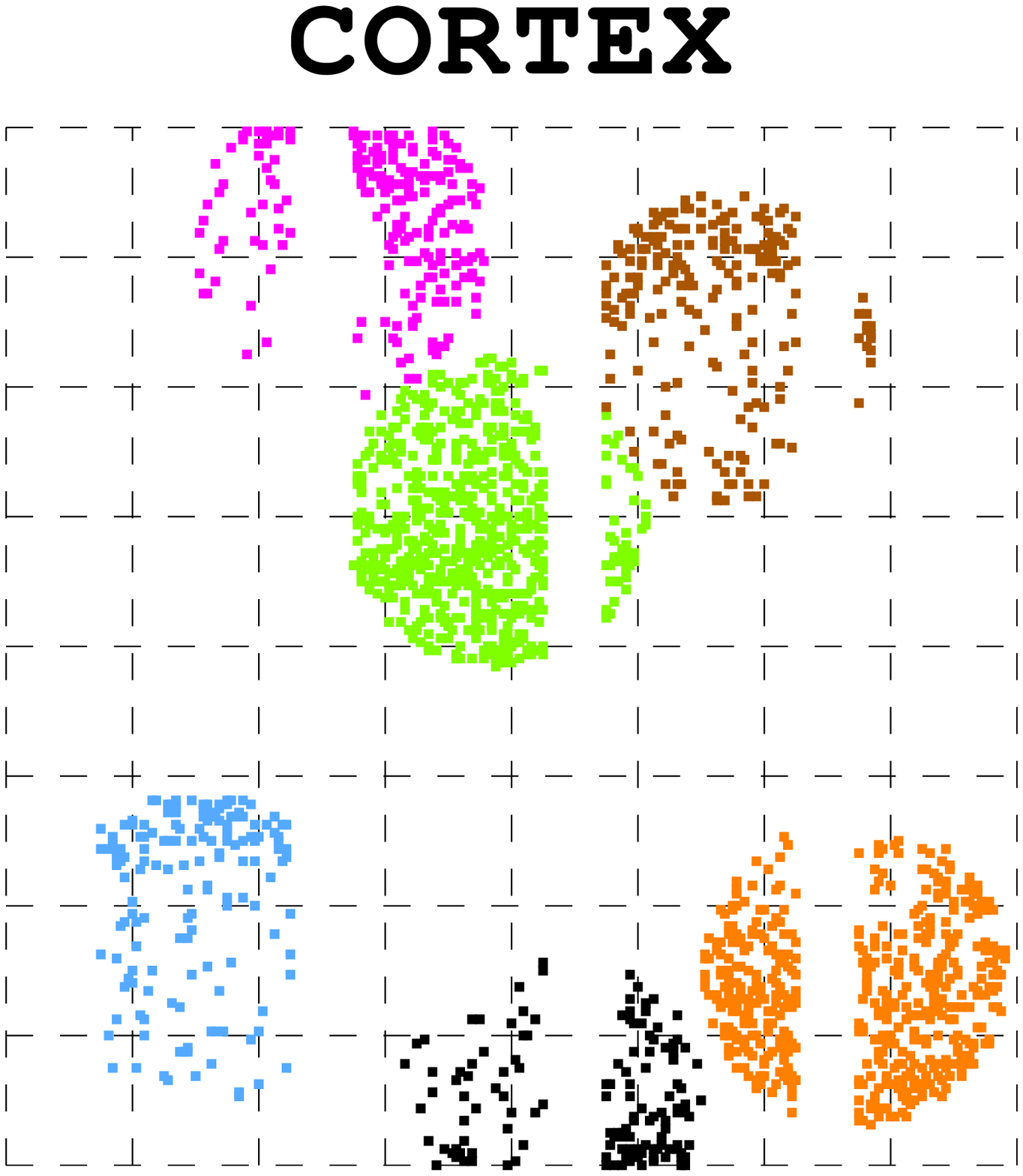}
\end{minipage}
\begin{minipage}{0.62\columnwidth}
\includegraphics[angle=270, origin=c,height=6.0cm,width=6.39cm]{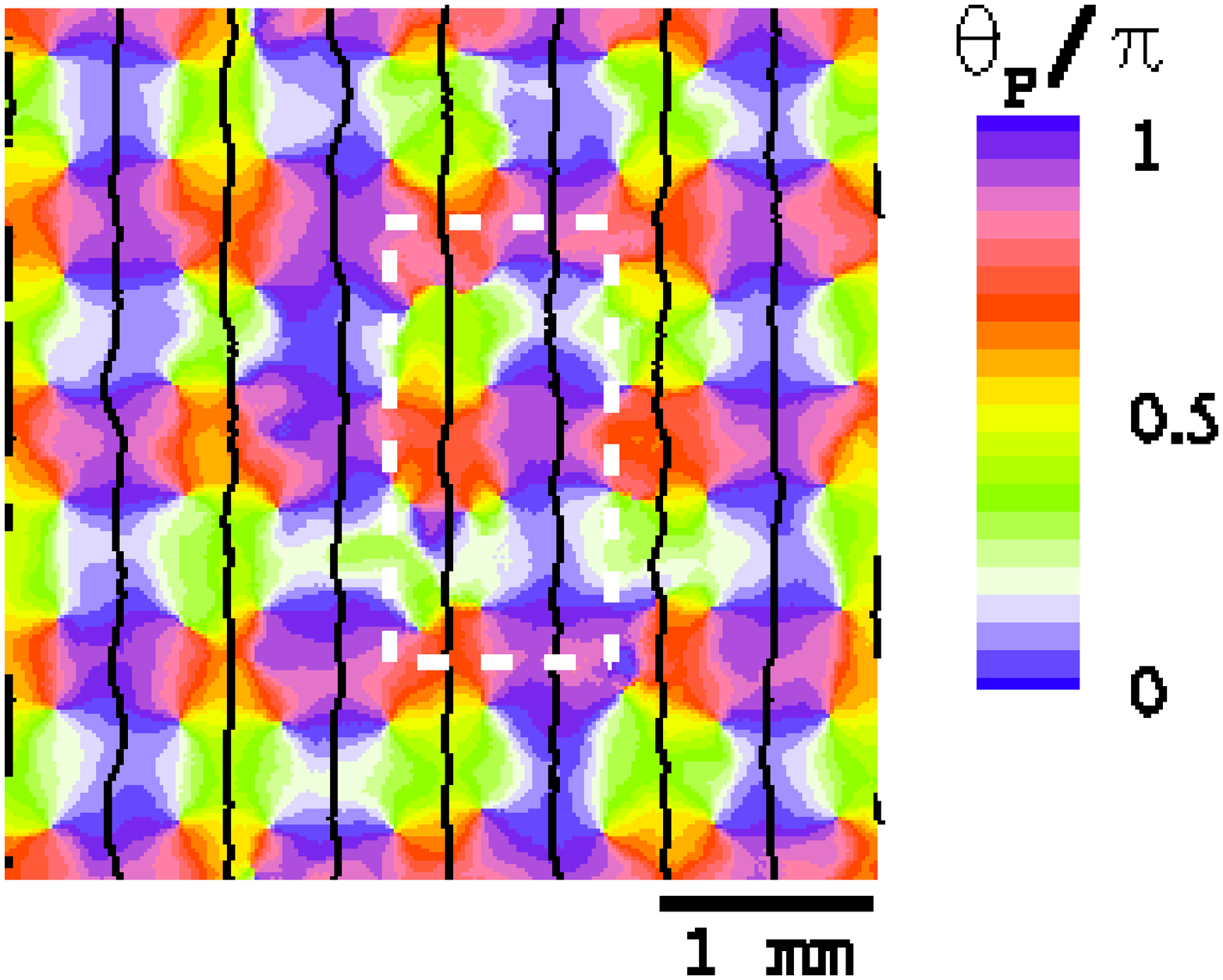}
\end{minipage}
\caption{\small Model architecture. (left top) A typical cluster of ON (blue circles) and OFF (red dots) M-LGN cells 
that feed into one cortical cell. Receptive field centers of LGN cells are organized on a square lattice (orange).
(left bottom) Some typical M-LGN axons in our model-V1.
Points of the same color are cortical cells that connect to the same LGN axon.
(right) Pinwheel structure and ocular dominance columns for M10 model, constructed from averaged responses in 
the spirit of optical imaging experiments\cite{bla92}. All sample cells used to study extraclassical
phenomena are taken from within the white dashed rectangle (see Supplementary Information for details).} 
\end{figure}

Some of the geometric differences (in the model as well as in the true biological situation) may be 
expressed by the dimensionless parameter $\Omega = \nu^{-1}\sigma_{c}\ell_{c}^{-1}$, where 
$\nu^{-1}$ is the cortical magnification factor, $\sigma_{c}$ is the LGN receptive field size
(center size), and $\ell_{c}$ is a characteristic length scale for the excitatory cortical connectivity.  
Substituting numerical values taken from experimental data, this parameter is 
$1$, $0.57$, $0.4$, and $0.25$, for M0, M10, P0 and P10 respectively. At $30^{\circ}$ eccentricity, the
experimental data suggests values for this parameter not very different from  its values at $10^{\circ}$ 
($\Omega = 0.5$ for M30 and $\Omega = 0.25$ for P30). 

\subsection{Visual stimuli and data collection}
The stimulus used in this paper to analyze the phenomena of surround suppression and receptive field 
growth at low contrast is a drifting grating confined to a circular aperture, surrounded by a blank
(mean luminance) background. The luminance of the stimulus is given by 
$I(\vec{y},t) = I_{0} (1+\epsilon \cos (\,\omega t - \vec{k}\cdot\vec{y}+\phi))$ for $||\vec{y}|| \leq r_{A}$
and $I(\vec{y},t) = I_{0}$ for $||\vec{y}|| > r_{A}$, with average luminance $I_{0}$, contrast $\epsilon$, temporal frequency $\omega$, spatial wave vector $\vec{k}$, phase $\phi$, and aperture radius $r_{A}$.
The aperture is centered on the receptive field of the cell and varied in size, while the other 
parameters are kept fixed and set to preferred values.  All stimuli are presented monocularly.
As the aperture size increases the response of a V1 cell to such
stimuli typically reaches a maximum, after which it settles down to some steady level.
The aperture size for which the response reaches its maximum is sometimes referred to as the 
``classical'' receptive field size \cite{dea94,lev97,sce99}. We will simply refer to the minimum aperture radius for which the response $f(r_{A})$ is $>95\%$ of its maximum as the receptive field size ($r$).
We define the surround size ($R$) as the minimum aperture radius $>r$ for which the suppression 
$f_{s}(r_{A})=f_{max}-f(r_{A})$ is $>95\%$ of its maximum. We define the asymptotic response $f_{\infty}$
as the average response beyond $R$. We define the suppression index ($SI_{1}$) as the relative 
surround suppression,
\begin{equation}
SI_{1}=\frac{f_{max}-f_{\infty}}{f_{max}-f_{0}}\; ,
\end{equation}
where $f_{0}$ is the response to a blank stimulus.
The suppression index $SI_{1}$ is similar to the one used in \cite{cav02}, but 
different from the integrated suppression index used in \cite{sce99}.

The primary data, i.e. responses and conductances as a function of aperture size for single eye
stimulation, are obtained for samples of approximately 200 cells for each configuration, containing about an equal number of simple and complex cells. Each stimulus was presented for 3 s and preceded by a 1 s blank stimulus. The procedure was repeated five times with different initial conditions and noise realizations.
Standard errors in cycle-trial average responses and conductances are negligibly small. The experiments were performed at ``high'' contrast, $\epsilon =1$, and ``low'' contrast, $\epsilon =0.3$. More precise definitions and further details are in Supplementary Information.
\subsection{DOG \& ROG models} 
In the Difference-of-Gaussians (DOG) model \cite{dea94,sce99,sce01},
the response $f(r_{A})$ is fit to
\[
f(r_{A})=
\]
\begin{equation}
f_{0}
+\frac{2}{\sqrt{\pi}}\left[
K_{E}\int_{0}^{r_{A}}e^{-(y/\sigma_{E})^{2}}dy
-K_{I}\int_{0}^{r_{A}}e^{-(y/\sigma_{I})^{2}}dy\right] \; .
\end{equation}
In this model the response is assumed to arise from a summation of background
activity ($f_{0}$), ``excitation'' (spatial scale $\sigma_{E}$) and
``inhibition'' (spatial scale  $\sigma_{I}$).
The integrated suppression index $SI_{2}$ is defined as
\begin{equation}
SI_{2}=\frac{K_{I}\sigma_{I}}{K_{E}\sigma_{E}} \:\: .
\end{equation}
As is true for $SI_{1}$, $SI_{2}$ can be larger than one, indicating surround suppression
beyond the background response.

Given the validity of the rectification model (Eq. \ref{eq:fr}), we can ``derive'' the DOG model
by the substitutions $g_{E,I}\sim \int_{0}^{r_{A}}e^{-(y/\sigma_{E,I})^{2}}dy$.
Obviously, identification of the terms in the DOG model with
the actual excitatory and inhibitory inputs can be little more than symbolic.

The Ratio-of-Gaussians (ROG) model \cite{cav02} is defined by
\begin{equation}
f(r_{A})=f_{0}
+\frac{k_{c}\left(\frac{2}{\sqrt{\pi}}
\int_{0}^{r_{A}}e^{-(y/w_{c})^{2}}dy\right)^{2}}
{1+k_{s}\left(\frac{2}{\sqrt{\pi}}
\int_{0}^{r_{A}}e^{-(y/w_{s})^{2}}dy\right)^{2}}\; .
\end{equation}
In this model the response beyond the background response is assumed
to arise from a division of center activity (``excitation'',
spatial scale $w_{c}$, gain $k_{c}$) and surround activity
(``inhibition'', spatial scale $w_{s}$, gain $k_{s}$).

From the slaving of the membrane potential (Eq. \ref{eq:vmem}), it is a simple matter to ``derive''
the  ROG model from the standard rectification model. Equation (\ref{eq:vmem}) can be rewritten 
such that the numerator (N) and denominator (D) represent a rectified weighted difference
of the excitatory and inhibitory conductances, and the total conductance $g_{T}$, respectively.
The ROG model used in \cite{cav02} is then obtained by the substitutions $N,D-1\sim
\left[\int_{0}^{r_{A}}e^{-(y/w_{c,s})^{2}}dy\right]^{2}$.
As is also true here, identification of the terms in the ROG model with actual excitatory and 
inhibitory inputs, can be little more than symbolic.

\section{Results}
\subsection{Classical response properties}
Classical responses do not specifically address
size effects of the stimulus. These are response properties such as orientation tuning, spatial and temporal frequency tuning, distribution of response modulations for drifting grating stimuli (simple \& complex cells) etc..
One of our model's strong accomplishments is that it produces, with the same fixed parameters, a wide range 
of classical responses as well as the two extraclassical responses which are the focus of this paper. 

Classical response properties are important because they set the context in which extraclassical 
responses occur. First, because extraclassical responses (responses evoked from outside the classical 
receptive field) are not know to occur without sufficient stimulation of the classical
receptive field. Second, extraclassical responses, in particular, depend strongly on how the cell's 
environment in the cortex is responding.
The responses of the cells that make up this environment will display an enormous diversity to 
any particular fixed stimulus. A cell's relevant cortical environment is generally made up of cells 
with, for instance vastly different orientation, spatial and temporal tuning widths and preferences.
A reasonable response of a cell's environment is thus accomplished if a model's  classical responses 
are realistic, i.e. agree with experimental data.

Some classical response properties of the model are illustrated in Figures 2 \& 3. All plots 
are for the M0 configuration (see Methods) but the other configurations yield similar results.
The sharpness and spatial distribution of orientation tuning in the model is illustrated in 
Figure 2A. As a measure for orientation tuning we used the circular variance ($CV$),
\begin{equation}
\label{eq:cv}
CV=1-\left|\frac{\int r(\theta)\exp(2i\theta)d\theta}{\int
r(\theta)d\theta}\right| \; .
\end{equation}
Here $r(\theta)$ is the mean firing rate and $\theta$ the grating angle.
The smaller the $CV$, the sharper the orientation tuning. Cells with $CV=0$
respond at just one angle, and hence are very sharply tuned.
Cells with $CV=1$ respond identically at all angles, and hence are not tuned
for orientation.
In Figure 2A we color coded the $CV$ for all cells within the white dashed rectangle
in Figure 1 of the paper. The stimulus (drifting grating) was presented to one eye, the other
eye received no visual input (see Materials and Methods for details). Pixels 
colored black indicate cells that do not show a significant response 
under visual stimulation and are mostly cells that receive their input from the other eye.
Notice that our model cortex is filled with sharply tuned cells, moderately tuned 
cells, and untuned cells, as is the primary visual cortex of macaque.
Notice also that there is no particular spatial organization of the sharpness of
orientation tuning.

\begin{figure}[here]
\centering
\begin{minipage}{0.35\columnwidth}
\includegraphics[height=5cm,width=2.cm]{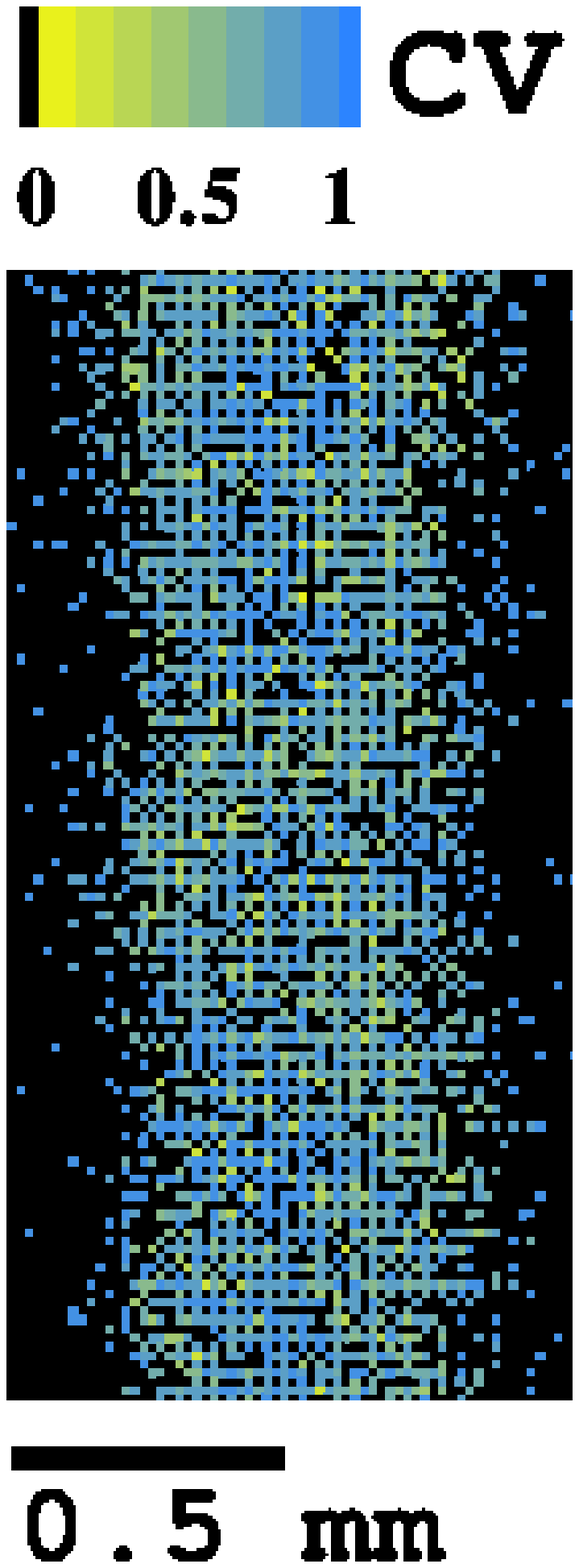}
\end{minipage}
\begin{minipage}{0.62\columnwidth}
\centering
\vspace{0.3cm}
\includegraphics[height=2.5cm,width=2.5cm]{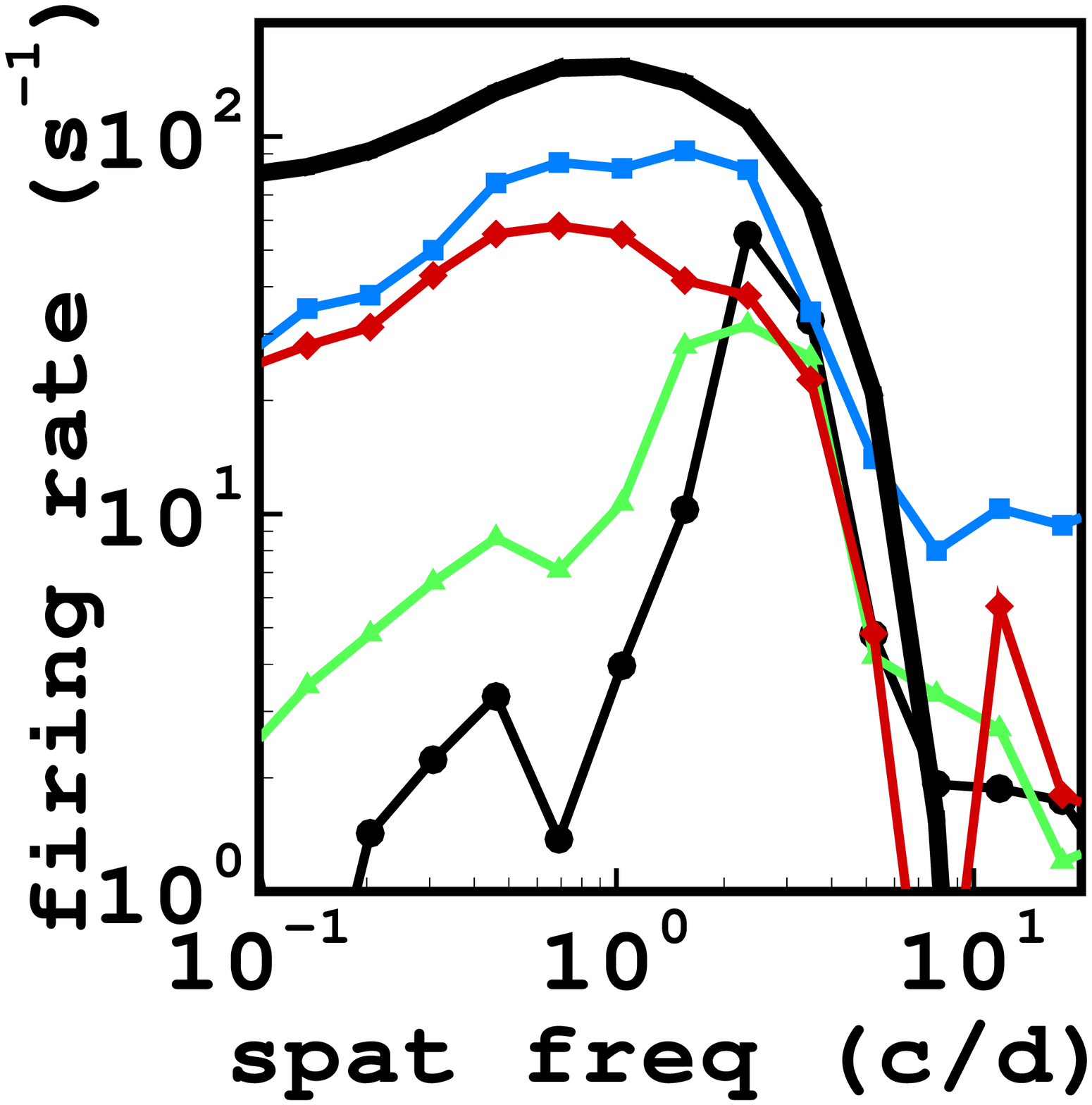}
\includegraphics[height=2.5cm,width=2.5cm]{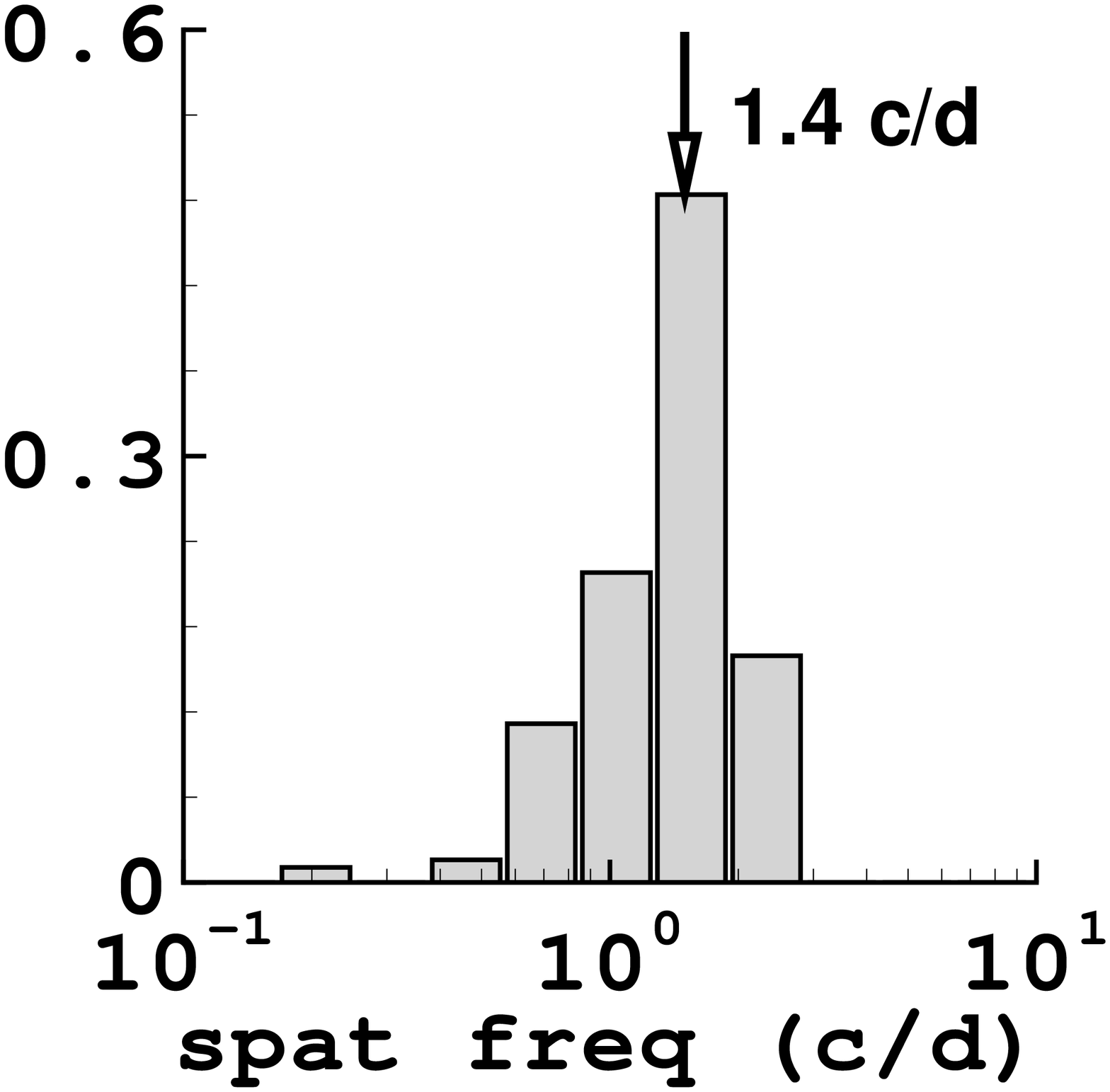}
\includegraphics[height=2.5cm,width=2.5cm]{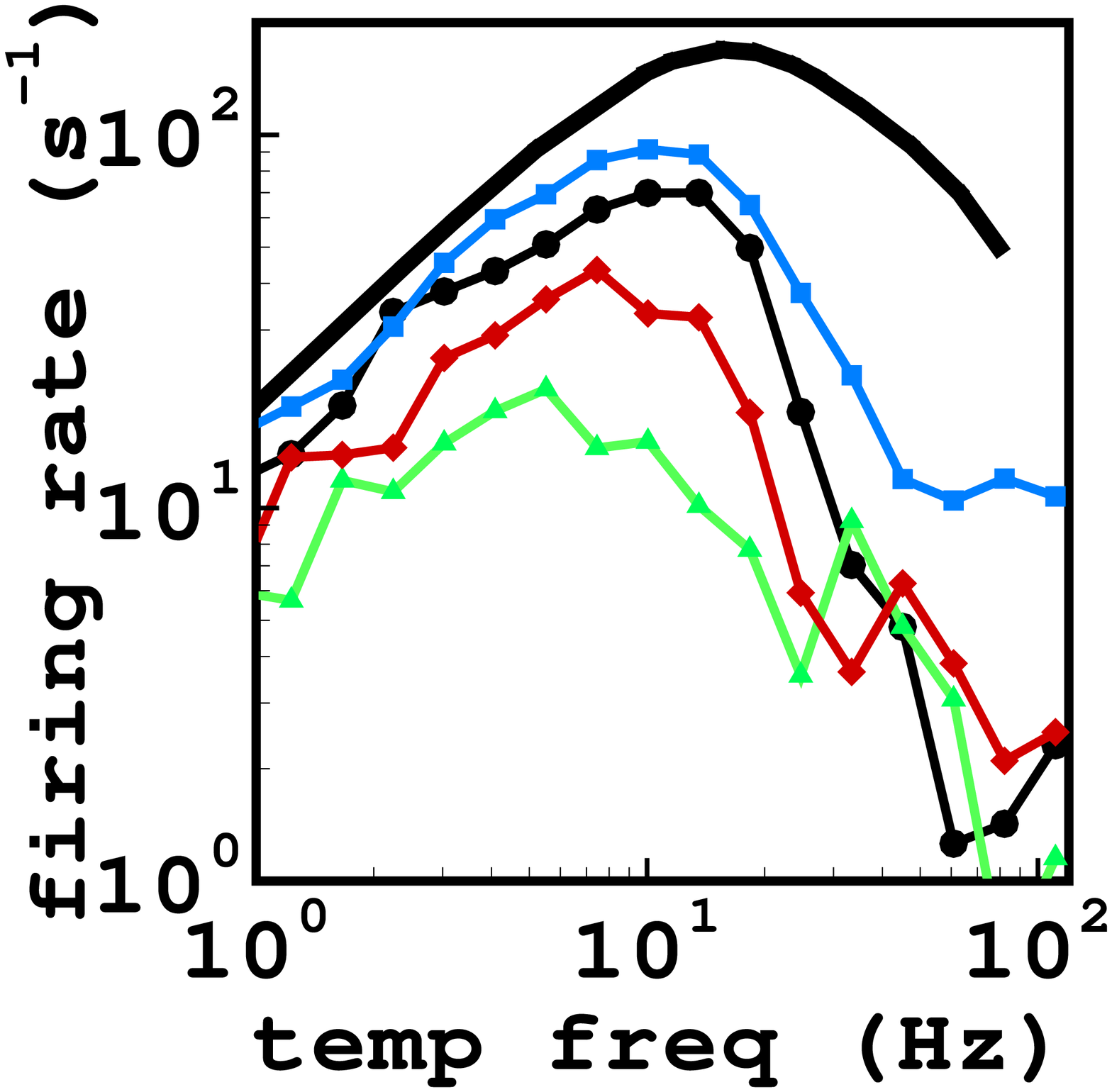}
\includegraphics[height=2.5cm,width=2.5cm]{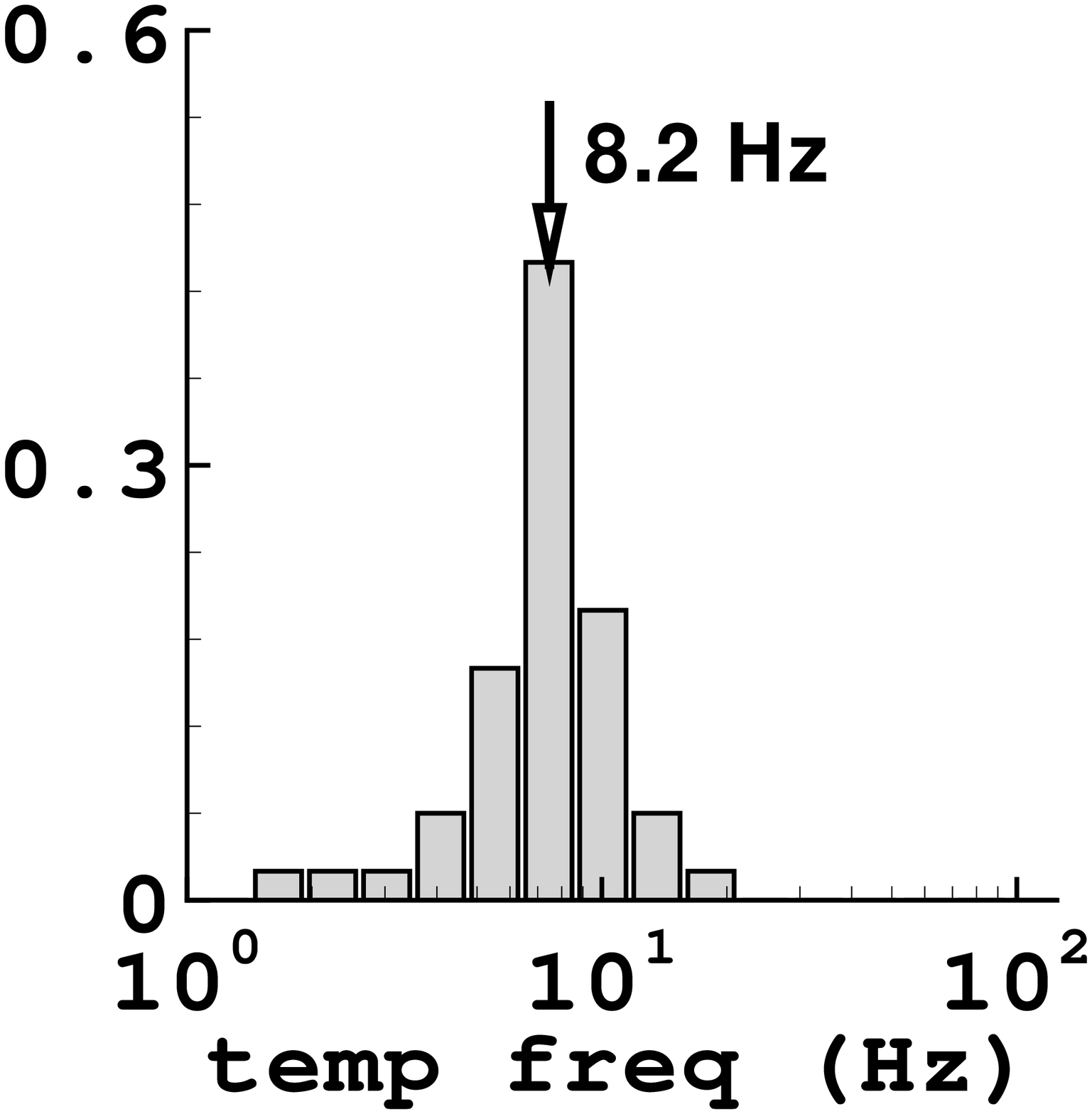}
\end{minipage}
\caption{\small Some classical response properties (selectivity and tuning) of the M0 version
of the model, other configurations yield similar results. (left) Spatial distribution of the sharpness of orientation tuning, expressed in
the circular variance (CV, color coded), for cells within the
white rectangle in Figure 1 of paper. Black pixels are cells that do not show sufficient
response under single eye stimulation.
(right) Temporal and spatial frequency tuning curves of some M0 sample cells and histograms
of preferred temporal and spatial frequencies for cells in the M0 sample set. Thick black curves refer
to the LGN cells.}
\end{figure}

Our model yields a realistic diversity in spatial and temporal frequency tuning properties.
This is illustrated in Figure 2B-E. In observing this diversity in the model it is important 
to note that all LGN cells in a particular configuration are identical by construction.
Their spatio-temporal tuning properties are indicated by the thick black curves in 
(B) and (D). Notice that, as in reality, the preferred temporal frequencies of our cortical 
cells are smaller than the preferred temporal frequency of our LGN cells.
The main reason for this is the inclusion of slow (NMDA) excitation in the model.
The diversity seen in the spatial frequency tuning of our cortical cells, particularly
in the bandwidth, is mostly cortical in origin but also partially results from the spatial diversity 
in the feedforward (LGN) connections.

Our first example of classical spatial summation properties of the model is provided
in Figure 3A, B. Shown are averaged response waveforms of spike train and membrane potential 
in response to a drifting grating in Figure 3A. These are responses of a simple and 
a complex cell in the model for several angles of the grating at their preferred spatial and temporal 
frequencies. These modulations in the spike train at the preferred angle are frequently used to classify 
simple and complex cells in V1. 
A cell is called complex whenever $F1/F0<1$, and simple otherwise, where F1 is the first harmonic of 
the response and F0 the average. The distribution of the spike train modulation index F1/F0 over our cell 
population is shown in Figure 3B (top left). Our model cortex contains about an equal amount of simple and 
complex cells.  The bimodal shape of the distribution of the spike train modulation index agrees well with
experimental data \cite{rin02}.

\begin{figure}[here]
\centering
\begin{minipage}{1\columnwidth}
\centering
(A)\hspace{2.in}(B) \\
\includegraphics[height=4.5cm,width=1.75cm]{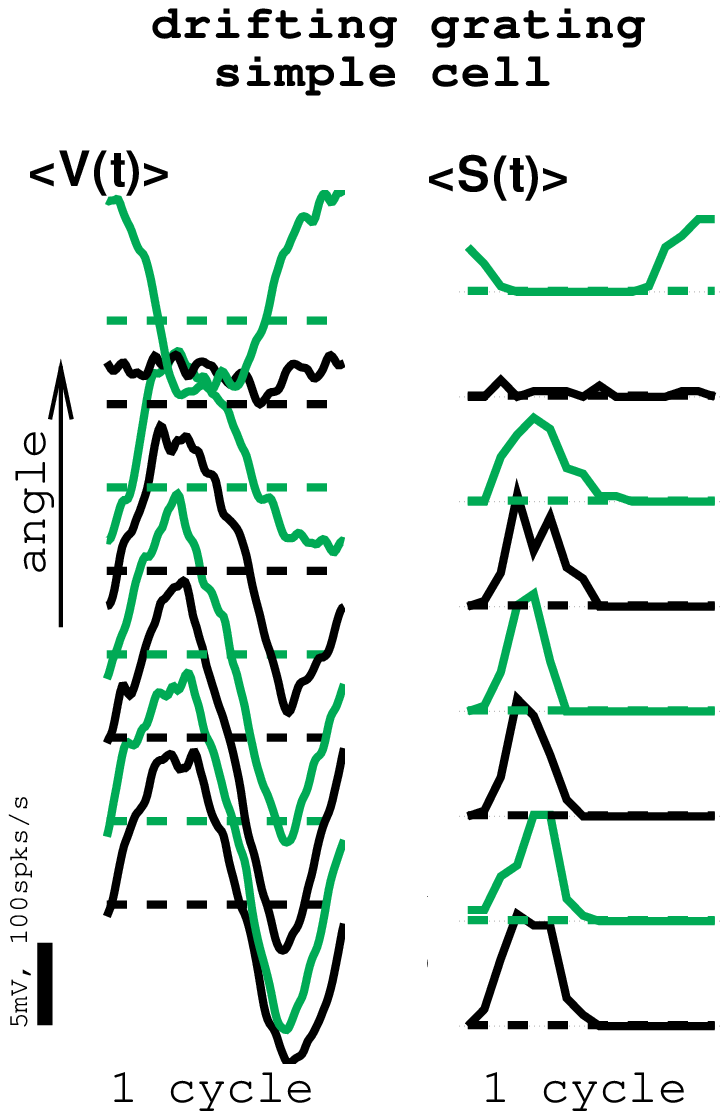}
\includegraphics[height=4.5cm,width=1.75cm]{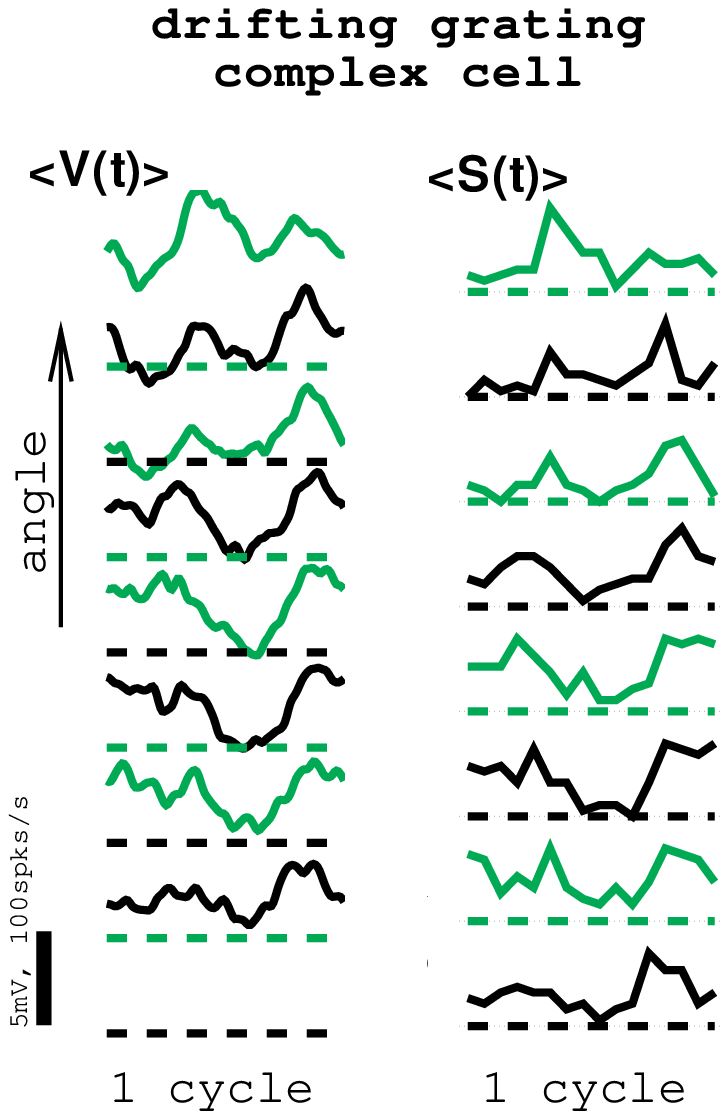}
\includegraphics[height=4cm,width=3cm]{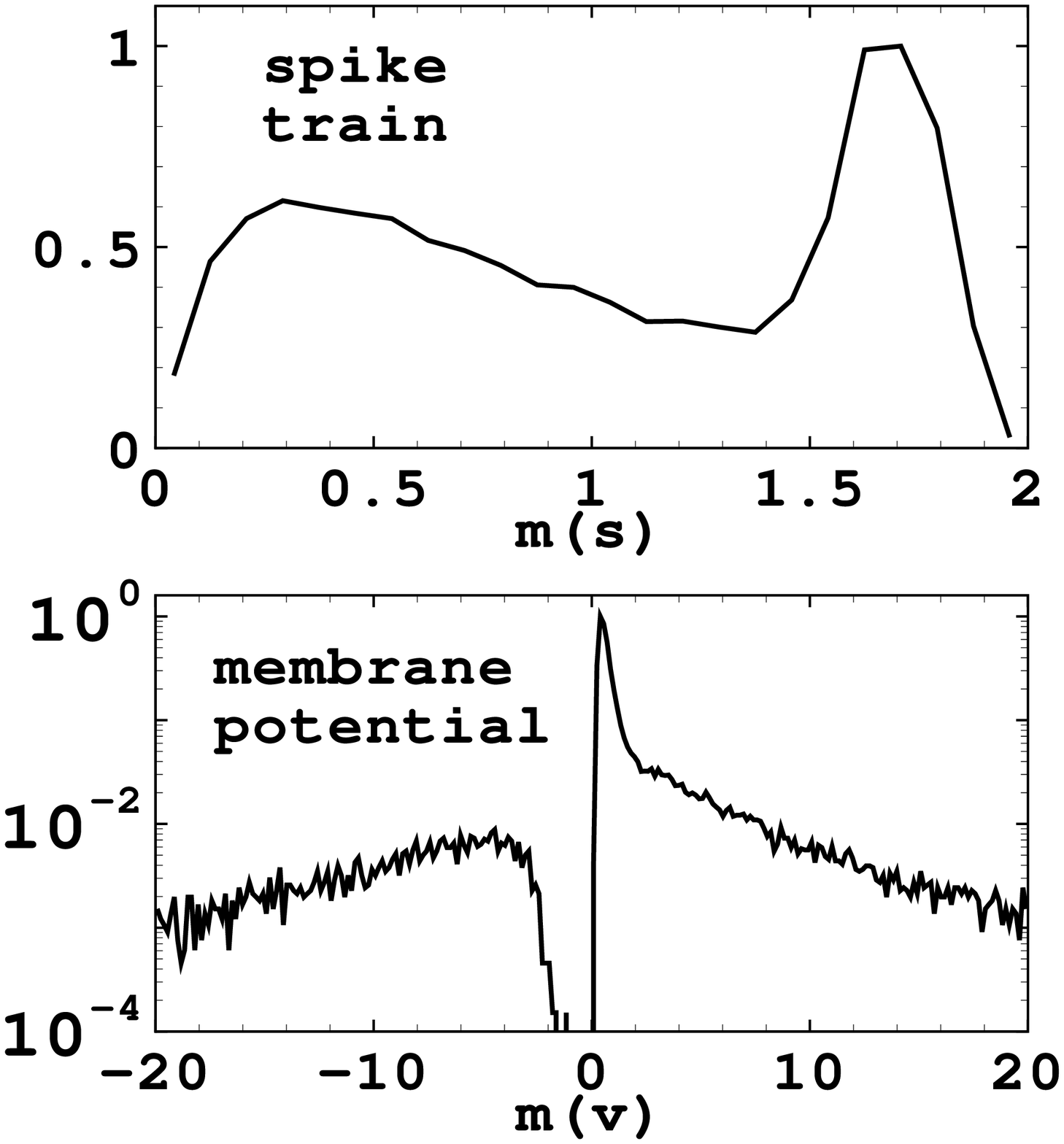}
\includegraphics[height=4cm,width=2cm]{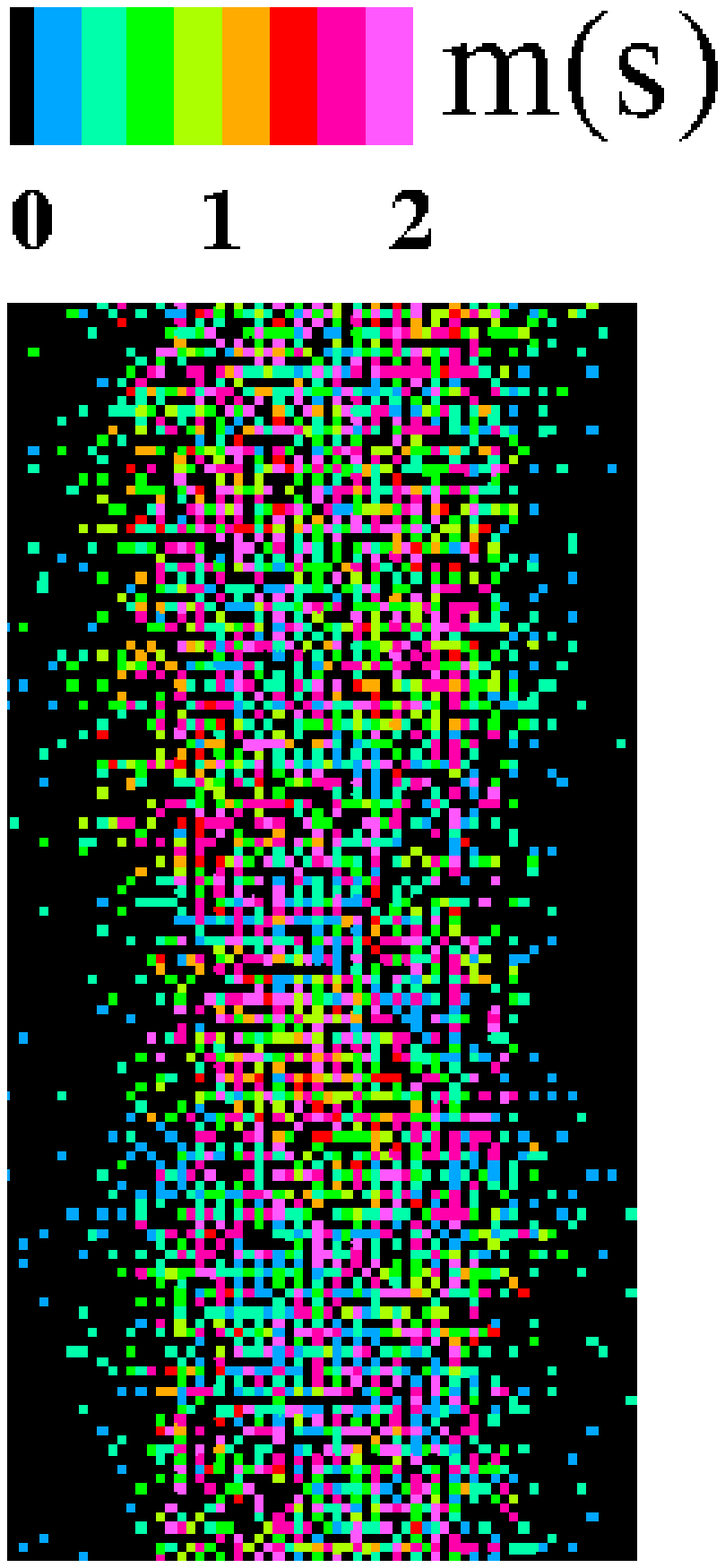}
\\
(C)\hspace{2.in}(D) \\
\includegraphics[height=4.5cm,width=1.75cm]{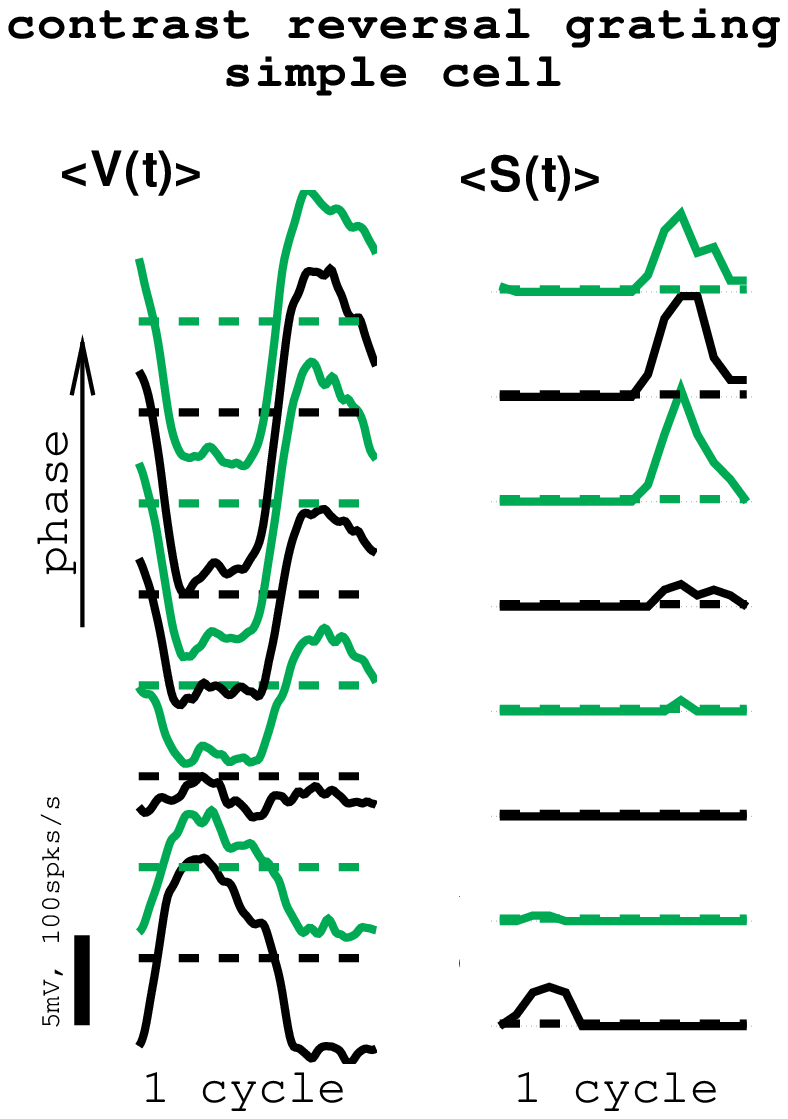}
\includegraphics[height=4.5cm,width=2cm]{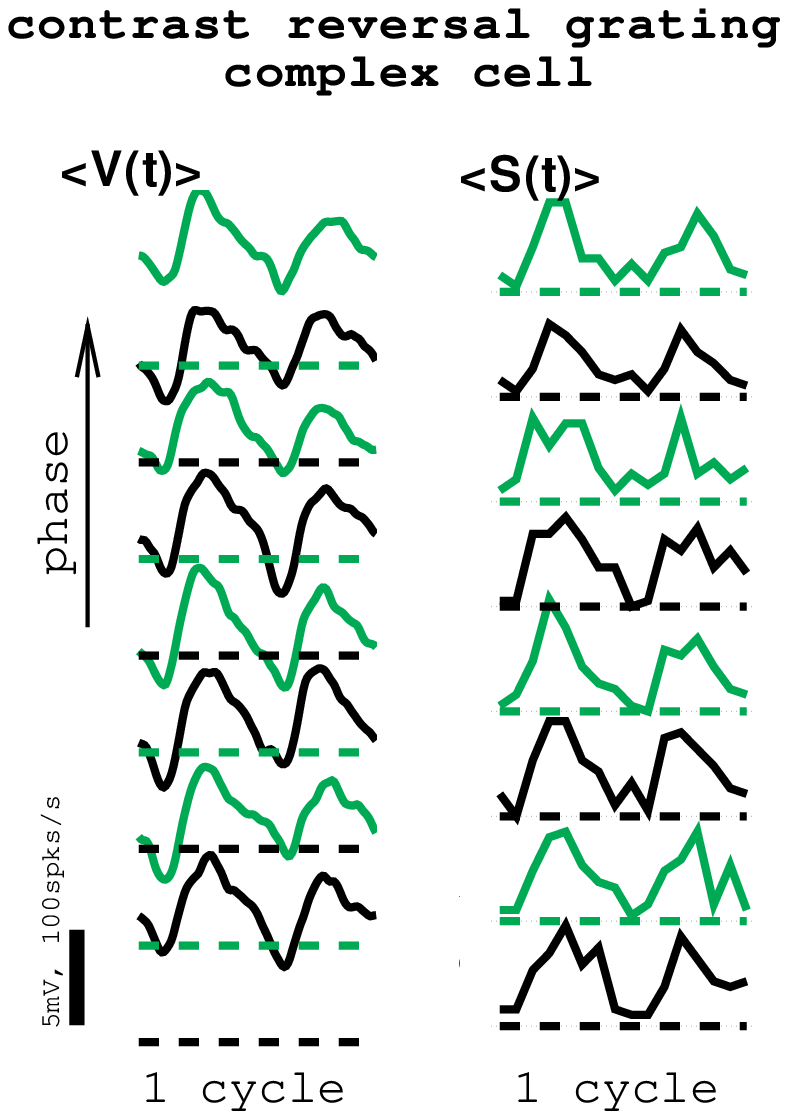}
\includegraphics[height=5cm,width=4cm]{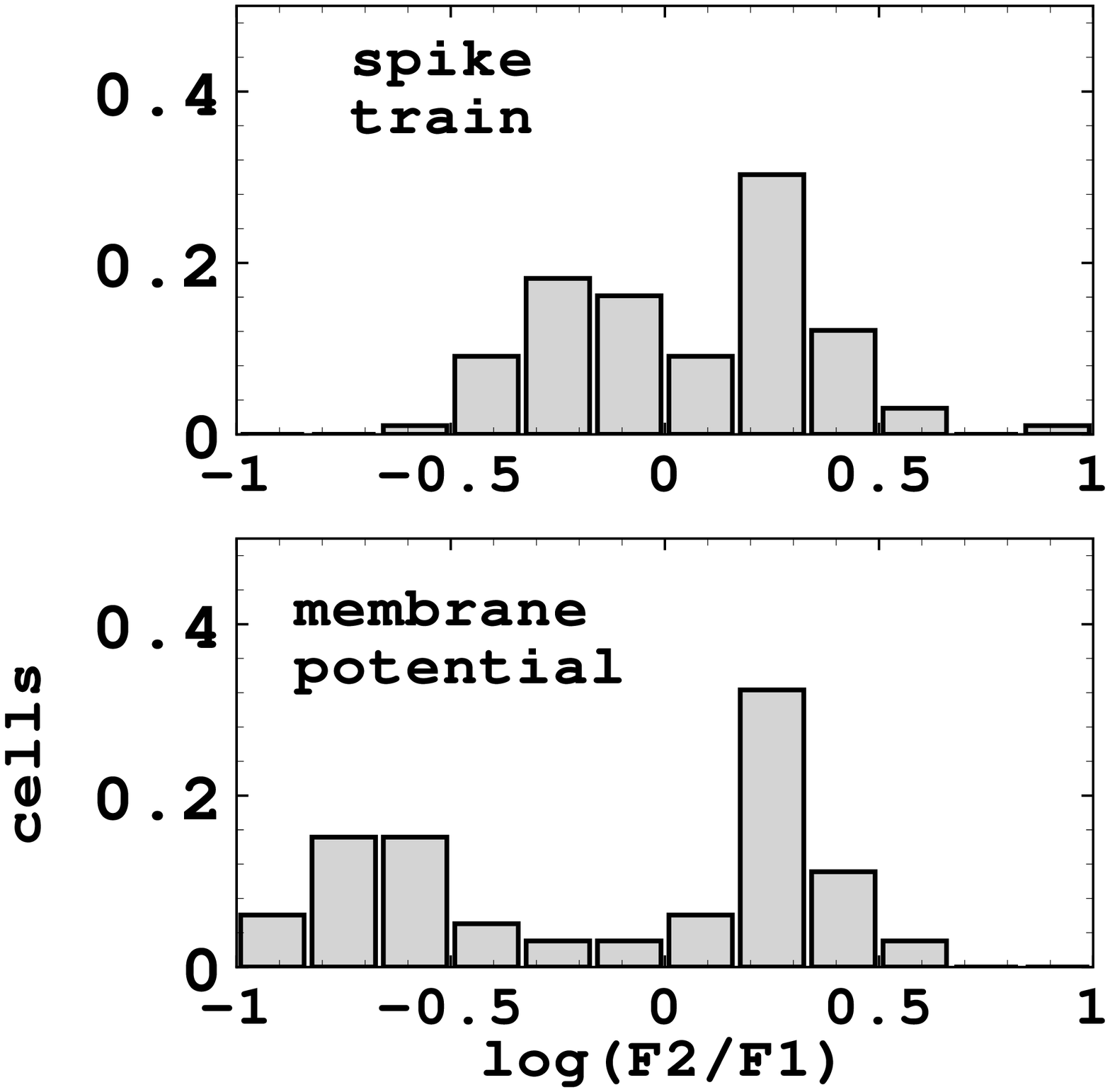}

\end{minipage}
\caption{\small Classical spatial summation properties of the P0 version
of the model, other configurations yield similar results. (A) Response waveforms for a simple 
and a complex cell in response to a drifting grating, for a number of different angles.
(B) Distributions (normalized to peak value one) of modulation index (F1/F0) in spike train 
(top left) and membrane potential (bottom left), and spatial distribution of the modulation index 
(spike train) for cells within the white rectangle in Figure 1 of the paper. Black pixels are cells that do 
not show sufficient response under single eye stimulation (right).
(C) Response waveforms for a simple and a complex cell in response to a contrast
reversal grating at the preferred angle, for a number of different spatial phases.
(D) Distributions of the phase averaged F2/F1 ratio for spike train (top) and membrane 
potential (bottom) for responses to a contrast reversal grating at the preferred angle.}
\end{figure}

It is easy to understand how the diversity in response modulations occurs in our model. The modulations
enter our model cortex via the LGN input received by 30\% of the cortical cells. The phases of these
LGN inputs into the different cortical cells vary randomly on $[0,2\pi]$. This is so because of the receptive 
field off-sets of the clusters of LGN cells connected to different cortical cells, the difference in 
shape (symmetry) of the clusters themselves, and the diversity in temporal delays in the LGN kernels.
A cell receives input from many other cells, thus a cell's excitatory and inhibitory inputs will show 
stronger or weaker modulations dependent on its specific environment in the network and whether
or not it receives LGN input. 
Interplay between the strengths and phases of the modulations in these inputs and cell specific 
parameters ultimately determine the modulation in the cell's spike and membrane potential response. 
Most cells that receive LGN input are simple cells (80\% in our model) and most cells that do not 
receive LGN input are complex (70\% in our model).
As mentioned earlier, strength parameters have been set so that the distribution of these modulations in 
the spike responses is in agreement with experimental data.

The distribution of modulations in the membrane potential (with respect to a blank stimulus) is 
shown in Figure 3B (bottom left).
Notice that the bimodality present for spike train modulations is not present in the 
modulation index distribution for the membrane potential. However, our model predicts
(not shown explicitly) that the classification of simple and complex cells can equally 
well be made in terms of the membrane potential modulation index distribution, the two modes 
in this case being its ``core'' (complex cells) and its ``tails'' (simple cells).
Also notice that our model predicts a ``gap'' in the distribution at small negative values.
The distribution of modulations in the membrane potential has not yet been observed 
experimentally in macaque. Some data for cat has recently been published \cite{pri04} and they
do not contradict the predictions based on our model. 
The spatial distribution of the modulation index (spike train) across all cells within the white 
dashed rectangle in Figure 1 of the paper is shown in Figure 3B (right). 
The stimulus was again presented to one eye, and pixels colored black indicate cells that do not show a significant response under visual stimulation. The figure shows that simple and complex cells are randomly 
distributed across space in our model cortex, i.e. there is no particular spatial organization 
of the modulation index.

Another example of classical spatial summation properties in our model is provided 
in the remainder of Figure 3.
Averaged response waveforms of spike train and membrane potential in response to a
contrast reversal grating at the preferred angle are shown in Figure 3C. Shown are the 
responses of a simple and a complex cell in the model for several spatial phases of the 
grating.
Simple cells perform an approximately linear spatial summation, that is, their responses 
contain a dominant F1 component and the spatial phase dependence of their response waveform is 
similar to the spatial phase dependence of the stimulus.
Complex cells respond nonlinearly, their response waveform is relatively insensitive to 
spatial phase and contain a dominant (frequency doubling) F2 component.
The distribution of the phase-averaged F2/F1 for our model cells is shown in Figure 3D.
For what concerns the spike train waveforms (top), this quantity displays a weak bimodality
and its behavior for our model cells agrees with experimental data \cite{haw87}, complex 
cells having mostly $F2/F1>1$ and simple cells $F2/F1<1$.
Note that this property of our model cells follows naturally, without any parameter adjustments, 
after the strength parameters have been set to achieve essentially only orientation tuning and a 
proper distribution of response modulations in response to a drifting grating (Fig. 3B), as 
mentioned earlier (see also \cite{wie01}).

It is easy to understand how the diversity in F2/F1 occurs. As explained in 
\cite{wie01}, for a contrast reversal grating stimulus each total LGN input into a cortical 
cell has in general a dominant F1 component with a phase close to either $0$ or $\pi$, determined by 
the relative positions of the ON and OFF subfields of the corresponding cluster.
The cortical excitatory and inhibitory inputs in a cell will thus have a relative strong F2 
component since they arise from many other cells. The actual strengths of F1 and F2 components in a 
cell's excitatory and inhibitory inputs thus depends on the cell's specific environment in the 
network and on whether it receives LGN input or not. Interplay of these inputs and cell specific 
parameters determine the F2/F1 ratio in the cell's spike and membrane potential response.
Clearly, most cells that receive LGN input (simple) will have $F2/F1<1$ and most cells that 
do not receive LGN input (complex) will have $F2/F1>1$.

No experimental data is available for the distribution of F2/F1 of the membrane potential
waveforms. Our model's prediction is shown in Figure 3D, bottom. Our model predicts that, 
quite contrary to the situation for the modulation index F1/F0, the (weak) bimodality of the 
distribution of F2/F1 for spike waveforms is not eliminated, but rather, becomes 
more pronounced in the F2/F1 distribution for membrane potential waveforms.  
This, in fact, can be understood quite simply from a standard rectification model, in 
which the membrane potential waveforms are subjected to a threshold to give the 
spike waveforms. For complex cells, both the membrane potential and spike responses will 
contain a strong F2 component. Hence in this case practically all of the membrane potential 
waveform will be above threshold, so that evaluation of F2/F1 will yield about the 
same result for spike waveforms as for membrane potential waveforms. This is also apparent in 
Figure 3D: the $F2/F1>1$ section of the two distributions (in top and bottom panels) is similar. 
For simple cells, the membrane potential and spike responses will contain a dominant F1 
component, and for both responses about an equally small F2 component. Because of the 
rectification, the F1 component in the membrane potential waveform is substantially reduced
in the spike waveform. Hence, F2/F1 will turn out substantially smaller when evaluated for 
membrane potential responses. This is again apparent in Figure 3D: the $F2/F1<1$ (simple cells) 
section of the distribution for spike trains (top) is shifted to the left in the F2/F1
distribution for membrane potential waveforms (bottom).

\subsection{Extraclassical spatial summation}
In this section we summarize the extraclassical results for our model and compare them with experimental 
data.
Quite contrary to classical response properties, the two extraclassical responses we discuss in this paper are, as in experimental data, not substantially different for simple and complex cells and results that follow are not type specific.
An example of the surround suppression and contrast dependent receptive field size observed in 
our model is shown in Figure 4A, for a cell from the para-foveal $4C\alpha$ model (M0) (see Methods).
Shown are responses for both firing rate and membrane potential, at high (solid) and low (dashed) contrast. 

Distributions of receptive field and surround sizes for the $4C\beta$, $10^{o}$-eccentricity model (P10)
are shown in Figure 4B \& C. The distributions for the other model configurations are given in 
Supplementary Information. Receptive field sizes and surround sizes in our model show excellent 
agreement with experimental data\cite{sce01,cav02}. This is true for the mean values, for the diversity,
as well as for their dependence on eccentricity from para-foveal to $10^{o}$ and $30^{o}$ eccentricity\cite{cav02}.

The distribution of surround suppression and receptive field growth for the M0 configuration 
of our model is given in Figure 4D \& E.
In agreement with experimental data, the shape of the distribution of the suppression index $SI_{1}$ 
(Methods) is skewed to low suppression\cite{cav02}. (Cells without surround suppression have $SI_{1}=0$, 
cells with fully suppressed response for large stimuli have $SI_{1}=1$.)
We also see a slight increase of the average suppression for low contrast\cite{sce99,cav02}.
The average suppression index (all eccentricities, see Methods) is $SI_{1}\sim 0.2$ and this is about 
half of what is observed experimentally \cite{cav02}.
The receptive field and surround growths (Fig. 4E) are expressed as ratios, $r_{-}/r_{+}$ and $R_{-}/R_{+}$
respectively.
We observe an average growth by about a factor of two in both receptive
field size ($\overline{r_{-}/r_{+}}\sim 2$) and surround size ($\overline{R_{-}/R_{+}}\sim 2$).
This receptive field growth is a little less than what is observed in experiments \cite{kap99,cav02}.

We fitted our data with the Difference-of-Gaussians (DOG) and Ratio-of-Gaussians (ROG)\cite{sce99,cav02}
models (Supplementary Information).
We obtain for the integrated suppression index $SI_{2}\sim 0.4$.
Average growth ratios for the excitatory space constant is 
$\overline{\sigma^{-}_{E}/\sigma^{+}_{E}}\sim 1.5$ (both DOG and ROG, all eccentricities).
The suppression index and growth ratios are again less than what is seen 
experimentally (0.6 and 2.3 respectively\cite{sce99}).
In agreement with experimental studies\cite{sce99,cav02} there is no systematic dependence 
of suppression on contrast in either index $SI_{1}$ or $SI_{2}$. 
This observation is illustrated in Figure 5A.

All of the above findings are based on spike responses. Membrane potential responses yield qualitatively 
similar results, but, due to the spike threshold, suppression in the membrane potential is systematically smaller. This is illustrated in Figure 5B. The same observation has also been made experimentally 
in cat\cite{and01}.

A more extensive summary of our model data for different eccentricities and including receptive
field sizes and surround sizes is given in Supplementary Information.

\subsection{Mechanisms of surround suppression}
The DOG and ROG models are phenomenological models and do not provide much insight into the neural 
mechanisms of the phenomena. Both models miss an essential feature of the excitatory 
and inhibitory inputs, which is that these inputs generally themselves show surround suppression\cite{and01}. 
In our model we similarly observe a significant suppression in both conductances, an example is shown
in Figure 5C. This cell shows that, unlike suggested by the DOG and ROG picture, surround 
suppression in the spike response takes place entirely in the region of decreasing synaptic inputs 
(conductances). We can say that the surround suppression of this cell is caused by a 
decrease of excitation, since the decrease of inhibition could not by itself suppress the 
cell's response. This cell is not atypical in our model and the above scenario is indeed how surround
suppression works in about 50\% of the cells.

\begin{figure}[here]
\label{examples}
\centering
\begin{minipage}{1\columnwidth}
\centering
(A) \\
\includegraphics[height=3.cm,width=4.5cm]{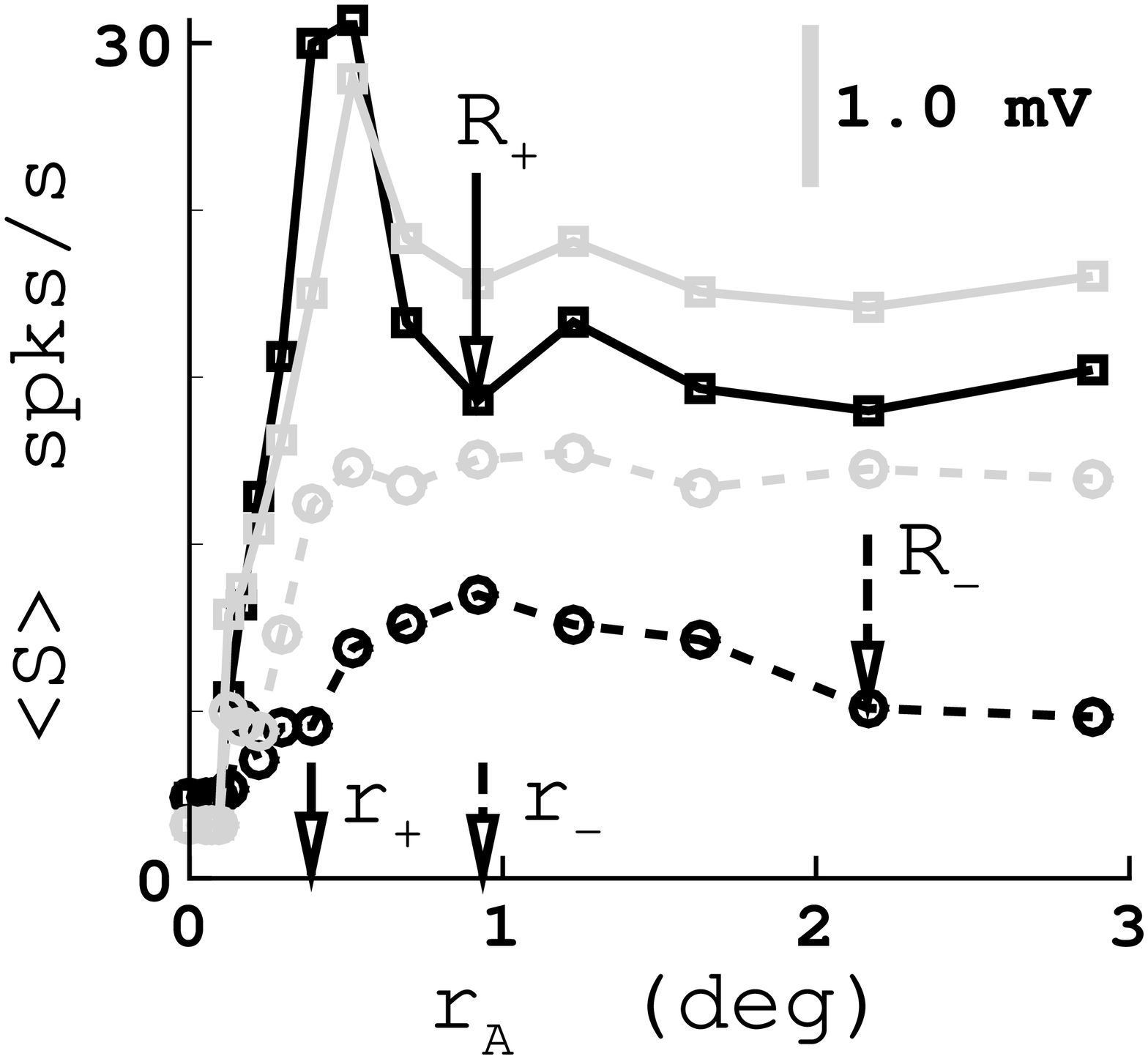}
\end{minipage}
\begin{minipage}{1\columnwidth}
\centering
(B)\hspace{0.35\columnwidth}(C) \\
\includegraphics[height=2.5cm,width=4cm]{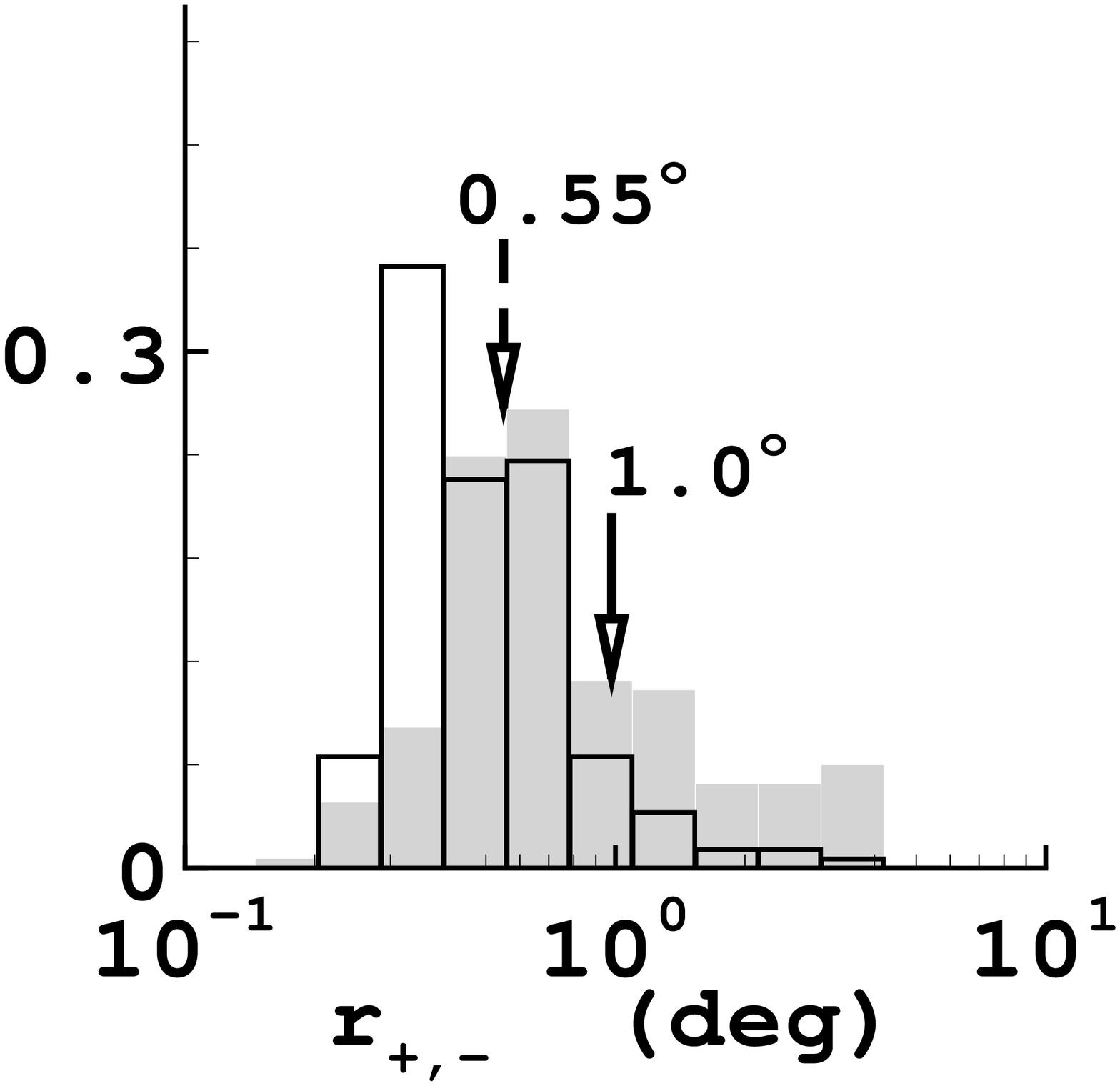}
\includegraphics[height=2.5cm,width=4cm]{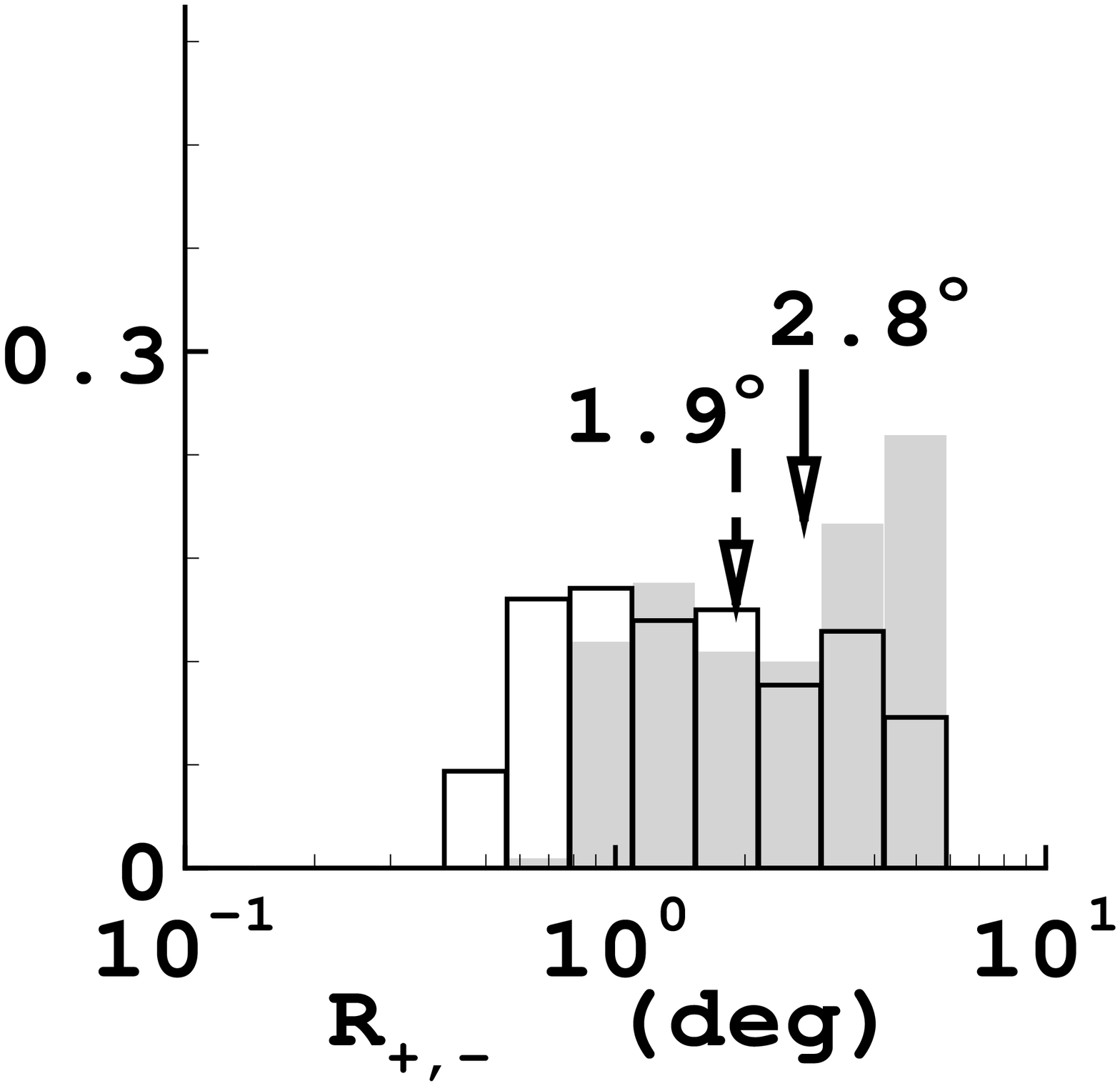}
\end{minipage}
\begin{minipage}{1\columnwidth}
\centering
(D)\hspace{0.35\columnwidth}(E) \\
\includegraphics[height=2.5cm,width=4cm]{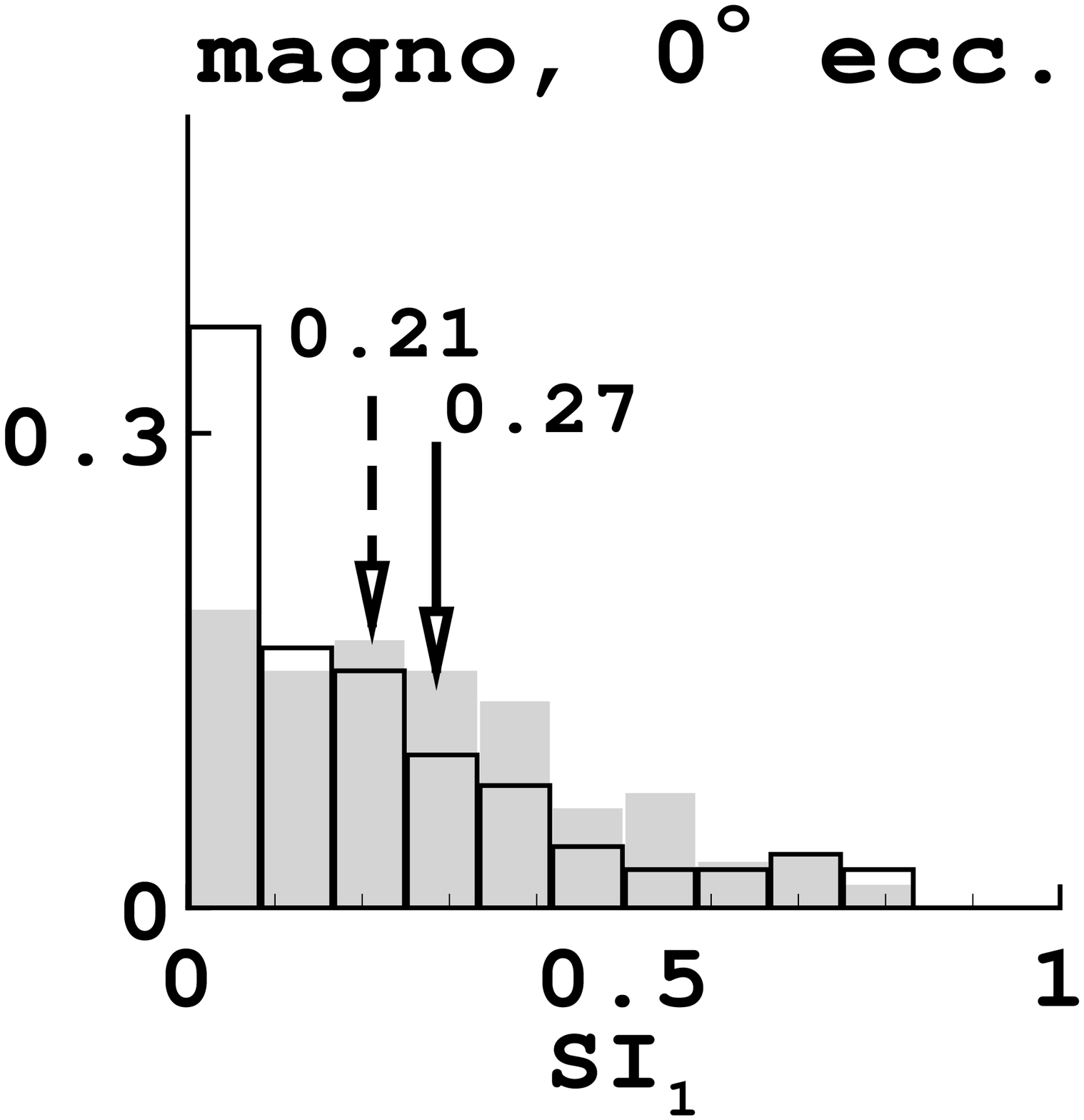}
\includegraphics[height=2.5cm,width=4cm]{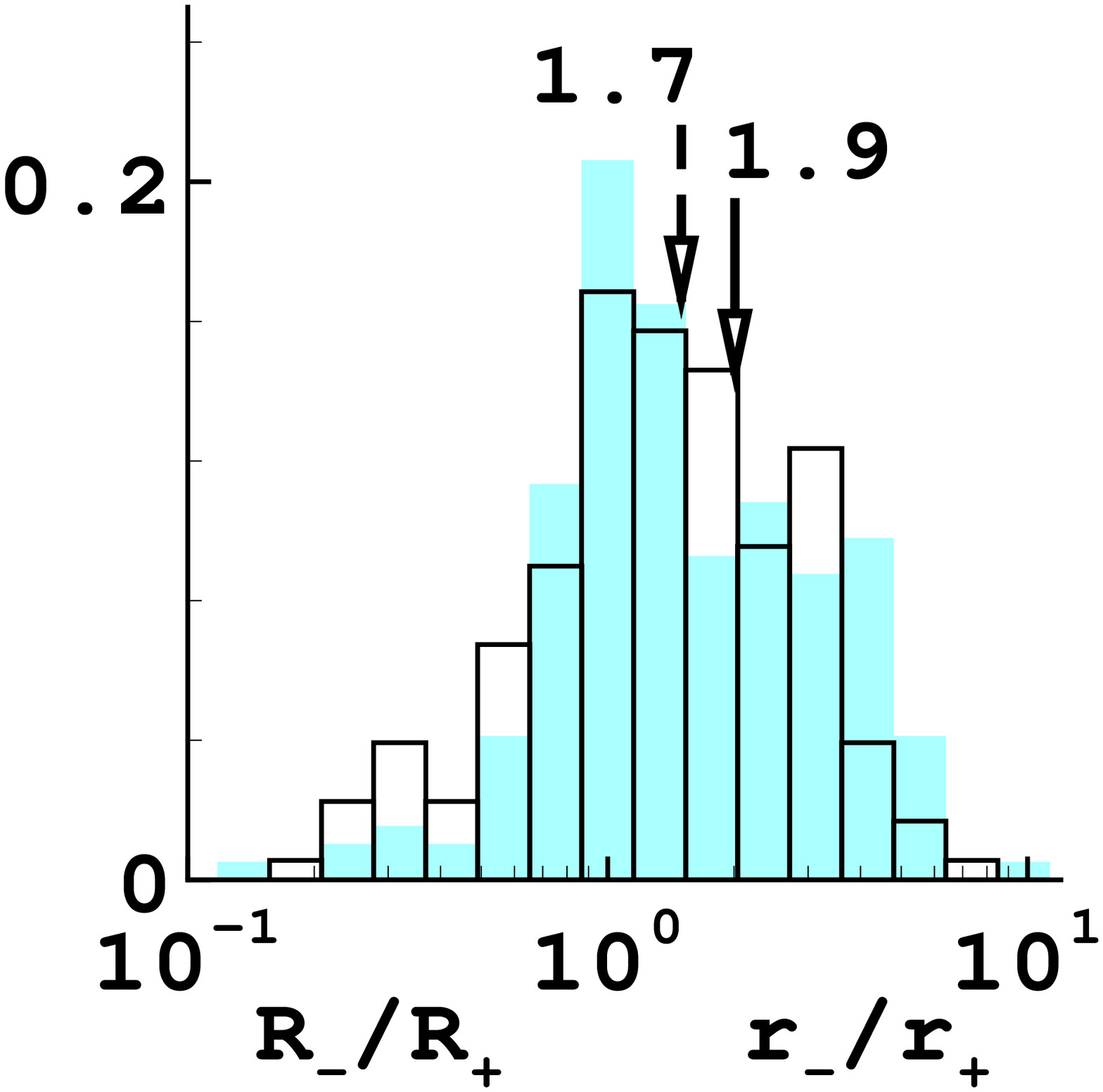}
\end{minipage}
\caption{\small Summary of extraclassical spatial summation in the model. (A) Response as function of 
aperture sized for a cell from the M0 model (see Methods). Shown are firing rate (black) and membrane potential (gray)
  for high (squares) and low (circles) contrast. Standard errors are negligibly small.  
  (B \& C) Receptive field and surround sizes for the P10 model at high (unfilled) and low
  (shaded) contrast. The diversity of responses produced by the model is similar to what is seen in  
  experimental data\cite{sce01,cav02}. 
  (D) Distribution for the M0 model of the suppression index $SI_1$ at high
  (unfilled) and low (green shaded) contrast. 
  All suppression is exclusively due to short-range cortical connectivity. (E) Distributions for the 
  M0 model of the ratios of the receptive field and surround sizes at low and high contrast, $r_{-}/r_{+}$
  (blue shaded) and $R_{-}/R_{+}$ (unfilled). (Wilcoxon test on ratio larger than unity: $p<0.001$ 
  for both receptive field and surround growth).
  For a more complete summary of our model data see Supplementary Information.}
\end{figure}

Analysis of the surround suppression in our model is based on the fact that the average
membrane potential $\left<v_{k}(t,r_{A})\right>$ and instantaneous firing rate
$\left<{\cal S}_{k}(t,r_{A})\right>$
(of the $k$-th neuron) are well-approximated by\cite{wie01}
\begin{equation}
\label{eq:vmem}
\left<v_{k}(t,r_{A})\right>\; \approx\; V_{k} \; \equiv \;
\frac{\left<I_{D,k}\right>}{\left<g_{T,k}\right>}\; ,
\end{equation}
and
\begin{equation}
\label{eq:fr}
\left<{\cal S}_{k}(t,r_{A})\right>\; \approx\; f_{k}\; \equiv \; \delta_{k}\left[ \left<I_{D,k}\right>
-\left<g_{T,k}\right>-\Delta_{k}\; \right]_{+} \; .
\end{equation}
Here $[x]_{+}=x$ if $x\geq 0$ and $[x]_{+}=0$ if $x\leq 0$, and for most cells good approximations
are obtained with a gain $\delta_{k}$ and threshold $\Delta_{k}$ that do not depend on the 
aperture radius $r_{A}$ nor time.
The total conductance $g_{T,k}$ and difference current $I_{D,k}$ are given by
\begin{equation}
\label{eq:gtot}
g_{T,k}=g_{L} + g_{E,k} + g_{I,k}
\end{equation}
\begin{equation}
\label{eq:diffc}
I_{D,k} = g_{E,k}\;V_{E} - g_{I,k}\;\left| V_{I}\right| \; .
\end{equation}
Equations (\ref{eq:vmem}) and (\ref{eq:fr}) allow us to base our analysis directly on the 
(cycle-trial averaged) conductances as a function of the aperture radius $r_{A}$ and time.
In what follows we drop the averaging notation $\left<\cdot \right>$, assuming it unless stated otherwise.
Given Equations (\ref{eq:vmem}) and (\ref{eq:fr}) there are three ways that surround 
suppression of spike train and membrane potential could arise, namely (A) 
$\partial g_{E,k}/\partial r_{A} \geq 0$
and $\partial g_{I,k}/\partial r_{A} > 0$, (B) $\partial g_{E,k}/\partial r_{A} < 0$
and $\partial g_{I,k}/\partial r_{A} \leq 0$, or (C) $\partial g_{E,k}/\partial r_{A} < 0$
and $\partial g_{I,k}/\partial r_{A} > 0$.
In other words, surround suppression is caused by (A) an increase in the inhibitory conductance, 
or (B) a decrease in the excitatory conductance, or (C) both (A) and (B) simultaneously.

\begin{figure}[here]
\centering
\begin{minipage}{1\columnwidth}
\centering
(A)\hspace{2.75in}(B) \\
\includegraphics[height=3cm,width=4cm]{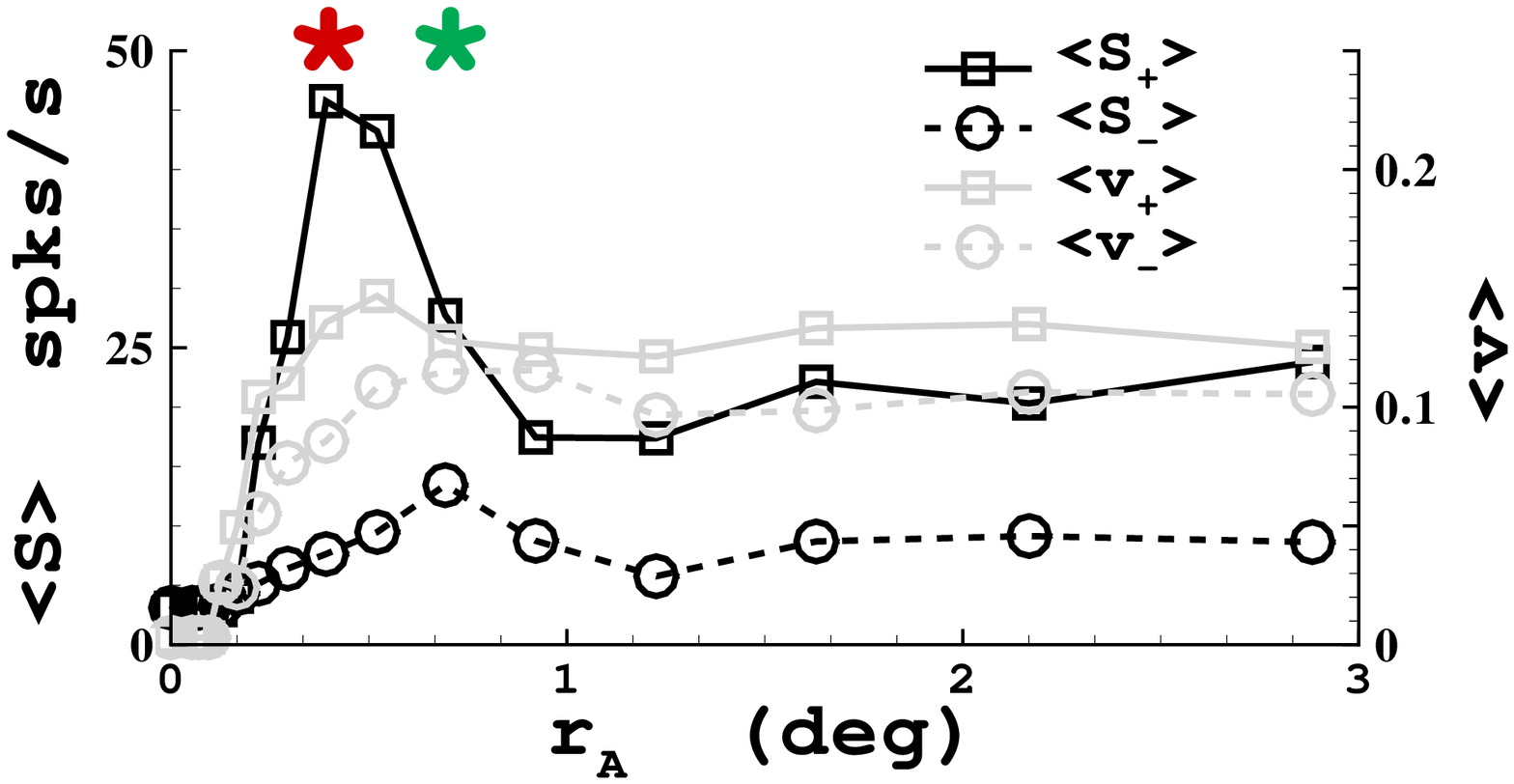}
\includegraphics[height=3cm,width=4cm]{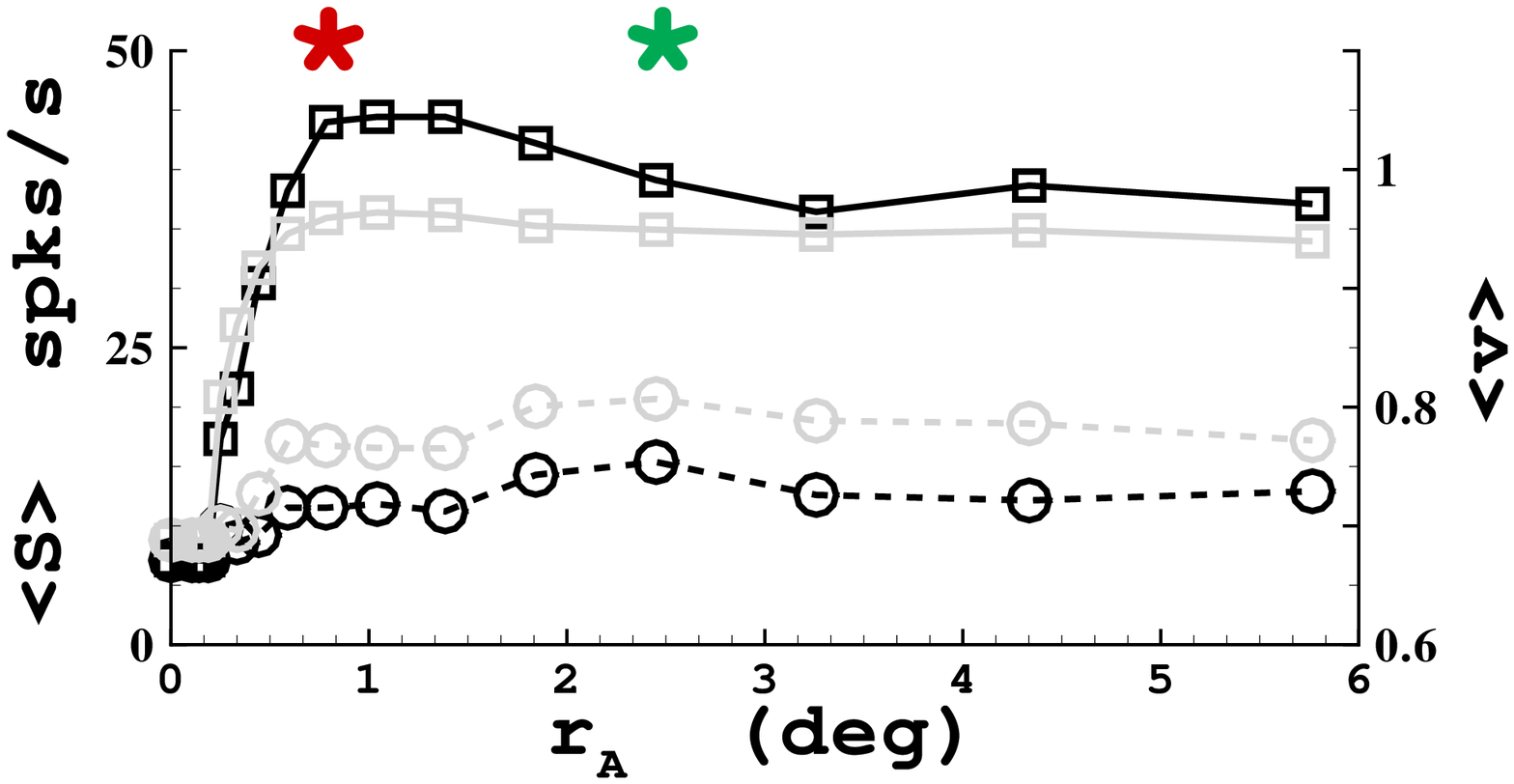}
 \\
(C)\hspace{2.7575in}(D)\\
\includegraphics[height=3cm,width=4cm]{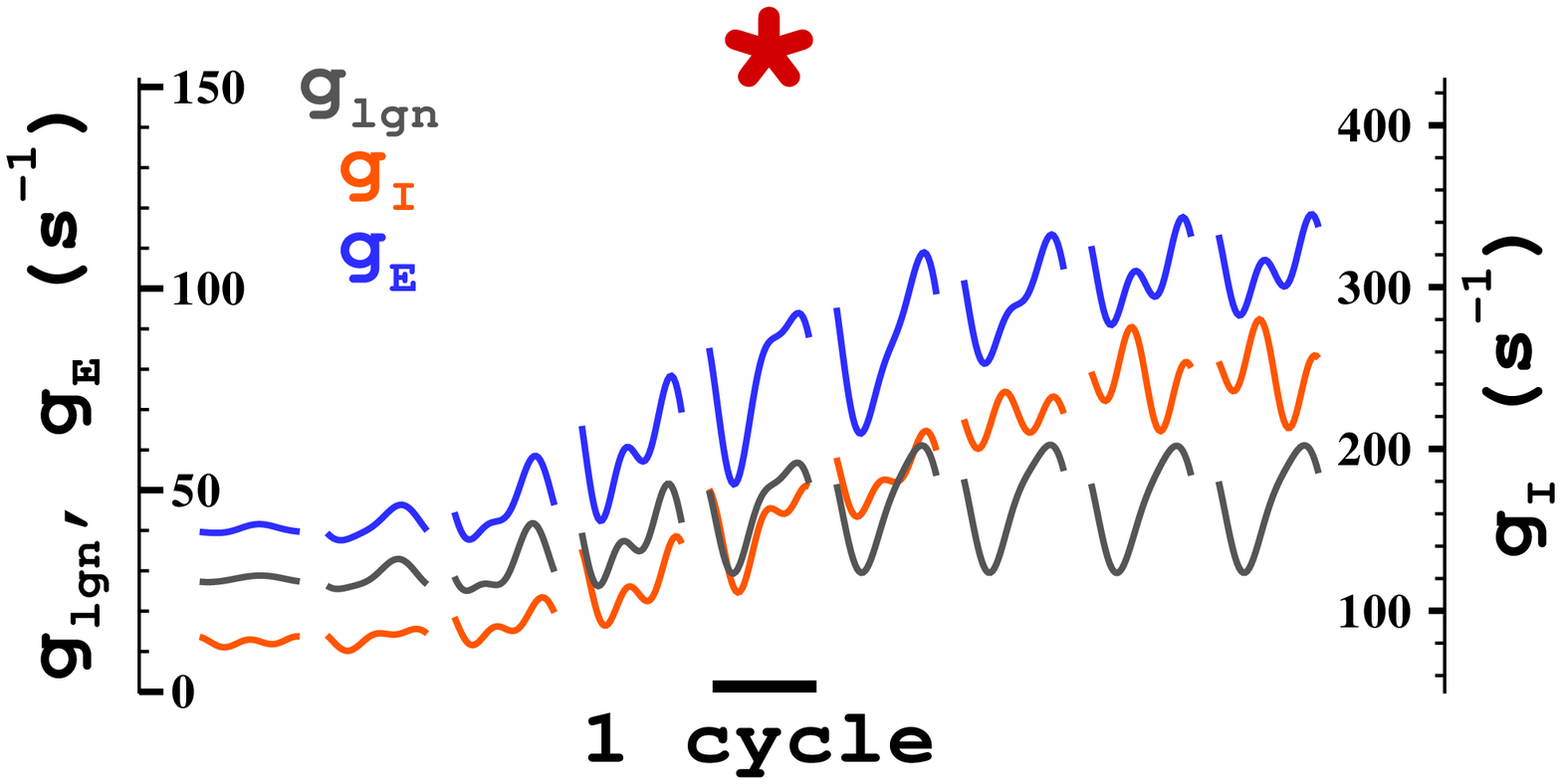}
\includegraphics[height=3cm,width=4cm]{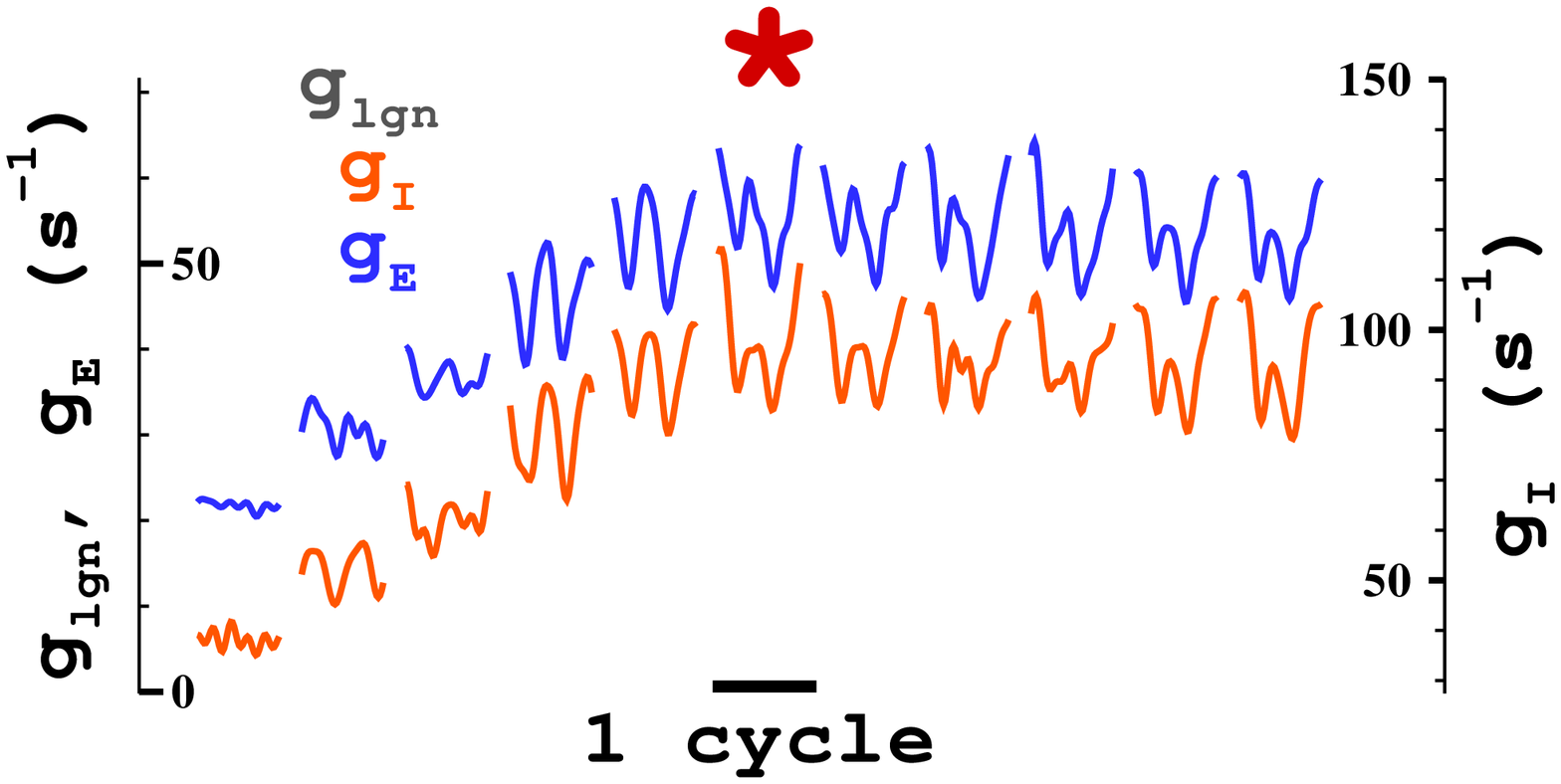}
 \\
(E)\hspace{2.7575in}(F)\\
\includegraphics[height=3cm,width=4cm]{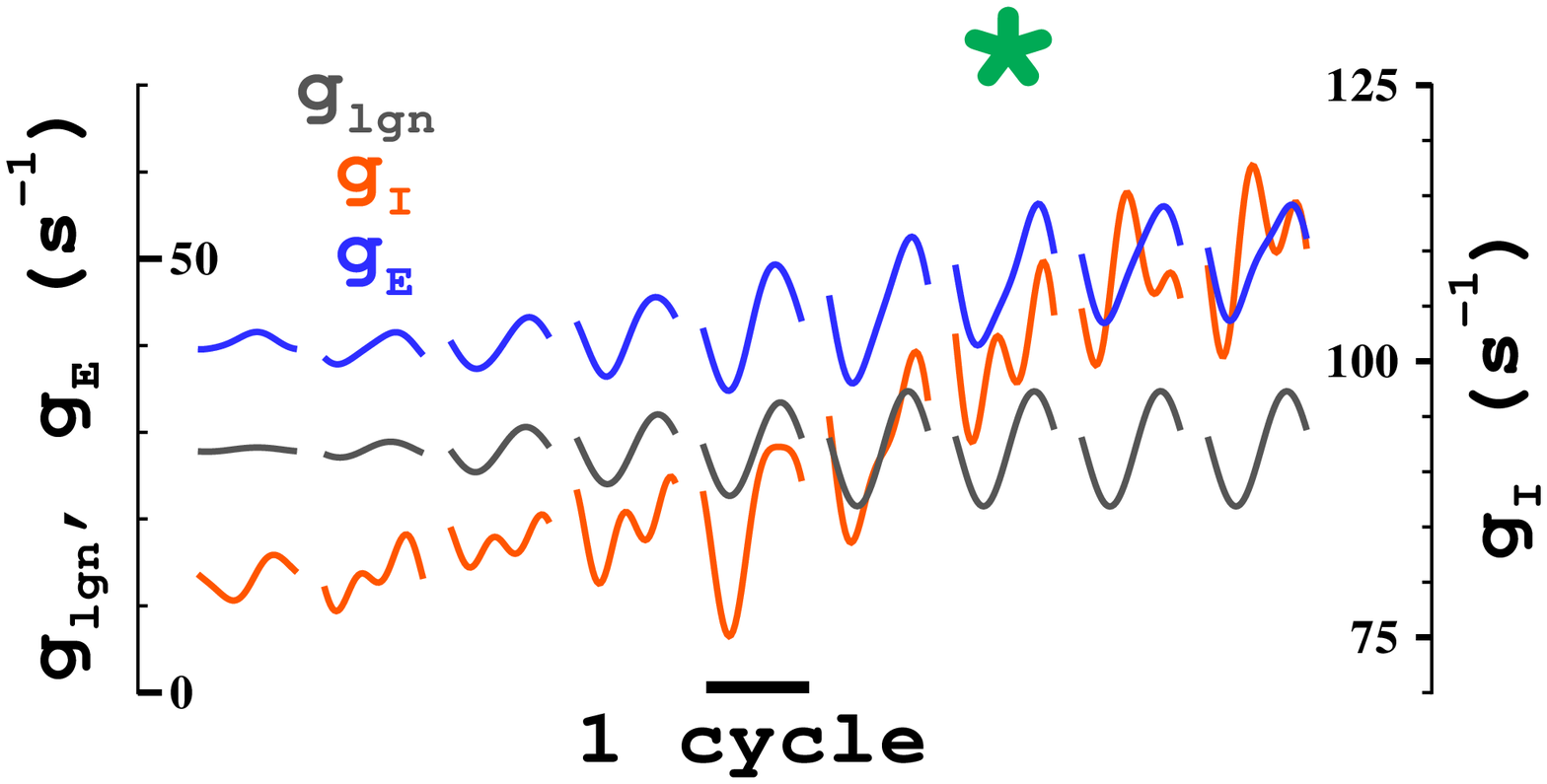}
\includegraphics[height=3cm,width=4cm]{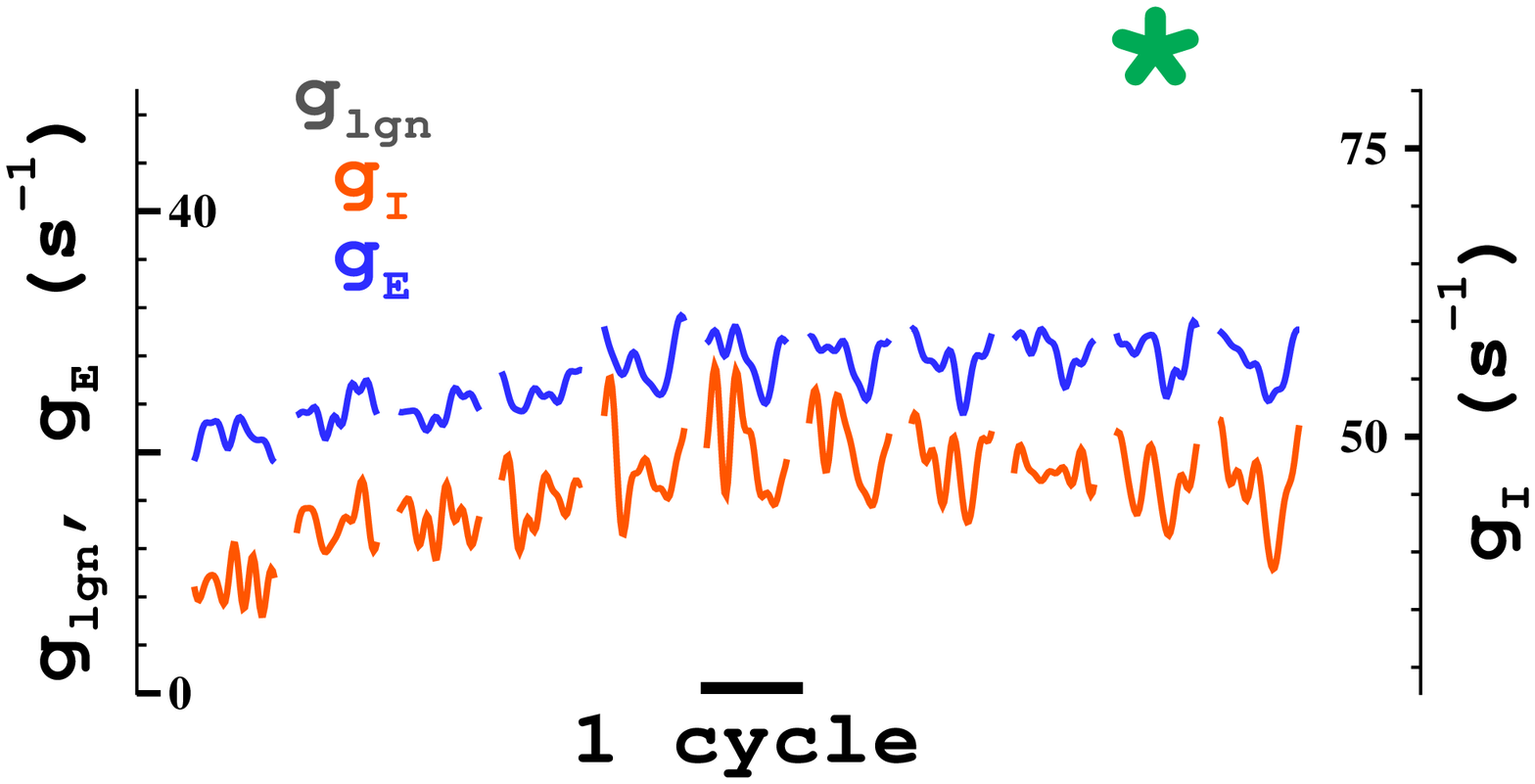}
\end{minipage}
\caption{\small Two example cells, an M0 simple cell which receives LGN input (left)
  and an M10 complex cell which does not receive LGN input (right).
  (A \& B) Responses as function of aperture size. Mean responses are plotted for the complex cell,
  first harmonic for the simple cell. Apertures of maximum of responses (i.e. receptive field sizes) 
  are indicated with asterisks (red$=$high contrast, green$=$low contrast). 
  (C \& D) Conductances for high contrast at apertures near the maximum responses. 
  (E \& F) Conductances for low contrast at apertures near the maximum responses.  
  Panels C-F each consist of seven sub-panels giving the cycle-trial averaged conductances as function
  of time (relative to cycle) and aperture size. Asterisks indicate corresponding apertures of 
  maximum response in A-B.}
\end{figure}

Examples of this analysis for a (simple) cell receiving LGN input and a (complex) cell that 
does not receive LGN input are given in Figure 6. The cycle-trial averaged conductances for
apertures around the aperture of maximum response (marked by an asterisk) are shown in Figure 6C-F.
For example, by comparing the conductances for aperture ``asterisk'' and the aperture for which the 
suppression is completed, we see that at high contrast the suppression mechanism for the simple cell 
is (A) and for the complex cell it is (B). At low contrast the suppression mechanisms are (C) and (B) respectively.

We observe all three mechanisms A, B, and C in our model. 
Typically, different mechanisms act sequentially as the aperture size $r_{A}$ 
increases from receptive field size $r$ to surround size $R$, while in some cases we find that 
different mechanisms are active during different times in the stimulus cycle. 

\begin{figure}[here]
\label{examples}
\centering
\begin{minipage}{1\columnwidth}
\centering
(A)\hspace{0.25\columnwidth}(B)\hspace{0.25\columnwidth}(C) \\
\includegraphics[height=3cm,width=3cm]{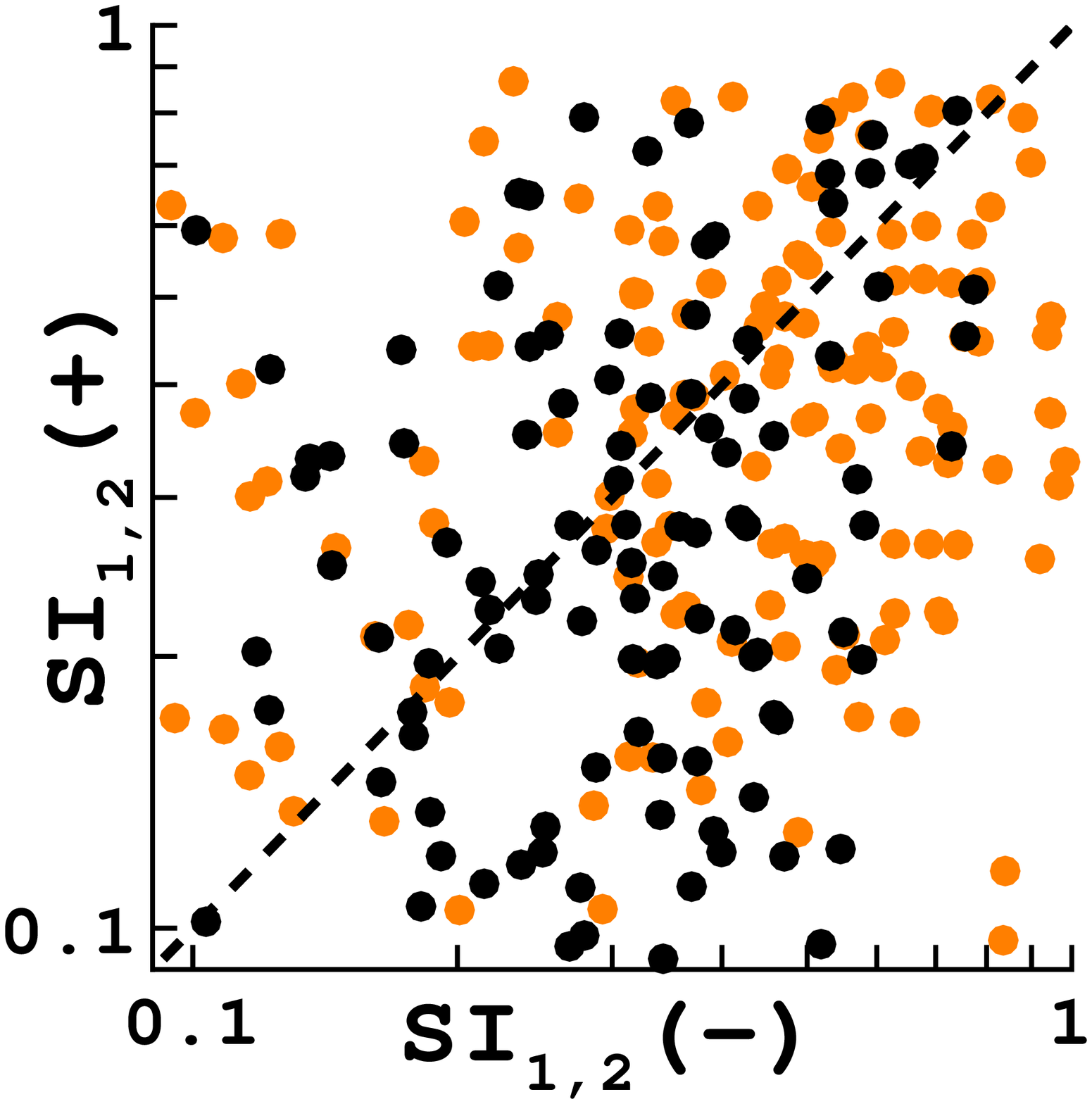}
\includegraphics[height=3cm,width=3cm]{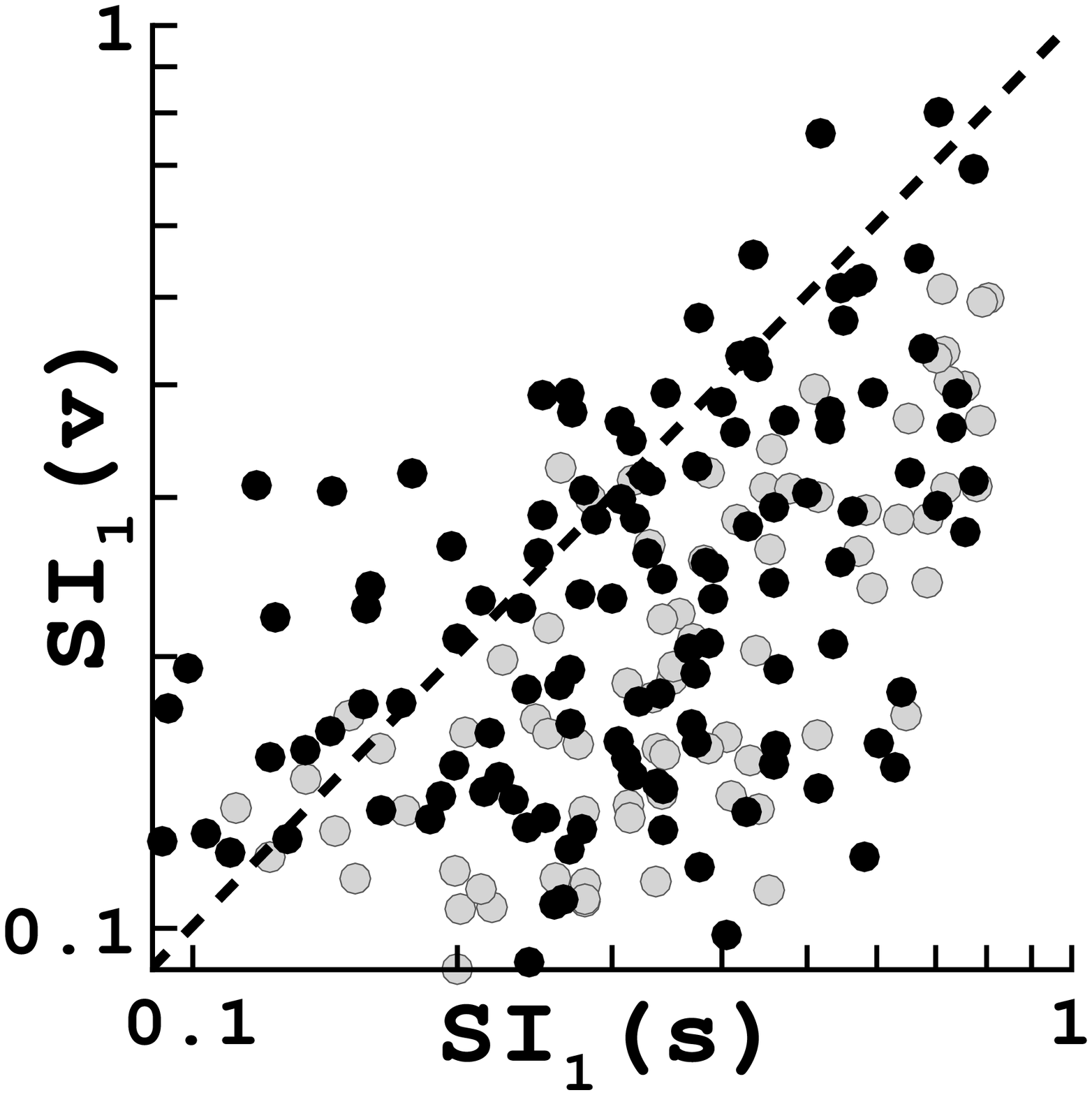}
\includegraphics[height=3cm,width=3cm]{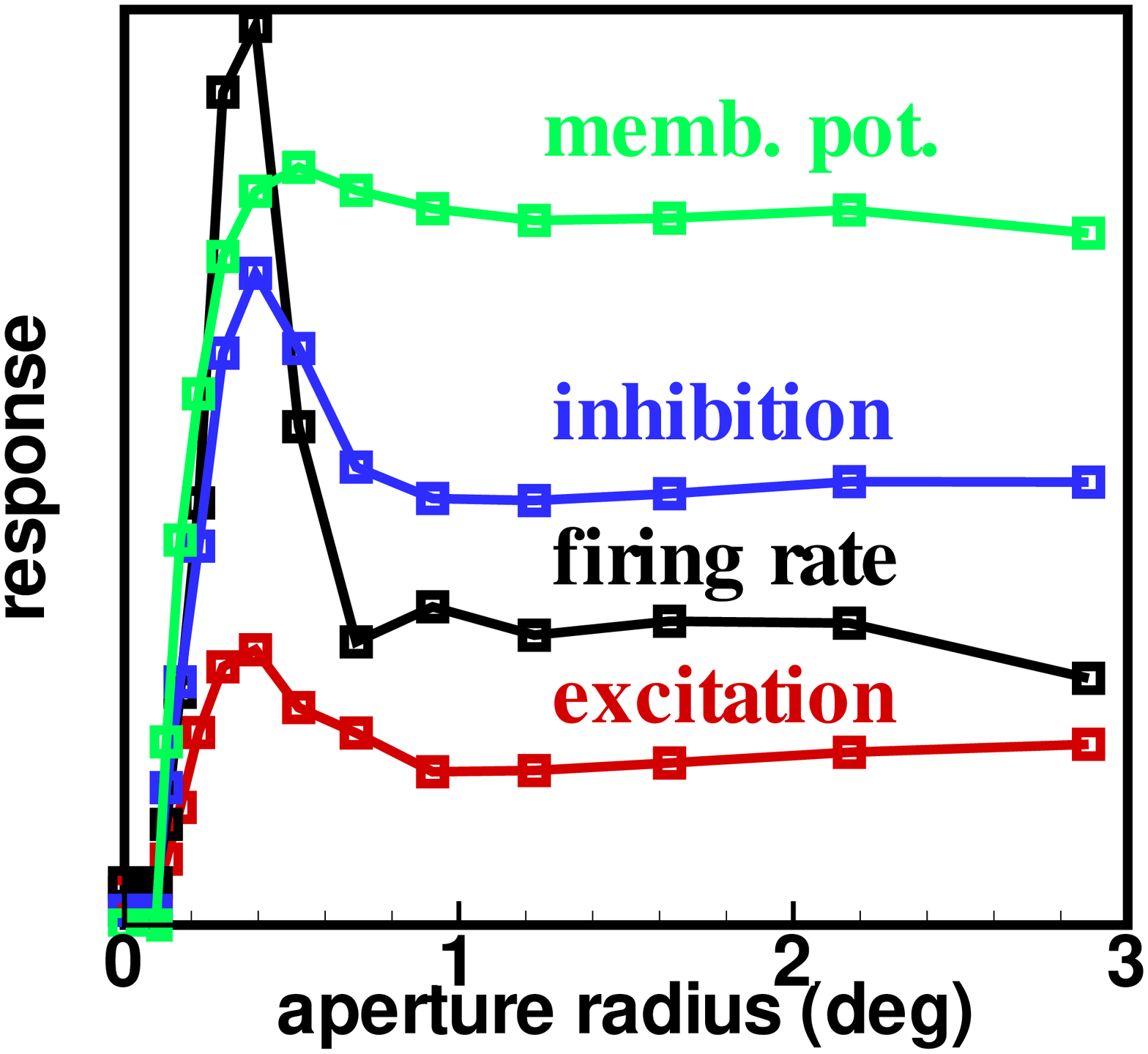}
\end{minipage}
\caption{\small Relations between some key response measures for the M0
configuration of the model, other cases yield qualitatively similar results.
(A) Scatterplot of surround suppression at low and high contrast
expressed in the two different suppression indexes $SI_{1}$ (black) and $SI_{2}$
(orange). 
(B) Scatterplot of surround suppression in spike train and membrane potential
at high (black) and low (gray) contrast.
(C) Spike responses, membrane potential responses, and cortical conductances as function of
aperture size for a model cell which shows about 50\% surround suppression (in spike train).
Notice the surround suppression of the conductances.}
\end{figure}

As may be clear from Figure 6, identifying the mechanisms for surround suppression based on 
Eq. (\ref{eq:vmem}) and (\ref{eq:fr}) can be rather more subtle than just comparing the mean (F0) 
conductance, its first harmonic (F1) or the peak conductance ($\sim$F0+F1). However, we find 
that for most cells, an analysis using the F0+F1 components of the conductances allows 
identification of the suppression mechanisms. Comparing conductances at $r_{A}=r$ and at 
$r_{A}=R$ in this way, we find that at low contrast all three mechanisms are about equally
prevalent, while at high contrast mechanism A is somewhat more likely than B and C.

\subsection{Mechanisms of contrast dependent receptive field size}
The DOG model suggests that growth in receptive field size at low contrast is due to 
an increase of the spatial summation extent of excitation\cite{sce99} ($\sigma_{E}$). 
This was partially confirmed experimentally in cat primary visual cortex\cite{and01}.
Although it has been claimed\cite{cav02} that the ROG model would explain receptive field growth solely from a
change in the relative gain parameter $k_{s}$, we believe this is incorrect. Since there is a one-one relation between $k_{s}$ and the surround suppression, this would imply that contrast dependent receptive 
field size simply results from contrast dependent surround suppression, which contradicts experimental 
data\cite{sce99,cav02}. Thus, as does the DOG model, the ROG model predicts that contrast dependent 
receptive field size is due to contrast dependence of the spatial summation extent of excitation.
As we show below, our simulations confirm an average growth of spatial summation extent of excitation (and inhibition) at low contrast. However, this growth is neither sufficient nor necessary 
to explain receptive field growth.

From Eq. (\ref{eq:vmem}) and (\ref{eq:fr}) it follows that a change in receptive field size in general results
from a change in behavior of the relative gain parameter, defined as 
\begin{equation}
\label{eq:gain}
G(r_{A})=\frac{\partial g_{E}/\partial r_{A}}{\partial g_{I}/\partial r_{A}} \; .
\end{equation}
Note that this is a rather different parameter than the ``surround gain'' parameter $k_{s}$ used in the 
ROG model. (For example, unlike $k_{s}$, $G(r_{A})$ is not simply related to the 
degree of surround suppression.)
Qualitatively, the conductances show a similar dependence on aperture size as the membrane potential 
responses and spike responses in that they display surround suppression (Fig. 5C). Receptive field
sizes based on these conductances are a measure of the spatial summation extent of excitation and inhibition.

A change in the spatial summation extent of $g_{E}$ and/or $g_{I}$ is just one of the many ways to change the behavior of $G$ and consequently the receptive field size. For example, some other possibilities 
are illustrated by the two cells in Fig. 6.
These cells show, both in spike and membrane potential responses, a receptive field 
growth of a factor of 2 (left) and 3 (right) at low contrast.  However, for the left cell, 
the spatial summation extent of excitation at low contrast is one aperture less than at high contrast, and 
for inhibition at low contrast it is one aperture larger than at high contrast. For the cell in the right 
the spatial summation extents of both excitation and inhibition do not change with contrast. 
 
In a similar way as for spike train responses, we also obtained receptive field sizes for the conductances.
As shown in Figure 7, both excitation and inhibition also show on the average an increase in their spatial
summation extent as contrast is decreased, but the increase is in general smaller than what is seen for 
spike responses, particularly for cells that show significant receptive field growth, say ($r_{+}/r_{-}>1.5$).
(Wilcoxon test on ratio of growth ratios larger than unity: $p<0.05$ (all cells, excitation, Fig. 7B), 
$p<0.15$ (all cells, inhibition, Fig. 7C), $p<0.001$ (cells with receptive field growth 
rate $r_{+}/r_{-}>1.5$, both excitation and inhibition.)
Further, Figure 7B \& C show that, although some increase in the spatial summation extent of 
excitation and inhibition is in general the rule, this increase is rather arbitrary and bears not 
much relation with the receptive field growth based on spike responses. 
For cells in the sample with larger receptive field growths (factor of $\sim 2$ and greater) this growth 
is always considerably less than the growth based on their spike responses.
The same conclusions follow from membrane potential responses (not shown).

A more precise analysis based on the relative gain parameter is given in Supplementary Information. For cells
with significant receptive field growth, ($r_{+}/r_{-}>1.5$) we are able to identify a systematic mechanistic
property. We find that for more than 50\% of such cells, a transition takes place from a high contrast RF size less or equal to the spatial summation extent of excitation and inhibition, to a low contrast receptive field size which exceeds both.

\begin{figure}[here]
\label{examples}
\centering
\begin{minipage}{1\columnwidth}
\centering
(A)\hspace{0.25\columnwidth}(B)\hspace{0.25\columnwidth}(C) \\
\includegraphics[height=3cm,width=3cm]{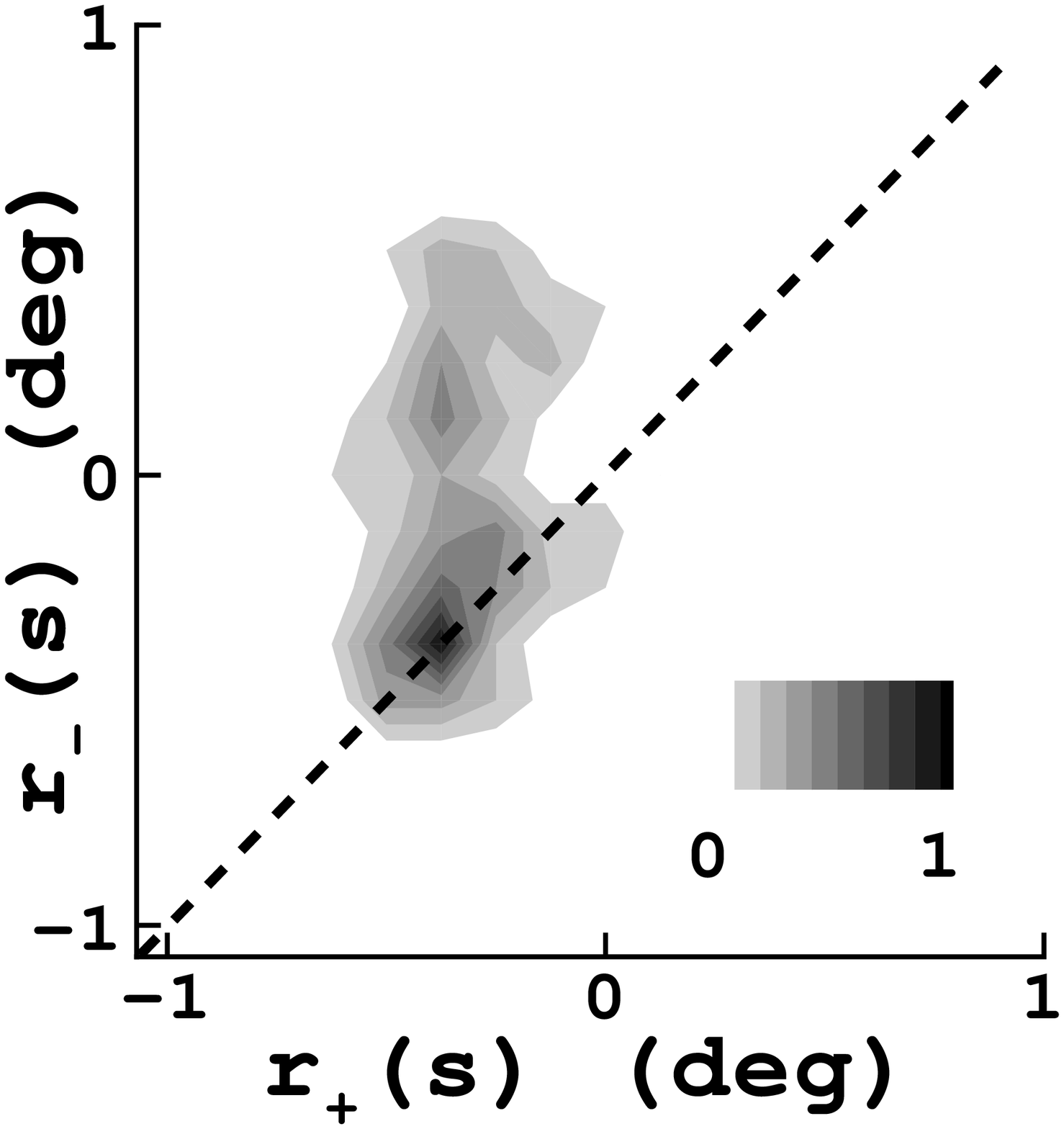}
\includegraphics[height=3cm,width=3cm]{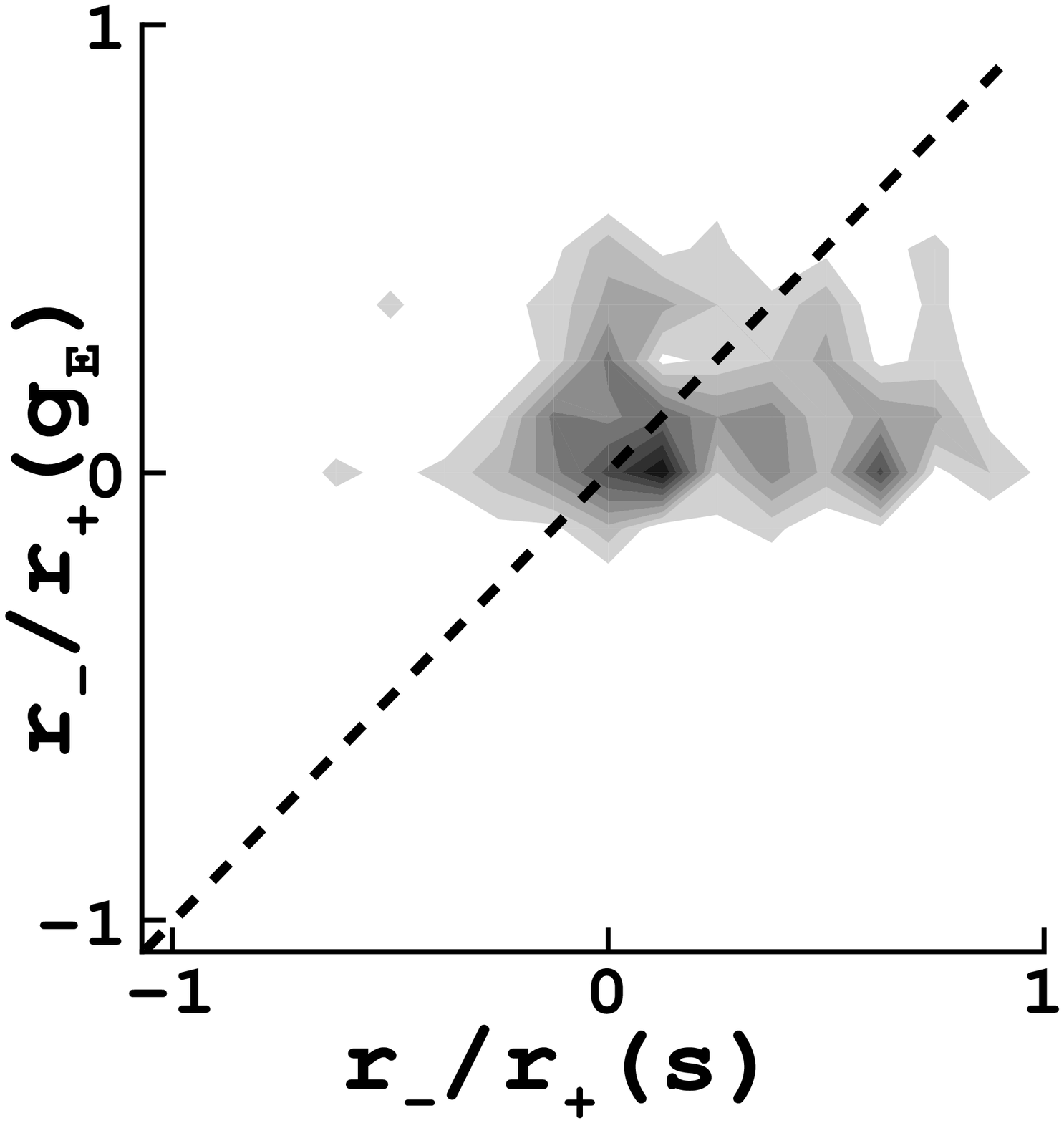}
\includegraphics[height=3cm,width=3cm]{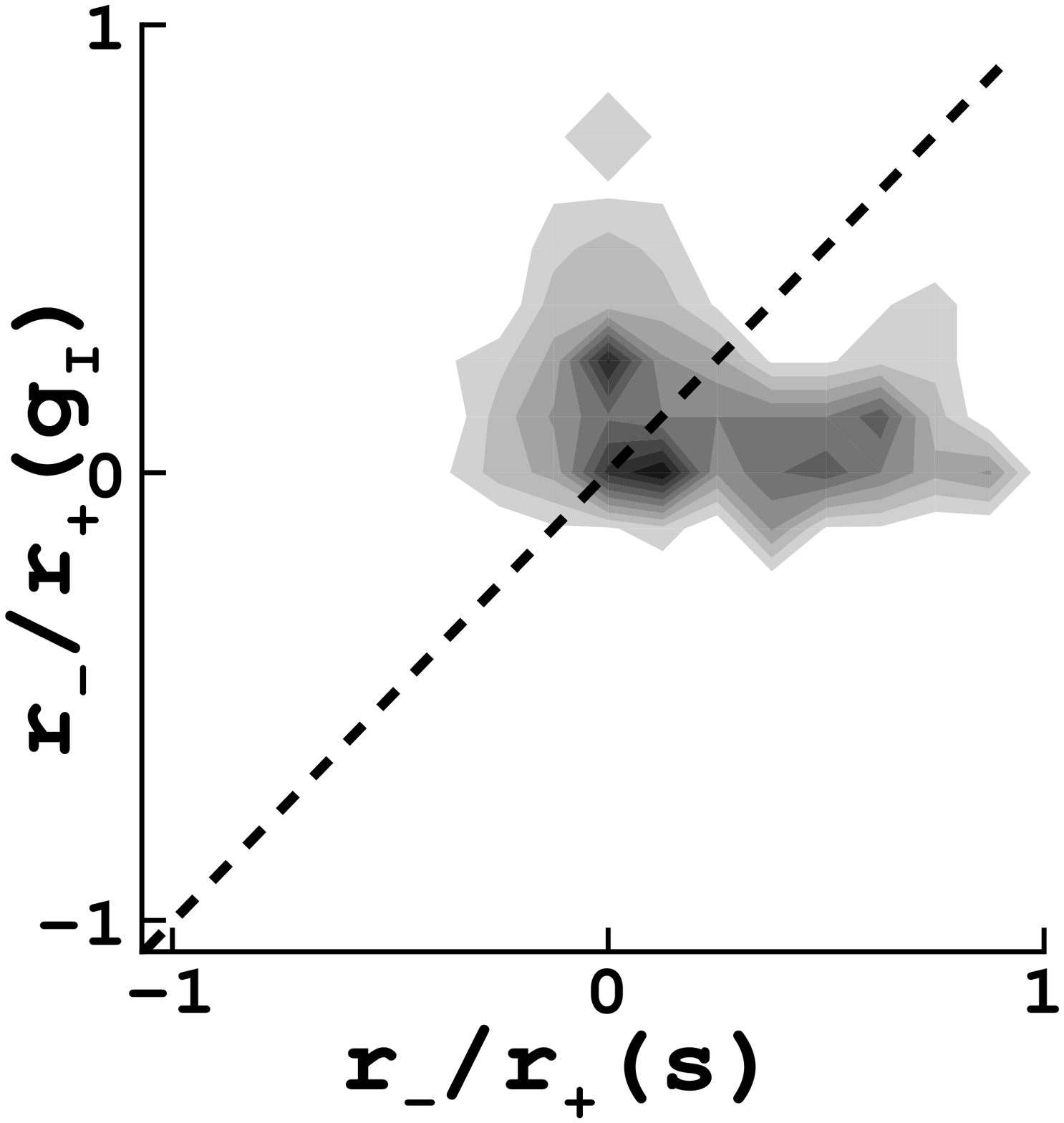}
\end{minipage}
\caption{\small (A) Joint distribution of high and low contrast receptive field sizes, $r_{+}$ and $r_{-}$,
  based on spike responses. All scales are logarithmic, base 10. All distributions are normalized to peak
  value one.
  Receptive field growth at low contrast is clear. Average growth ratio is 1.9 and is significantly 
  greater than unity (Wilcoxon test, $p<0.001$). 
  (B \& C) Joint distributions of receptive field growth and
  growth of spatial summation extent of excitation (B) and inhibition
  (C) (computed as ratios). There is no simple relation between receptive field
  growth and the growth of the spatial summation extent of excitatory or
  inhibitory inputs. For cells in the sample with larger receptive field growths (factor of $\sim
  2$ and greater) this growth is always considerably larger than the
  growths of their excitatory and inhibitory inputs.}
\end{figure}

\subsection{LGN contributions}
The extraclassical responses in our model discussed so far are, by construction, exclusively
the result of cortical interactions and not inherited from LGN inputs. This is due to our use of the 
standard center-surround model for LGN receptive fields (Supplementary Information).
Thus our LGN cells show neither surround suppression nor contrast dependent receptive field size.
For what concerns the latter, this is true at all spatial frequency. Surround suppression, however, is 
absent at the relevant (optimal) spatial frequencies for our cortical cells ($SI_{1}(g_{lgn})<0.01$), but
does appear at lower spatial frequencies. We used this fact to study the transfer of LGN surround 
suppression to cortical cells. 
Surround suppression of our LGN cells for different spatial frequencies is illustrated in Figure 8A. 
No suppression occurs for cortical optimal spatial frequencies $k_{C}$, suppression starts at 
spatial frequencies of 
about $0.5k_{C}$, it becomes stronger for smaller spatial frequencies and is about 25\% 
at $0.25k_{C}$. Further, we find that at $0.25k_{C}$ the surround is not yet able to evoke responses 
on its own, and the suppression thus in some sense qualifies as ``extraclassical''.

\begin{figure}[here]
\label{examples}
\centering
\begin{minipage}{1\columnwidth}
\centering
(A)\hspace{0.25\columnwidth}(B)\hspace{0.25\columnwidth}(C) \\
\includegraphics[height=3cm,width=3cm]{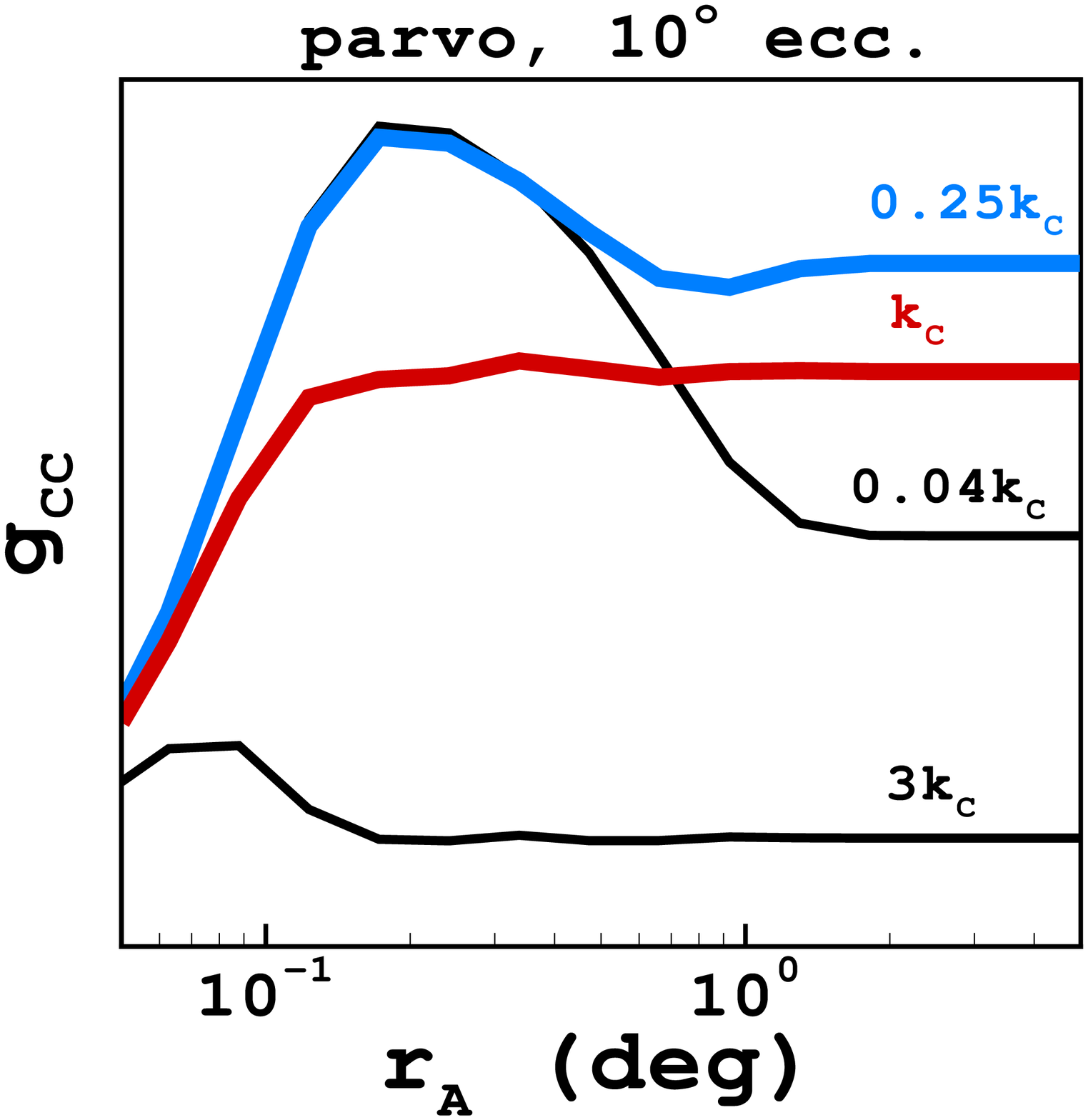}
\includegraphics[height=3cm,width=3cm]{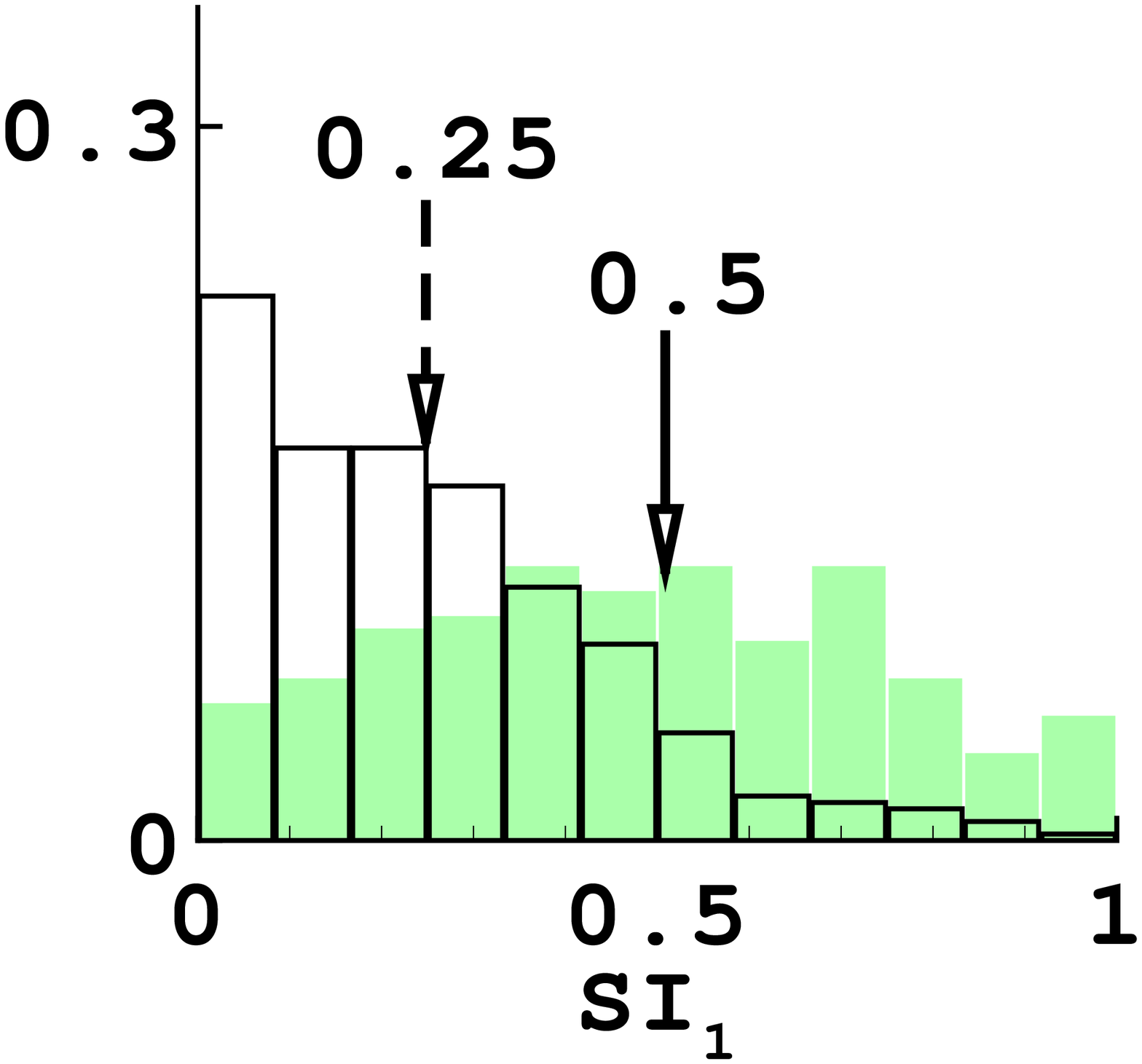}
\includegraphics[height=3cm,width=3cm]{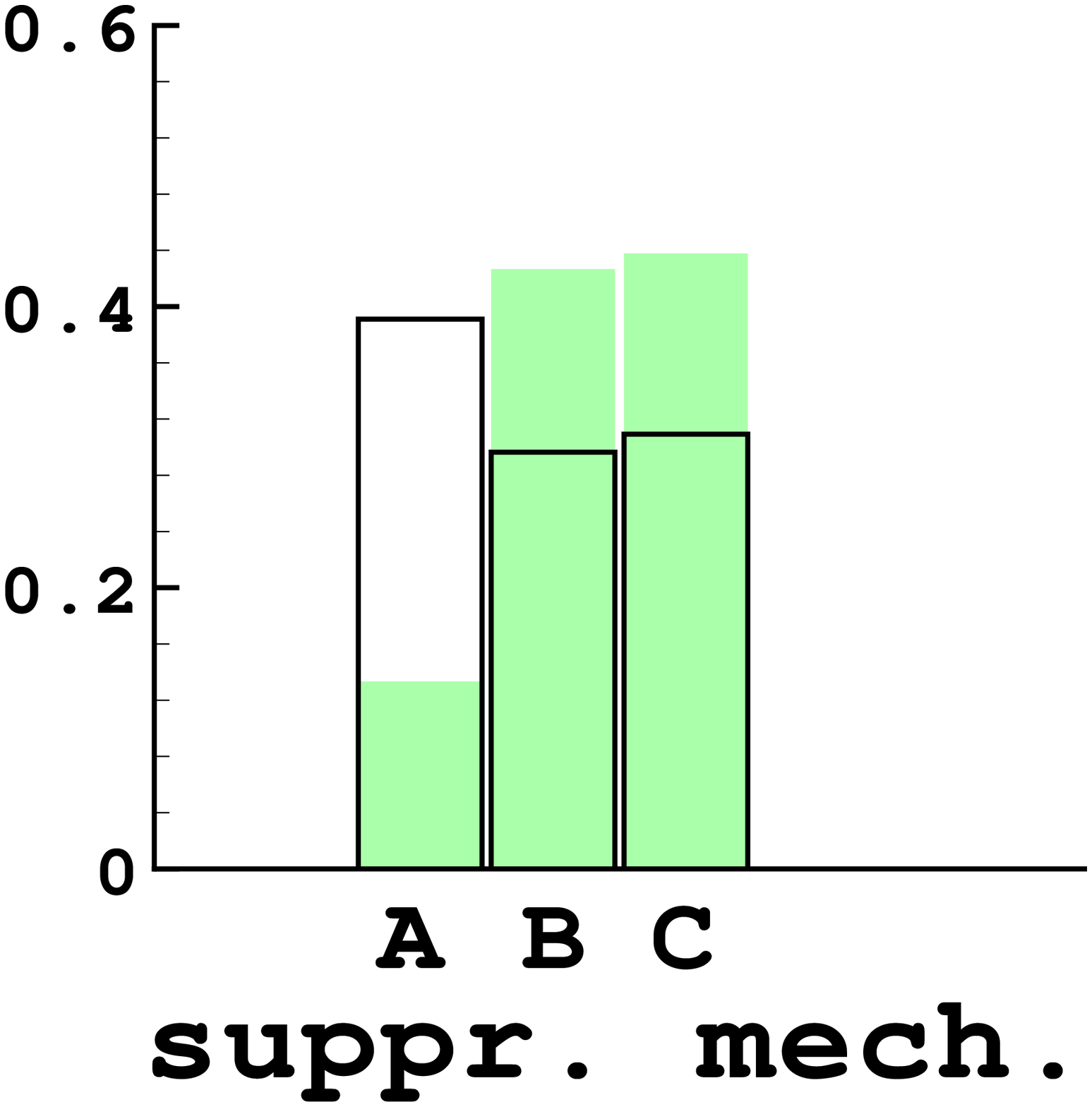}
\end{minipage}
\caption{\small Transfer of LGN surround suppression to cortical cells. (A) Responses (F1) of the P10 model
LGN cells as function of aperture size, at different spatial frequencies (all LGN cells in a particular
configuration are identical). Average cortically preferred spatial frequency is $k_{C}$ (red), at which 
our LGN cells show no surround suppression. The LGN cells show about 25\% (classical) surround suppression 
at a spatial frequency $k=0.25k_{C}$ (blue).
(B \& C) Surround suppression and prevalence of suppression mechanisms at spatial frequencies
$k=k_{C}$ (unfilled) and $k=0.25k_{C}$ (green shaded).}
\end{figure}

Our simulations show that the cortical contributions to surround suppression and contrast dependent 
receptive field size account for a large fraction, though not all of the magnitude of the phenomena observed experimentally. This leaves room for contributions from the LGN input. It seems reasonable to assume that 
LGN cells in macaque will display both extraclassical phenomena, although, somewhat surprisingly, this has 
to our knowledge not been verified yet.
Surround suppression of LGN cells at cortical optimal spatial frequencies has been observed in 
marmoset\cite{sol02} and cat\cite{oze04}. Contrast dependent receptive field growth of LGN 
cells has been observed in marmoset and an average growth ratio of 1.3 was reported\cite{sol02}.
Transfer of receptive field growth of LGN cells to V1 cortical cells seems unavoidable because it 
simply introduces an overall scaling factor on the entire visual input
in V1. Inheritance of LGN surround suppression by V1 cells is not so obvious.
Consistent with the Hubel and Wiesel view (the way our model is constructed), LGN input
arrives in a V1 cell essentially as output of small clusters of about 10-20 LGN cells.
It is thus not immediately clear if and how much LGN surround suppression can be transferred to V1 
cortical cells. We addressed this issue by repeating our simulation, originally performed at $k_{C}$, at 
the smaller spatial frequency of $0.25k_{C}$. 

Results of this simulation are shown in in the remainder of Figure 8. We see (Fig. 8B)
that practically all LGN surround suppression is transferred to cortical cells.
Given that an average surround suppression of $SI_{1}\sim 0.4$ is observed in 
macaque \cite{cav02}, our results show that cortical short-range connections together 
with surround suppression present in LGN cells can easily explain the degree of 
surround suppression seen experimentally. 
We also see that the presence of LGN surround suppression also has 
consequences for the prevalence of the different mechanisms by which cortical suppression
is achieved. When LGN surround suppression is present, the prevalence of the 
suppression mechanisms is substantially altered in favor of mechanisms B and C (which 
require a reduction of excitation) at the expense of mechanism A (increased direct inhibition).  
Results for contrast dependent receptive field size are practically unaltered when LGN
suppression is included (not shown). 

\subsection{The cortical magnification factor}
As mentioned in the Introduction, arguments in favor of involvement of long-range connections and/or extrastriate feedback in extraclassical phenomena, are indirect and all rely on the cortical magnification factor as a key ingredient. Receptive field size and scatter are systematically ignored. It is 
argued that surround sizes would be too big to result from local short-range connections. We have 
already shown that, on the contrary, through polysynaptic interactions in the network it is 
possible to create the surround sizes observed experimentally with only local short-range connections.

One naturally wonders to what extent our findings depend on the actual value of the cortical 
magnification factor.
Intuitively, a smaller cortical magnification factor is not beneficial for the
role of short-range connections in the creation of extraclassical receptive field phenomena, 
since these connections cover less visual space.
However, the minimum amount of visual space covered is set by the receptive field size and scatter. 

To check whether this minimum visual range of cortical short-range connections is in itself sufficient to 
generate surround suppression and contrast dependent receptive field size we repeated our simulations 
with an infinite cortical magnification factor, $\nu^{-1}=\infty$ (geometric parameter $\Omega = \infty$,
all else unchanged, see Methods,). 
The results are shown in Figure 9 for the P0 case (M0 yields similar results). 
Clearly, the finite receptive field scatter by itself, 60\% (M0) and 30\% (P0) of the average 
receptive field size (Methods), is sufficient to generate both extraclassical phenomena 
to practically the same degree as it does in the presence of a realistic cortical 
magnification factor. Other properties of our model discussed in this paper are also not qualitatively different for infinite cortical magnification factor, with one exception. As is apparent from Figure 9A, 
it is now more difficult to connect the LGN axons in such a manner that the organization of orientation preference and ocular dominance displays the same level of order as seen for a realistic cortical magnification factor (Fig. 1). (This does of course not imply that order could not be improved with more specific connections than the ones used).

Arguments ruling out local short-range connections and LGN input as the origins of extraclassical 
phenomena based on the cortical magnification factor are inherently weak, since it is a macroscopic measure 
and thus inferences based on it regarding which cells could influence which cell and on what time scale, 
cannot be very precise.
Our simulations in this section further challenge such arguments, by showing that receptive field size and 
scatter by themselves, regardless of the cortical magnification factor, can be a determinative factor for extraclassical receptive field phenomena.

\begin{figure}[here]
\centering
\begin{minipage}{0.62\columnwidth}
\includegraphics[angle=270, origin=c,height=6.0cm,width=6.39cm]{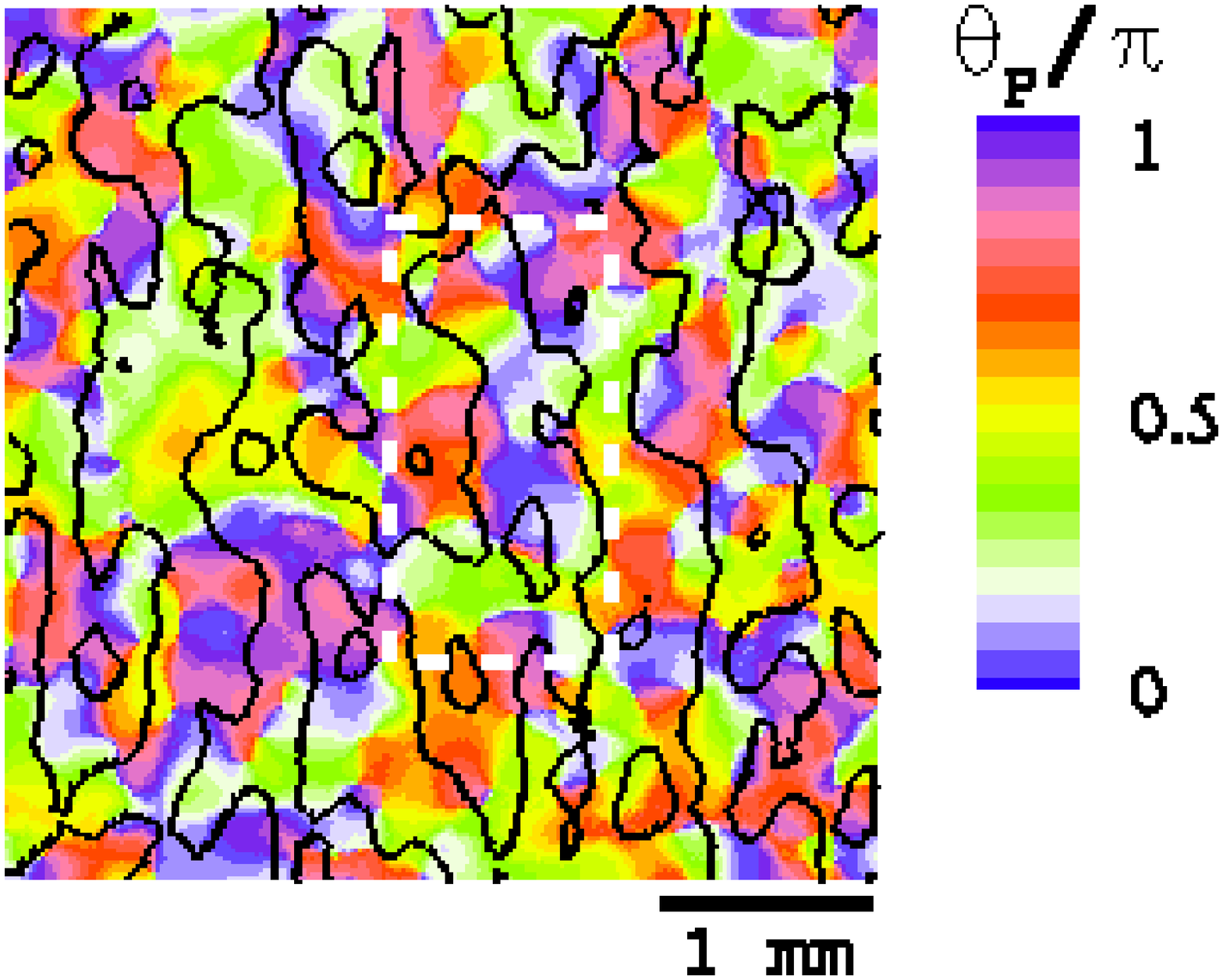}
\end{minipage}
\begin{minipage}{0.35\columnwidth}
\centering
\vspace{0.3cm}
\includegraphics[height=2.75cm,width=2.75cm]{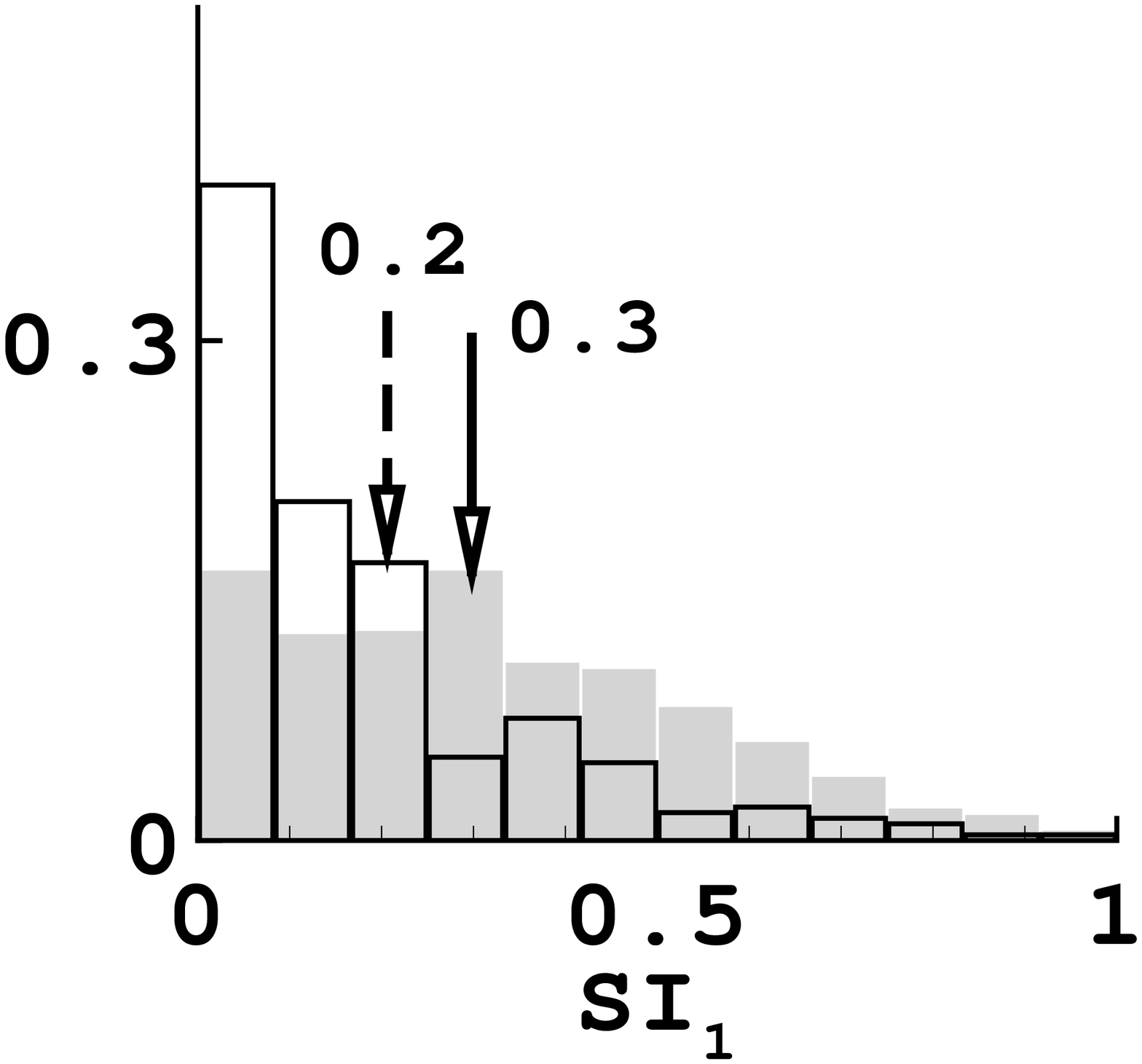}
\includegraphics[height=2.75cm,width=2.75cm]{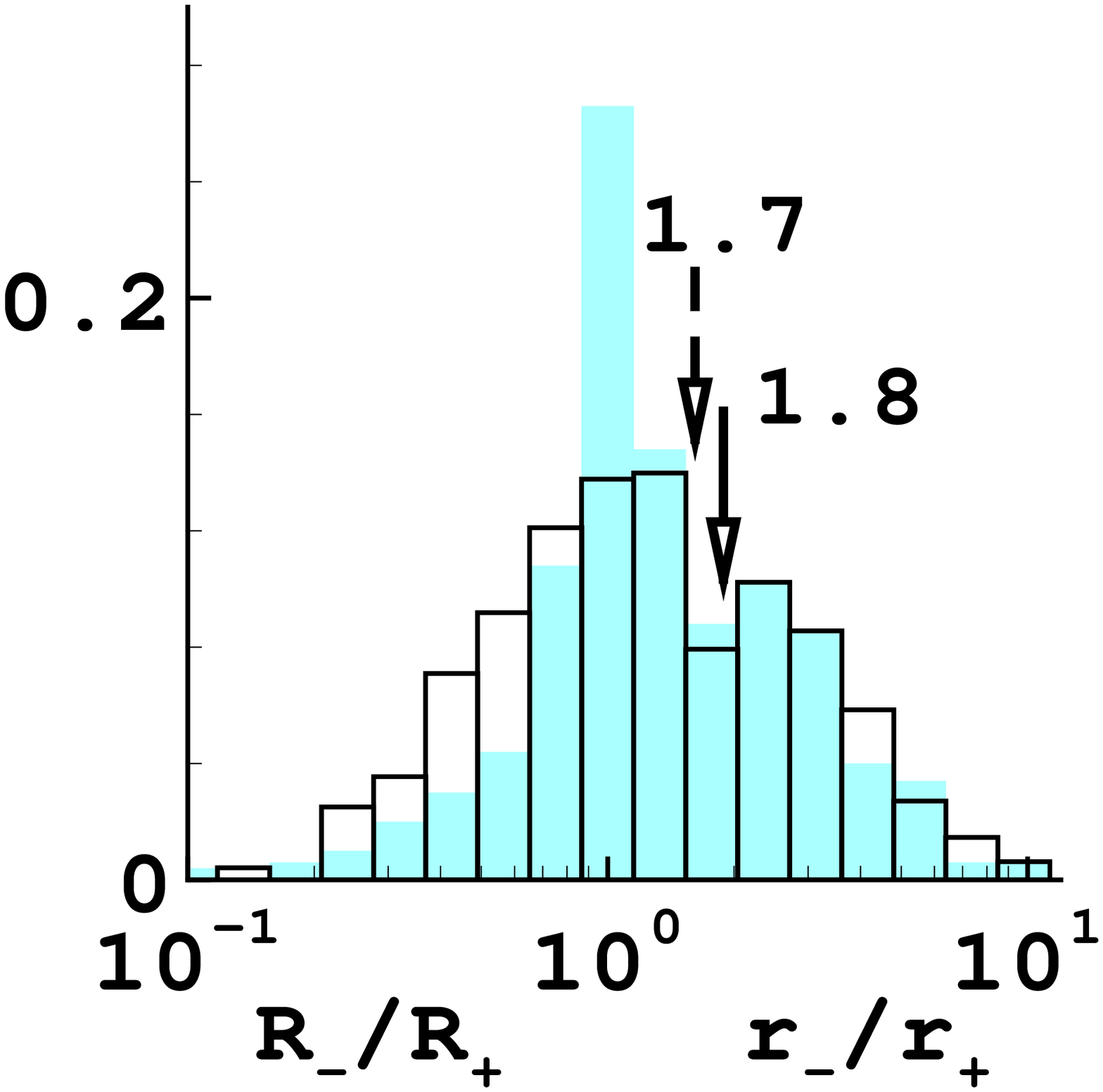}
\end{minipage}
\caption{\small Results for infinite cortical magnification factor ($\Omega = \infty$) for
the P0 configuration, M0 configuration yields similar results. (left) Simulated optical
image of ocular dominance and orientation preference, in the spirit of optical imaging
experiments \cite{bla92} (compare Figure 1).
(right top) Suppression index $SI_{1}$ at high (unfilled) and low (green shaded) contrast (compare Figure 2).
(right bottom) Receptive field and surround growth ratios, $r_{-}/r_{+}$ (blue shaded) and $R_{-}/R_{+}$
(unfilled).} 
\end{figure}

\section{Discussion}
There is considerable debate over the origins of extraclassical receptive field phenomena such as
surround suppression and contrast dependent receptive field size. 
A primary reason is that there are no experimental data yet available that directly point to 
mechanisms of these phenomena. This requires that any data needs to be interpreted through some sort 
of model or theoretical framework. At the same time, such a model or theory needs to be sufficiently sophisticated, for example, it needs to adequately address classical response properties since they set 
the context for the extraclassical phenomena. We believe our model is the first anatomically and physiologically realistic model to simultaneously address classical and extraclassical spatial summation.

We know of only one attempt in the published literature\cite{som98} to develop a
spiking neural network model to address the extraclassical response
phenomena discussed in this paper. That model does not address classical responses, and 
neural mechanisms for extraclassical phenomena, by construction, arise from long-range
connections. Further, in that model contrast dependent receptive field size is achieved via 
contrast dependent surround suppression, that is, the surround suppression systematically decreases as a function of decreasing contrast, which contradicts experimental data\cite{sce99,cav02}.

We have shown that considerable surround suppression and contrast dependent receptive field 
growth can be spontaneously generated solely by short-range cortical connections in V1 without any 
contributions from other sources.
We demonstrated that surround suppression of LGN cells, if present, is easily transferred to 
V1. Our simulations thus provide rigorous computational support for the intriguing hypothesis that the 
LGN input and cortical short-range connections in V1 are primarily responsible for the phenomena, with 
little or no contributions from long-range or extrastriate connections. 
We showed that with only 25\% suppression in the LGN cells, the cortical surround suppression in 
the model exceeds the suppression observed experimentally.
More radically interpreted, our results thus suggest that long-range lateral connections
and/or extrastriate feedback contribute {\em negatively} to surround suppression, 
that is, rather than being suppressive, our results suggest that their contributions are in fact 
{\em facilitatory}.

In our model all three neural mechanisms for surround suppression are active, in agreement with 
experimental observations in cat\cite{and01}.  
When LGN suppression is included, we observe strong contributions to cortical suppression from 
a reduction of recurrent cortical excitation, rather than from an increase in direct cortical 
inhibition. 
We find, on average, a growth of spatial summation extent of excitation and inhibition at low 
contrast, as predicted by DOG\cite{sce99} and ROG models.
But this growth bears no simple relationship with the receptive field growth seen in spike responses,
which usually involve other/additional changes in the relative gain of the 
excitatory and inhibitory inputs. Notably, significant receptive field growth is usually much 
larger than the growth of the spatial summation extent of excitation and inhibition (Fig. 7). 

As does the biological primary visual cortex, our model produces these properties 
in distinctly different geometric settings (with identical strength parameters), namely for 
the magno and parvo input layers, at para-foveal eccentricities and around $10^{o}$ eccentricity.
Dimensional observations (parameter $\Omega$) imply our model results also translate to $30^{o}$ 
eccentricity, modulo a geometric scaling factor.
The ubiquitous nature of these phenomena and their mechanisms in our model suggest that they are 
basic response properties of V1. 
Given realistic classical response properties, as discussed in this paper, they 
seem to require little more than receptive field scatter and isotropic short-range connectivity,
with perhaps some weak constraints on macroscopic organization of ocular dominance and orientation
preference. They do not seem to require more elaborate architecture or physiological properties, such as specific cortical connectivity, long-range connections within V1, extrastriate feedback, synaptic 
depression/facilitation etc..

Recent data for cat suggest that partial inheritance of surround 
suppression in V1 from LGN cells does indeed occur\cite{oze04}. 
Other recent experiments show that strong surround suppression is observed for drifting gratings 
having spatial and temporal frequencies outside the range at which most cortical cells typically 
respond, indicating these signals arise within the input layer of V1 or the LGN itself
\cite{web04}.
Given the general nature of our results, we may conclude that the presence of the phenomena 
in LGN cells could in principle be of the same origin as we have suggested here for V1 cells. 

There are various further aspects of surround suppression that we have not explicitly addressed in 
this paper.
Among them are orientation tuning of the surround and dynamics (timing) of the suppression.
Preliminary simulations indicate that orientation tuning of the surround is well-captured by our model. 
For what concerns timing of the suppression seen in our model, it is clear that through polysynaptic interactions in the network, delays of the onset of surround suppression could range anywhere from 0-20 ms, conservatively estimated. Indeed, preliminary simulations show that timing of the surround suppression in our model is consistent with recent experimental findings\cite{xin04}.
Our model has a rich dynamics and is well-suited to also yield relevant results regarding the
dynamics of surround suppression. This is one of our interests for future research.

\section*{Acknowledgments}
This work was supported  by grants from ONR (MURI program, N00014-01-1-0625) and NIMA 
(NMA201-02-C0012).

\bibliographystyle{pnas}
\bibliography{paper.bib}

\section{Supplementary Information}
\subsection{Model cortex}
Our model cortex consists of $N=256\times 256$ conductance based integrate-and-fire 
point neurons (one compartment), of which 75\% are excitatory and 25\% are inhibitory
and which are randomly distributed on a square grid.

Rescaling of Eq. (1) in the paper works as follows. Numerical values
for the biophysical parameters are, for the capacitance $C=10^{-6}$ F
$\mbox{cm}^{-2}$, the leakage conductance $g_{L}=50\times 10^{-6}$
$\Omega^{-1}$ $\mbox{cm}^{-2}$, the leakage reversal potential
$v_{L}=-70$ mV, the excitatory reversal potential $v_{E}=0$ mV, the
inhibitory reversal potential $v_{I}=-80$ mV, the spiking threshold
$v_{T}=-55$ mV, and the reset potential $v_{R}=-70$ mV. The equation 
is then divided by $C\star v_{T}$. Physiological conductances ($\Omega^{-1}$ 
$\mbox{cm}^{-2}$) and currents (Ampere $\mbox{cm}^{-2}$)  are obtained by multiplication
with $10^{-6}$ and $15\times 10^{-9}$ respectively.  

The functions ${\cal C}_{\mu^{\prime},\mu}^{k^{\prime},k}(r)$ in Eq. (3) of the paper describe the cortical
spatial couplings (cortical connections), they are constructed as follows. We
assume the availability of postsynaptic sites $N_{d}$ on a cell
(dendrites) to decay exponentially as a function of distance with
length scale $D_{\mu^{\prime},\mu}^{k^{\prime},k}$, that is,
$N_{d}\sim\exp [-(r/D_{\mu^{\prime},\mu}^{k^{\prime},k})^{2}]$, and
make a similar assumption for the presynaptic sites $N_{a}$ (axons),
$N_{a}\sim\exp [-(r/A_{\mu^{\prime},\mu}^{k^{\prime},k})^{2}]$.	 Then
the spatial coupling strength (assuming individual synapses have equal
strength) between two cells decays exponentially with length scale
$(\sigma_{\mu^{\prime},\mu}^{k^{\prime},k})^{2} =
(D_{\mu^{\prime},\mu}^{k^{\prime},k})^{2}+(A_{\mu^{\prime},\mu}^{k^{\prime},k})^{2}$
and can be written as
\begin{equation}
{\cal C}_{\mu^{\prime},\mu}^{k^{\prime},k}(r)=c_{\mu^{\prime},\mu}^{k^{\prime},k}
N_{\mu^{\prime},\mu}^{k^{\prime},k} \exp [-(r/\sigma_{\mu^{\prime},\mu}^{k^{\prime},k})^{2}] \;,
\end{equation}
with the normalization constants
\begin{equation}
N_{\mu^{\prime},\mu}^{k^{\prime},k}=\left\{ \sum_{i\in {\cal P}_{k}(\mu)}
\exp [-(||\vec{x}_{i}||/\sigma_{\mu^{\prime},\mu}^{k^{\prime},k})^{2}] \right\}^{-1} \; .
\end{equation}
In this way, the parameters $c_{\mu^{\prime},\mu}^{k^{\prime},k}$ are interaction strengths that define the 
density and length scale invariant contribution of population ${\cal P}_{k}(\mu)$ to the conductance of a cell in
population ${\cal P}_{k^{\prime}}(\mu^{\prime})$.
The change in membrane potential of cell $i\in {\cal P}_{k^{\prime}}(\mu^{\prime})$ due to a single spike of cell
$j\in {\cal P}_{k}(\mu)$ is proportional to $c_{\mu^{\prime},\mu}^{k^{\prime},k}
(\sigma_{\mu^{\prime},\mu}^{k^{\prime},k})^{-2}
(n_{k,\mu})^{-1} \exp [-(r_{i,j}/\sigma_{\mu^{\prime},\mu}^{k^{\prime},k})^{2}]$, where
$n_{k,\mu}$ is the cell density of population ${\cal P}_{k}(\mu)$
and $r_{i,j}=||\vec{x}_{i}-\vec{x}_{j}||$.

The temporal kernels $G_{\mu,j}(\tau)$ in Eq. (3) of the paper describe the synaptic dynamics of
cortical synapses, are normalized to unity,
$\int_{-\infty}^{\infty}G_{\mu,j}(\tau)d\tau = 1$, and are of the form
\[
G_{\mu,i}(\tau) = 
\]
\begin{equation}
\label{eq:cker}
\left\{ \begin{array}{ll}
0 & \mbox{if $\tau \leq 0$} \\
k_{\mu,i}\left(\tau e^{-\tau/a_{\mu,i}}\right)^{5} & \mbox{if $0< \tau < \Delta_{\mu}a_{\mu,i}$} \\
k_{\mu,i}\left(\Delta_{\mu}a_{\mu,i} e^{-\Delta_{\mu}}\right)^{5}
e^{-(\tau-\Delta_{\mu}a_{\mu,i})/b_{\mu}} & \mbox{if $\tau \geq \Delta_{\mu}a_{\mu,i}$}
\end{array} \right. \; .
\end{equation}
The time constants are based on experimental observations \cite{koc99}. The
excitatory kernels $G_{E,i}$ have a fast (AMPA) component defined by the peak times $a_{E,i}$, which
are drawn from a uniform distribution between $1$ ms and $4$ ms, and a
slow (NMDA) component which is defined by the decay time $b_{E}=15$ ms. Transition between the two
regimes is set by $\Delta_{E}=4/3$. Similarly, the inhibitory kernels $G_{I,i}$ have a fast
(GABA) component set by $a_{I,i}$, chosen from a uniform distribution between $3$ ms and
$6$ ms, and a slow component \cite{gib99} defined by $b_{I}=10$ ms, while $\Delta_{I}=3/2$. The
constants $k_{\mu,i}$ are normalization constants. These kernels imply a spike memory on
the order of $70$ ms for excitation and $50$ ms for inhibition.

The cortical spatial coupling length scales are taken to be in agreement with anatomical
data \cite{fit85,lun87,cal96,cal98}, axon and dendrite parameters are
$A_{\mu^{\prime},E}^{k^{\prime},k}=200$ $\mu m$ (axons of excitatory neurons),
$A_{\mu^{\prime},I}^{k^{\prime},k}=100$ $\mu m$ (axons of inhibitory neurons), and
$D_{\mu^{\prime},\mu}^{k^{\prime},k}=50$ $\mu m$ (dendrites of all neurons) respectively.
This implies that excitatory connections, both on excitatory and inhibitory cells, have characteristic
length scale $\sigma_{\mu^{\prime},E}^{k^{\prime},k}\sim 200$ $\mu m$ while inhibitory connections,
both on excitatory and inhibitory cells, have a characteristic length scale
$\sigma_{\mu^{\prime},I}^{k^{\prime},k}\sim 100$ $\mu m$.
So, in agreement with the anatomy our model has inhibitory connectivity which is of
a much shorter spatial range (a factor of two) than the excitatory connectivity.

Because of lack of experimental data, strength parameters in the model are free parameters.
They occur in the LGN input, noise levels, and the cortical interaction. In particular, the cortical 
coupling strength matrix $c_{\mu^{\prime},\mu}^{k^{\prime},k}$ consists entirely of such 
parameters. It is currently not experimentally feasible to directly determine the 
components of the cortical coupling strength matrix $c_{\mu^{\prime},\mu}^{k^{\prime},k}$
or the other strength parameters in the model. 
A natural question is ``which subsets of strength parameters from the 
set of all possible strength parameters yield physiologically realistic results?''.
A complete answer to the question is also not yet feasible, because of the size of the 
simulation and the number of parameters involved. 

We provide in this paper a set of strength parameters that yields physiologically 
realistic results. The set of strength parameters we provide 
is obtained by adhering to a few general principles regarding the LGN input and 
cortico-cortical interactions. These general principles are: (a) No distinction between
LGN input in the excitatory and inhibitory cell populations. (b) Cells with LGN input, both
excitatory and inhibitory cells, receive their excitation in about equal amounts from LGN 
input, cells with LGN input, and cells without LGN input. (c) Cells without LGN input, both
excitatory and inhibitory cells, receive most of their excitation from cells with LGN input. 
(d) Cells with (without) LGN input, both excitatory and inhibitory cells, receive most 
of their inhibition from cells with (without) LGN input.

Proceeding in this way, we considered a small number of classical response 
properties for setting the operating point (set of strength parameters). 
Specifically, these classical response properties are: (i) Absence of any global 
phase-locked oscillations and synchrony, both under visual stimulation
and without visual stimulation. This requirement limits the overall maximum size of the 
strength parameters. (ii) Distribution of activity (firing rates) over the 
cell population, with and without visual stimulation. This requirement constrains the 
overall balance between excitatory and inhibitory strength parameters. (iii) Distribution
(over cell population) of response modulations for drifting grating stimuli. This 
essentially is a requirement on the model's composition in terms of simple and complex 
cells. (iv) Orientation tuning of both simple and complex cells. This requirement results 
in large conductances, particularly, large inhibitory conductances. Note also that this 
requirement is a major constraint, since orientation tuning for drifting gratings is not 
easily achieved, e.g. see \cite{mcl00}.
This is so because of the fact that the average LGN input is practically 
untuned for orientation and inhibitory connections are of much shorter 
range spatially than excitatory connections. 

Numerical values for the strength parameters were obtained by performing many trial-and-error 
simulations, in a properly scaled-down version of the model, in search of a suitable operating point.
The coupling matrix resulting from this approach and used in the numerical simulations presented in the 
paper is 
\begin{equation}
\label{eq:cmat}
\left[ \begin{array}{cccc}
c_{E,E}^{0,0} & c_{E,I}^{0,0} & c_{E,E}^{0,1} & c_{E,I}^{0,1} \vspace{0.2cm} \\
c_{I,E}^{0,0} & c_{I,I}^{0,0} & c_{I,E}^{0,1} & c_{I,I}^{0,1} \vspace{0.2cm} \\
c_{E,E}^{1,0} & c_{E,I}^{1,0} & c_{E,E}^{1,1} & c_{E,I}^{1,1} \vspace{0.2cm} \\
c_{I,E}^{1,0} & c_{I,I}^{1,0} & c_{I,E}^{1,1} & c_{I,I}^{1,1}
\end{array} \right] =
\left[ \begin{array}{cccc}
1 & 4.5 & 10  & 2 \vspace{0.3cm} \\
1.5 & 6 & 11 & 2.5 \vspace{0.3cm} \\
3 & 5 & 2 & 14 \vspace{0.3cm} \\
3 & 5 & 3 & 14
\end{array} \right] \;\; .
\end{equation}

This coupling strength matrix (Eq. \ref{eq:cmat}) and the other strength parameters
(in noise levels and LGN input, to be discussed below), were chosen because they 
set the operating point of the model such that it reproduces the classical response 
properties (i)-(iv) in agreement with experimental data. Further support for 
the physiological relevance of this operating point is provided by the fact that,
in this setting, model properties accurately extrapolate to a variety of other 
known properties of the biological visual cortex beyond properties (i)-(iv). 
Among them are (v) spatial and temporal frequency tuning (vi) a cortex operating at high 
conductance levels, (vii) distributions of response modulations for contrast reversal 
gratings, (viii) dynamics of responses to stochastic stimuli (reverse correlation), and, of 
course, (ix) the extraclassical response properties which are the main topic of the paper. 
Furthermore, results obtained at this operating point are robust, that is, they remain 
quantitatively similar and thus physiologically realistic, when a random change $<10$\% 
is introduced in the set of strength parameters.

The external stochastic terms $\eta_{\mu,i}(t)$ in Eq. (1) of the paper are given by
\begin{equation}
\eta_{\mu,i}(t)=\eta^{0}_{\mu,i}\int_{-\infty}^{\infty}G^{P}_{\mu,i}(t-\tau)
{\cal S}^{P}_{\mu,i}(\tau)d\tau \; .
\end{equation}
Where the kernels $G^{P}_{\mu,i}$ have the same form as (\ref{eq:cker}) (with
randomly selected $a_{\mu,i}^{P}$'s) and ${\cal S}^{P}_{\mu,i}$ are Poisson
spike trains (mean firing rates $100$ spikes/s ($\mu =E$) and $125$ spikes/s ($\mu =I$))
belonging to neuron $i$ (different ones for each cell). The noise strengths
$\eta^{0}_{\mu,i}$ are drawn from a uniform distribution between 1 and 5 if
$i\in{\cal P}_{0}(E)$, are equal to 2 if $i\in{\cal P}_{1}(E)$, are drawn from a
uniform distribution between 0 and 30 if $i\in{\cal P}_{0}(I)$, and are drawn from a
uniform distribution between 16 and 46 if $i\in{\cal P}_{1}(I)$.

\subsection{Geometric parameters and LGN input} 
The sets $N^{LGN}_{Q,j}$ are constructed as follows.
Our 4x4 ${mm}^{2}$ modelcortex is partitioned into 8 parallel bands (0.5x4 ${mm}^{2}$),
which alternate representation between the two eyes.  Subsequently, initial 
retinotopic maps for each eye were defined as the identity
map plus scatter as follows.  The $30\%$ (members of ${\cal
P}_{1}(\mu)$) of the neurons that receive LGN input, were assigned
initial RF centers ($\vec{y}_{k}$) depending on their positions
($\vec{x}_{k}$) in the model cortex via $\vec{y}_{k}=\nu \vec{x}_{k} +
\gamma \vec{\rho}_{k}$, where $\nu$ is the inverse cortical
magnification factor, $\gamma$ is the initial RF scatter and
$\vec{\rho}_{k}$ are scatter vectors, components of which are drawn
from a uniform distribution on $[-1,1]$. The parameter $\nu$ is
$0.2^{\circ}/mm$ for $0^{\circ}$ eccentricity (M0,P0) and
$0.7^{\circ}/mm$ for $10^{\circ}$ eccentricity (M10,P10), which
correspond to cortical magnification factors at the lower end of their
experimentally observed range in macaque \cite{hub74,ess84,too88}.  The
scatter parameter $\gamma$ is $0.3^{\circ}$, $0.1^{\circ}$,
$0.35^{\circ}$ and $0.35^{\circ}$ for M0, P0, M10 and P10
respectively.  These scatter values are in the experimentally observed
range \cite{hub74,dea99} and moreover assure a more or less uniform
initial distribution of cortical receptive fields.

Next, the neurons that were assigned initial receptive field centers are
connected to LGN axons of the corresponding eye to set up their
orientation preference (pinwheels).  Besides an initial RF
center, each cell in ${\cal P}_{1}(\mu)$ (for each eye)
is assigned a template for the organization of the ON and OFF
subfields of its initial RF, which is randomly chosen from the 4
basic symmetry configurations seen experimentally \cite{dea99}.
LGN-cortical connections are initially made so as to best approximate
the assigned initial receptive field center and template for each cell. Then, the
LGN-cortical connections are rearranged by disconnecting and
reconnecting cells, so as to achieve LGN axon sizes that agree with
the anatomical findings for macaque \cite{bla83,fre89}.	For each
disconnection, a new connection is selected from all possible LGN axon
candidates (in cortex), such that the LGN cell belonging to this newly
selected axon has a receptive field closest to the receptive field of the LGN cell that was
disconnected. If no candidates were available no reconnection was
made. The constraint put on the sizes of the LGN axonal arbors in
agreement with their experimentally observed maximum sizes is important,
since we find that this puts considerable restrictions on the possible
connections, leading to a realistic local scatter in preferred angle 
(as observed experimentally \cite{mal97}).
 
In our description of the LGN input we use the term ``initial'' when referring to 
receptive fields and retinotopic maps, since these were only assigned to build the
connections between LGN cells and a fraction ($30\%$) of the cortical
cells. Actual receptive fields and retinotopic maps in our simulations depend on
the final connections. In addition, they are largely shaped by
the cortical interactions, since there are many cells (70\%) that do
not receive LGN input.  The actual retinotopic maps are approximately
identity maps, but with scatter somewhat smaller than in the
initial maps. We see a rich variety in shapes, sizes and
organization of ON and OFF subfields, much like what is seen
experimentally \cite{dea99}.  The receptive field sizes and spatial
and temporal frequency tuning properties of our model cells also agree
well with experimental data.

\subsection{Data collection}
Preferred angles, spatial and temporal frequencies are obtained by
using a drifting grating stimulus for one eye (other eye $I=0$), with
high contrast ($\epsilon =1$), in a large aperture ($r_{A} \sim$ 7-10
times the average receptive field size). Responses to contrast-reversal stimuli, i.e.
$\cos (\,\omega t - \vec{k}\cdot\vec{y}+\phi)$ of the drifting grating stimulus is replaced
by $\cos (\,\omega t)\cos (\vec{k}\cdot\vec{y}+\phi)$, are obtained in a similar way. 
The receptive field centers are mapped using a drifting grating stimulus for one eye
(other eye $I=0$), with high contrast ($\epsilon =1$), in a small
aperture ($r_{A} \sim$ 1/4th of the average receptive field size).  The apertures
are centered on a rectangular grid (5x5 or 6x6, grid spacing about half of the average
RF size) which itself is centered on the visual field covered by our model cortex.  The angle,
spatial and temporal frequencies of the grating are kept fixed during
this experiment. The temporal frequencies are set to the averaged preferred values
for the case under consideration.  The preferred parameters and the receptive field 
centers are from averaged spike responses, using the mean response for
complex cells and the first harmonic for simple cells.	Each stimulus
presentation (all parameters fixed except time) lasts for 4 s and is
preceded by a 1 s blank stimulus ($\epsilon =0$). This procedure is
repeated twice, starting with different initial conditions and different external noise realizations.  
Average responses are obtained by averaging over cycles and trials.

The primary data, i.e. responses and conductances as a function of aperture size for single eye
stimulation, are obtained with the temporal and spatial frequencies of the grating set
equal to the averaged preferred values for each case (M0, M10, P0, P10).
Data samples consist of cells that have their preferred angle equal to the grating angle,
their preferred temporal frequency within 2 Hz of the grating frequency, a preferred
spatial frequency $k_{p}$ that satisfies $\frac{1}{2}k<k_{p}<2k$ where $k$ is the grating
spatial frequency, a receptive field center that is less than $\frac{1}{20}$th of the average RF
size away from the aperture center, a maximum response at low contrast that
is greater than $f_{b}+5$ where $f_{b}$ is the mean blank response
(in spikes/s), and finally, a central cortical location confined to the dashed white rectangle in
Figure 1 of the paper. In this way we collect approximately 200 cells in each sample.

The experiments were performed at ``high'' contrast, $\epsilon =1$, and
``low'' contrast, $\epsilon =0.3$. In our model contrast dependence 
of the LGN cells is approximately linear.
Most cortical model cells, consequently, also have a fairly linear contrast response function.
Therefore, the numerical contrast values used in the model should not directly be compared with
numerical contrast values used in experiments with animals, but rather they should
be compared only with the linear part of the contrast response function, with $\epsilon =1$
being the ``high'' part and $\epsilon =0.3$ being the ``low'' part.
Note that this also means complications resulting from response saturation at high contrast are
not present in our model. Indeed, our findings presented in this paper are insensitive to using
instead of $\epsilon_{\mbox{high}} =1$ a lower value of $\epsilon_{\mbox{high}} =0.8$, for which 
high contrast responses are approximately 80\% of their saturation values.
Further, when we model the contrast response functions of M-LGN cells more realistically,
using experimental data from \cite{sol02}, and confine our experiments 
to the linear part of the contrast response functions, our findings remain unchanged.

\subsection{Simplified model equations} 
We rewrite Eq. (1) of the paper as \cite{wie01}
\begin{equation}
\label{eq:main}
\frac{dv_{k}}{dt} = -g_{T,k}(t,r_{A})\; v{_k} + I_{D,k}(t,r_{A}) \; ,
\end{equation}
where
\begin{equation}
\label{eq:gtotm}
g_{T,k}(t,r_{A})=g_{L} + g_{E,k}(t,[{\cal S}]_{E},r_{A}) + g_{I,k}(t,[{\cal S}]_{I},r_{A})
\end{equation}
\begin{equation}
\label{eq:diffcm}
I_{D,k}(r_{A},t) = g_{E,k}(t,[{\cal S}]_{E},r_{A})\;V_{E} -
g_{I,k}(t,[{\cal S}]_{I},r_{A})\;\left| V_{I}\right| \; .
\end{equation}
Recall that our model is in fact a system of many strongly coupled equations.
Together they determine the total conductance $g_{T,k}(t,r_{A})$ and
difference current $I_{D,k}(t,r_{A})$ which govern the behavior of a model cell.
Our model cortex operates at large conductance levels as part of the requirement to achieve 
a broad range of known properties of the biological visual cortex.
Indeed, several experiments have demonstrated large cortical conductances under {\em in vivo} visual 
stimulation, see e.g. \cite{des03} for a review.
A large total conductance implies that the membrane potential is well-approximated simply by
Ohm's law $v_{k}(t,r_{A})\approx I_{D,k}(t,r_{A})/g_{T,k}(t,r_{A})$. We also find in our model
that the cycle-to-cycle and trial-to-trial fluctuations in $g_{T,k}(t,r_{A})$ are considerably
smaller than the mean. Therefore, we find that to a good approximation \cite{wie01}
\begin{equation}
\label{eq:vmemm}
\left<v_{k}(t,r_{A})\right>\; \approx\; V_{k}(r_{A},t) \; \equiv\;
\frac{\left<I_{D,k}(t,r_{A})\right>}{\left<g_{T,k}(t,r_{A})\right>}\; .
\end{equation}
Further, we find that for our model cells the spike rates are well-predicted by a
rectification model which assumes the firing rate to be proportional to the rectified
total current at the spike threshold ($v_{k}=1$),
\[
\left<{\cal S}_{k}(t,r_{A})\right>\; \approx\; f_{k}(t,r_{A})\; \equiv\;
\]
\begin{equation}
\label{eq:frm}
 \delta_{k}\left[
\left<I_{D,k}(t,r_{A})\right> -\left<g_{T,k}(t,r_{A})\right>-\Delta_{k}\; \right]_{+} \; ,
\end{equation}
where $[x]_{+}=x$ if $x\geq 0$ and $[x]_{+}=0$ if $x\leq 0$, and where, to a good approximation,
the gain $\delta_{k}$ and threshold $\Delta_{k}$ do not depend on the aperture radius $r_{A}$
for most cells. With use of expression (\ref{eq:vmemm}) for the membrane potential,
our rectification model (Eq. \ref{eq:frm}) is equivalent to the standard rectification model \cite{gra63}
$\left<{\cal S}_{k}(t,r_{A})\right>\; \approx\; \tilde{\delta}_{k} \left[ \left<v_{k}(t,r_{A})\right>
- v_{T,k} \right]_{+}$, where the gain parameter $\tilde{\delta}_{k}$ and threshold $v_{T,k}$ however depend
on the total conductance, that is, on aperture size and time.
The standard rectification model with constant (cell specific) gain and threshold parameters has
been shown to work well in cat \cite{car00} for large drifting grating stimuli of varying
angles.
Our rectification model gives a significantly better fit for our stimuli than the
standard rectification model (with fixed gain and threshold). Some fits are shown in Figure 10.
It provides a relation between firing rate and excitatory and inhibitory conductances in
which these conductances act simply via addition and subtraction, respectively.
Since the parameters are constants it allows us to base our analysis of the firing rate
(as with the membrane potential) directly on the behavior of the excitatory and inhibitory
conductances as a function of aperture size and time.
\begin{figure}[here]
\centering
\begin{minipage}{1\columnwidth}
\centering
(A)\hspace{2.75in}(B) \\
\includegraphics[height=3cm,width=4cm]{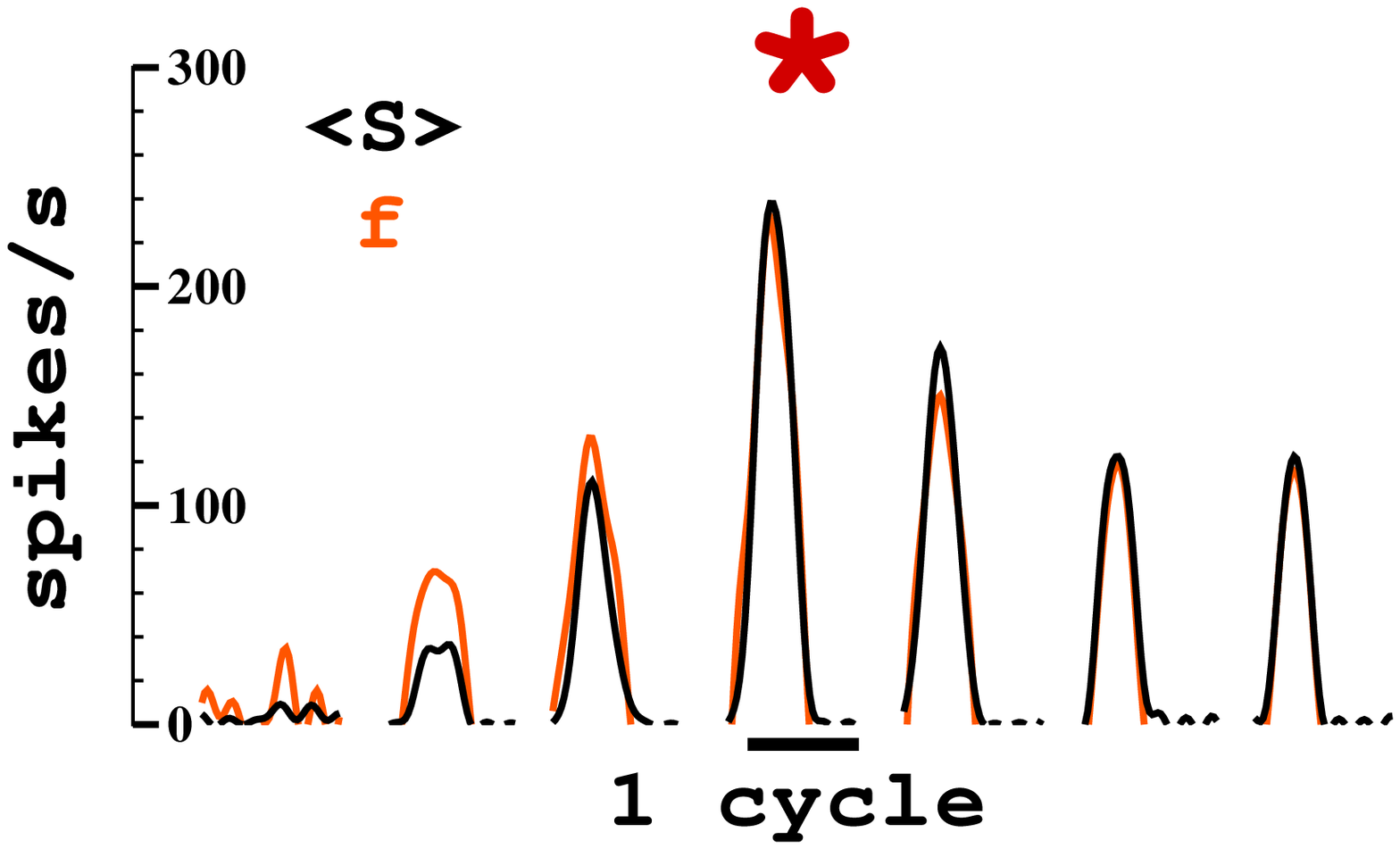}
\includegraphics[height=3cm,width=4cm]{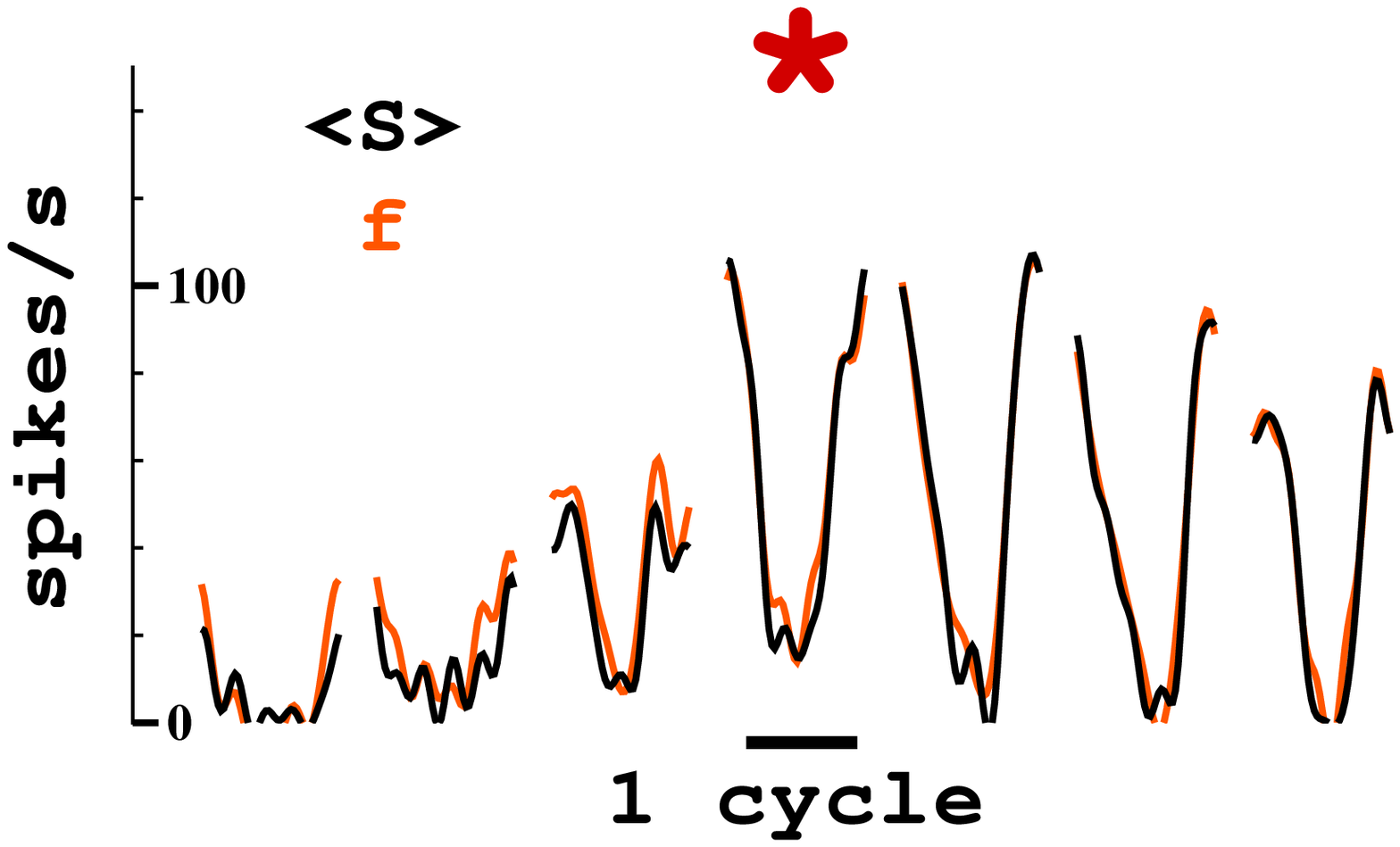}
 \\
(C)\hspace{2.7575in}(D)\\
\includegraphics[height=3cm,width=4cm]{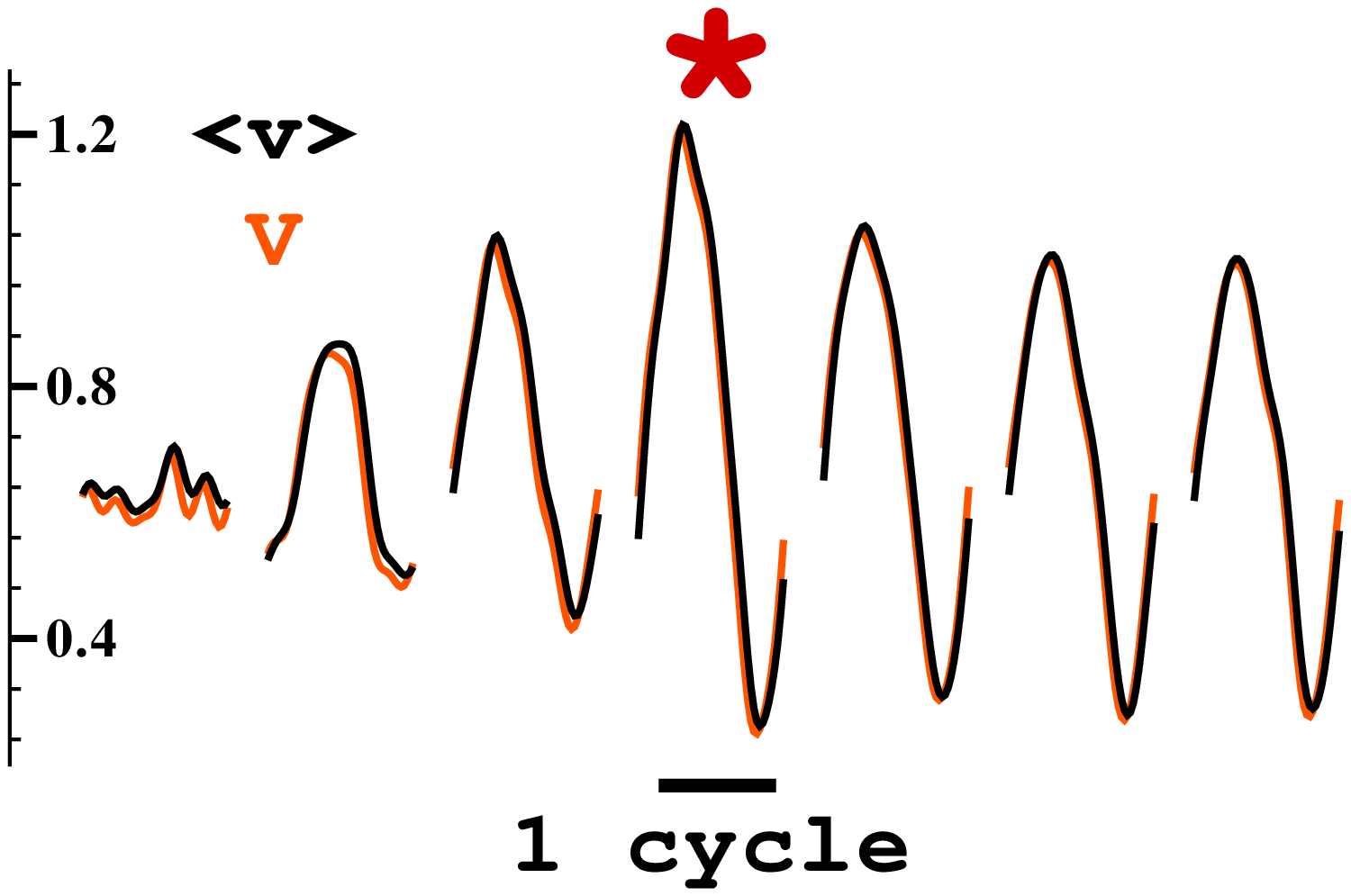}
\includegraphics[height=3cm,width=4cm]{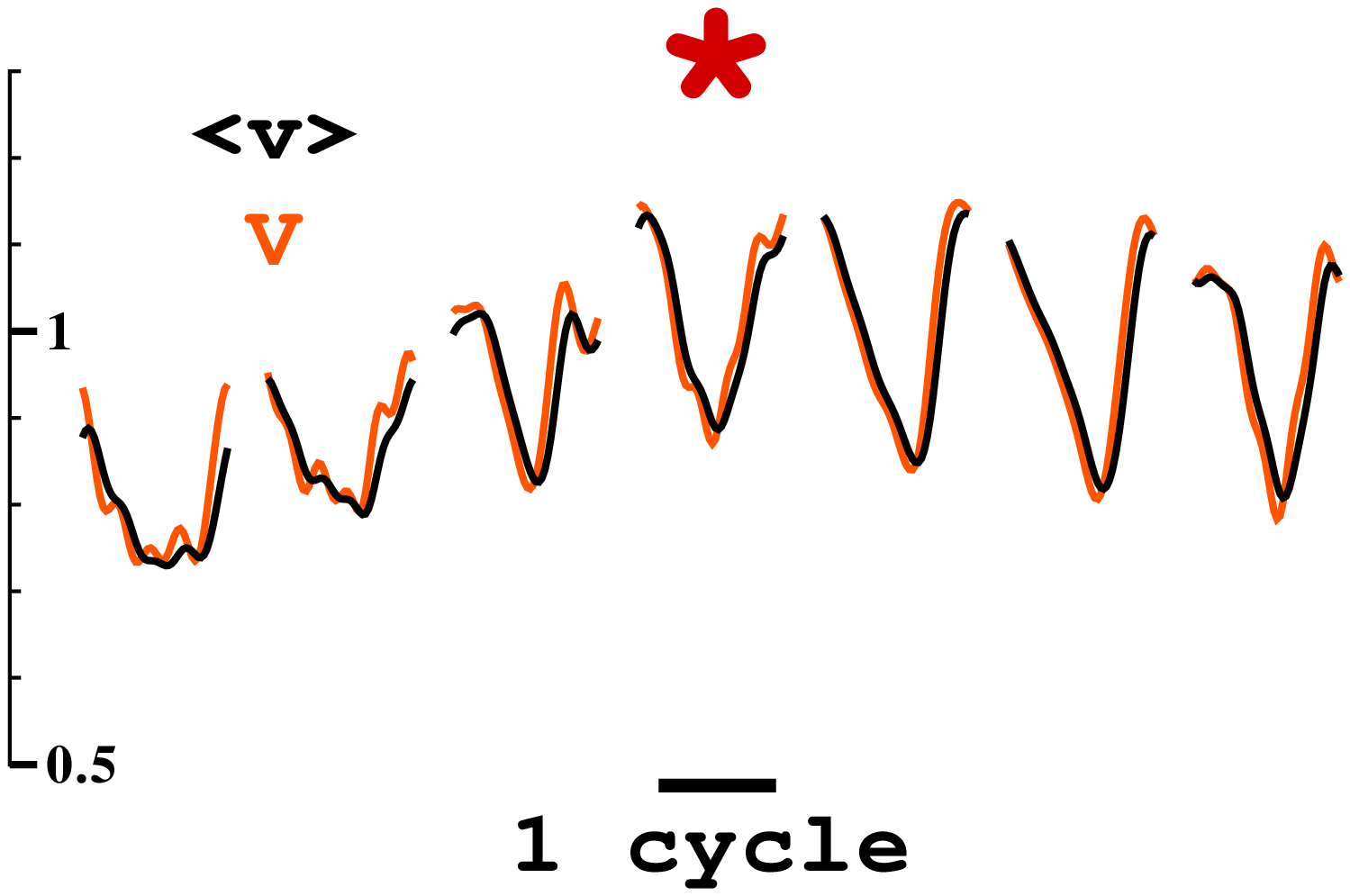}
\end{minipage}
\caption{\small Some examples of the approximations given by Eq. (6) and (7) of the paper for a
simple cell (left) and a complex cell (right) for apertures around the aperture of the
maximum response. Figure layout is as Figure 6 of the paper.
(A \& B) Comparison of the instantaneous firing rate $\left<{\cal S}(t,r_{A})\right>$ to
  the rectification model approximation  $f(t,r_{A})$ (Eq. \ref{eq:frm}). Note that the
  approximation is reasonable for all aperture sizes, but better for those that impinge on
  the suppressive surround. Fitting parameters are $\delta = 1.2$, $\Delta = -100$
  $s^{-1}$ (left) and $\delta = 1.8$, $\Delta = -25$ $s^{-1}$ (right).
  (C \& D) Comparison of model responses $\left<v(t,r_{A})\right>$ with the Ohmic (slaving)
  approximation $V(t,r_{A})$ (Eq. \ref{eq:vmemm}), indicating the model's operation at high
  conductance states.
 }
\end{figure}

\subsection{Simulated optical imaging}
To simulate optical imaging experiments \cite{bla92} for ocular dominance and orientation
preference, high contrast drifting gratings in a large aperture are presented
to the left and right eye separately, temporal and spatial frequencies of the grating are set equal
to averaged preferred values.
Stimulus presentation is as discussed for finding the preferred angles etc..
We assume the optical signal at pixel $\ell$ arising from a single eye (L or R) stimulation with
a grating with angle $\theta_{k}$ to be
\begin{equation}
r_{\ell,Q}^{\mbox{opt}}(\theta_{k} ; n) \propto\:
\frac{1}{n} \sum_{j\in{\cal N}_{n}(\ell)} r_{j,Q}(\theta_{k})  \; ,\hspace{1cm} Q=\mbox{L or R} \; ,
\end{equation}
where $r_{j,Q}(\theta_{k})$ is the mean (spike) response of cell $j$ and
${\cal N}_{n}(\ell)$ is the neighborhood of the nearest $n$ cells to pixel $\ell$.
Using this assumption we can produce an arbitrary ``optical'' imaging result for our model
by further processing of $r_{\ell,Q}^{\mbox{opt}}(\theta_{k} ; n)$ in the same spirit as
in optical imaging experiments.
We compute ocular dominance ${\cal O}_{\ell}(n)$ for a given pixel
of our model cortex as
\begin{equation}
{\cal O}_{\ell}(n) = \left\{
\begin{array}{ll}
\mbox{L} & \mbox{if}\:\: \sum_{k}r_{\ell,L}^{\mbox{opt}}(\theta_{k} ; n)\: \geq
\:\sum_{k}r_{\ell,R}^{\mbox{opt}}(\theta_{k} ; n)\\
\mbox{R} & \mbox{else}
\end{array}\right. .
\end{equation}
Orientation preference ${\theta}_{P,\ell}(n)$ (preferred angle of pixel $\ell$) is computed from the
``averaged difference vector'' similarly as in actual experiments \cite{bla92},
\begin{equation}
{\theta}_{P,\ell}(n) = \frac{1}{2} \arg \sum_{k}
\left\{ \sum_{Q=L,R} r_{\ell,Q}^{\mbox{opt}}(\theta_{k} ; n) \right\} e^{2i\theta_{k}}\; .
\end{equation}
Results in Figures 1 and 9 of the paper are obtained with $n=75$.

\subsection{Extraclassical spatial summation}
\label{sec:ss}
A summary of our data for the four configurations M0, M10, P0, and P10 is provided
in Figure 11. The figure is organized in columns and rows, each column corresponding to a particular 
configuration, and each row corresponding to a particular response measure. All results shown are 
based on spike train responses.
The second row shows the distributions of receptive field size for high 
and low contrast ($r_{+,-}$), the third row shows the distribution of the surround size at 
both contrasts ($R_{+,-}$).
The distributions of receptive field size and surround size show good agreement with 
experimental data \cite{cav02}. Notice the growth of both receptive field size and 
surround size for low contrast.
Also, note the considerable diversity in receptive field and surround sizes in the 
model. This is not a result of any variability in the LGN receptive field sizes, all LGN cells
for a particular configuration have identical receptive field size. Instead, the diversity
seen in the receptive field and surround sizes of the cortical cells is a 
result of the cortical interactions and realistic constraints on LGN axon sizes. 
\begin{figure}[here]
\label{data}
\centering
\begin{minipage}{1\columnwidth}
\centering
\includegraphics[height=2.5cm,width=2cm]{fig2_1L.eps}
\includegraphics[height=2.5cm,width=2cm]{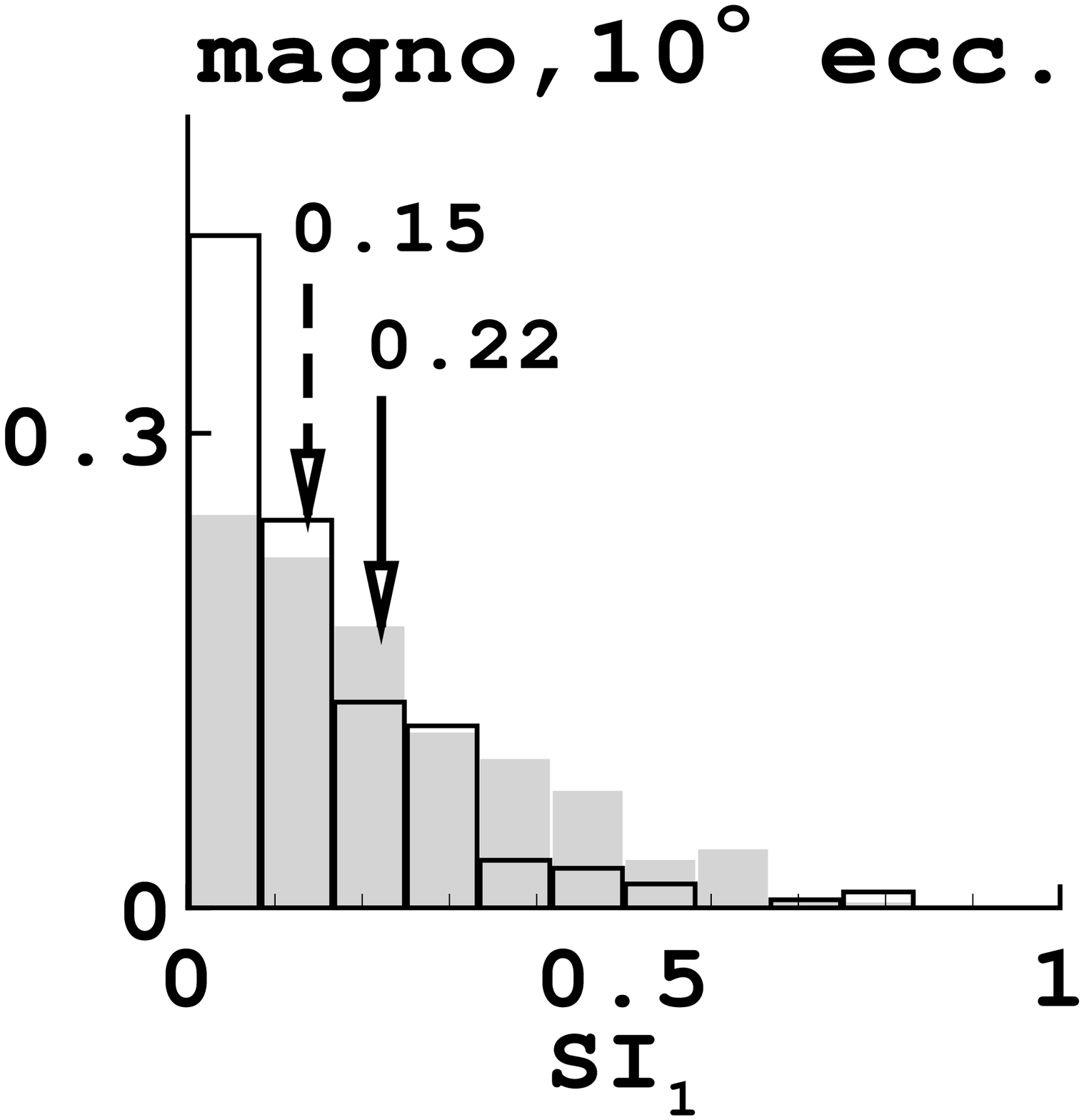}
\includegraphics[height=2.5cm,width=2cm]{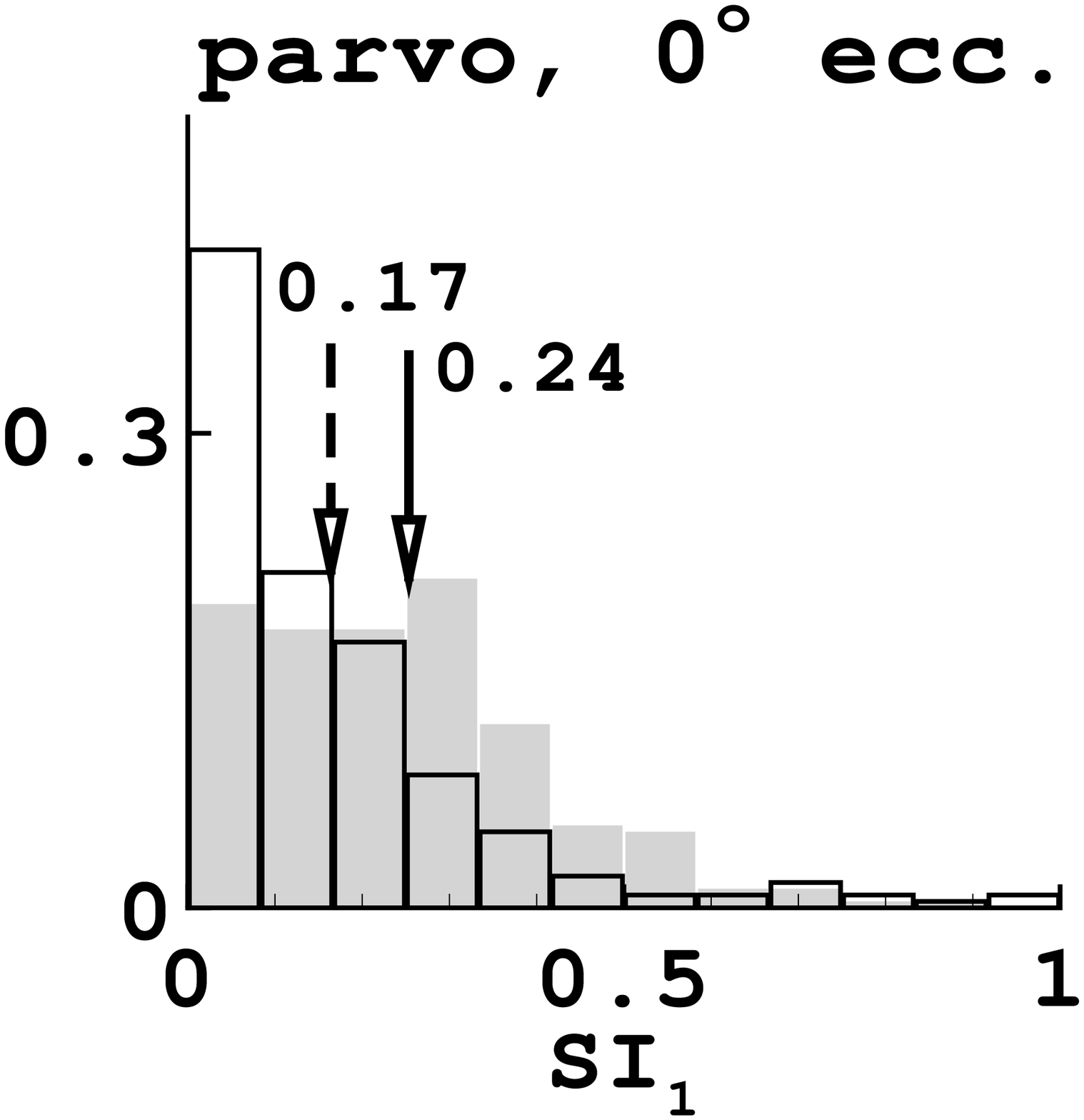}
\includegraphics[height=2.5cm,width=2cm]{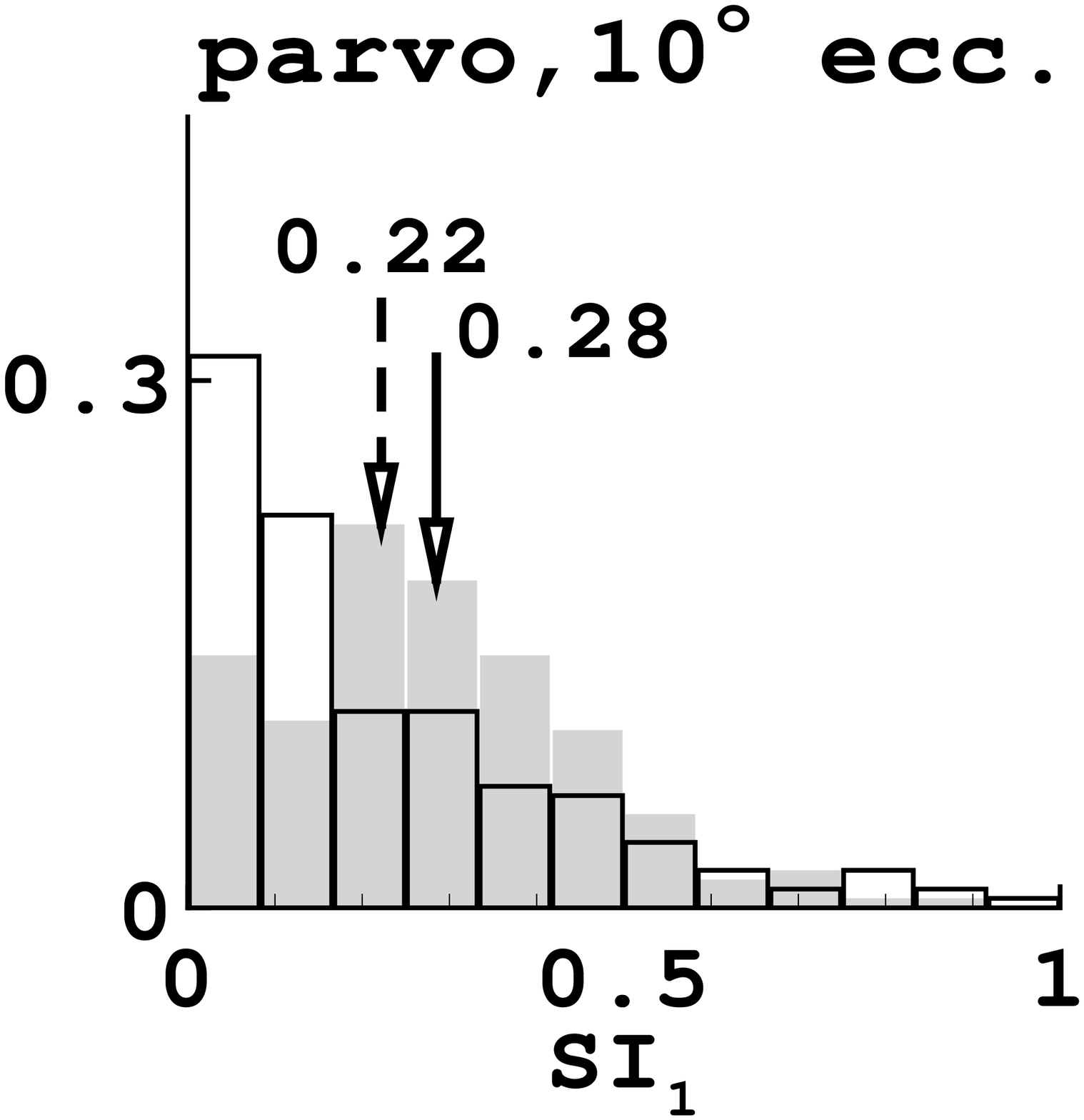}
\includegraphics[height=2.5cm,width=2cm]{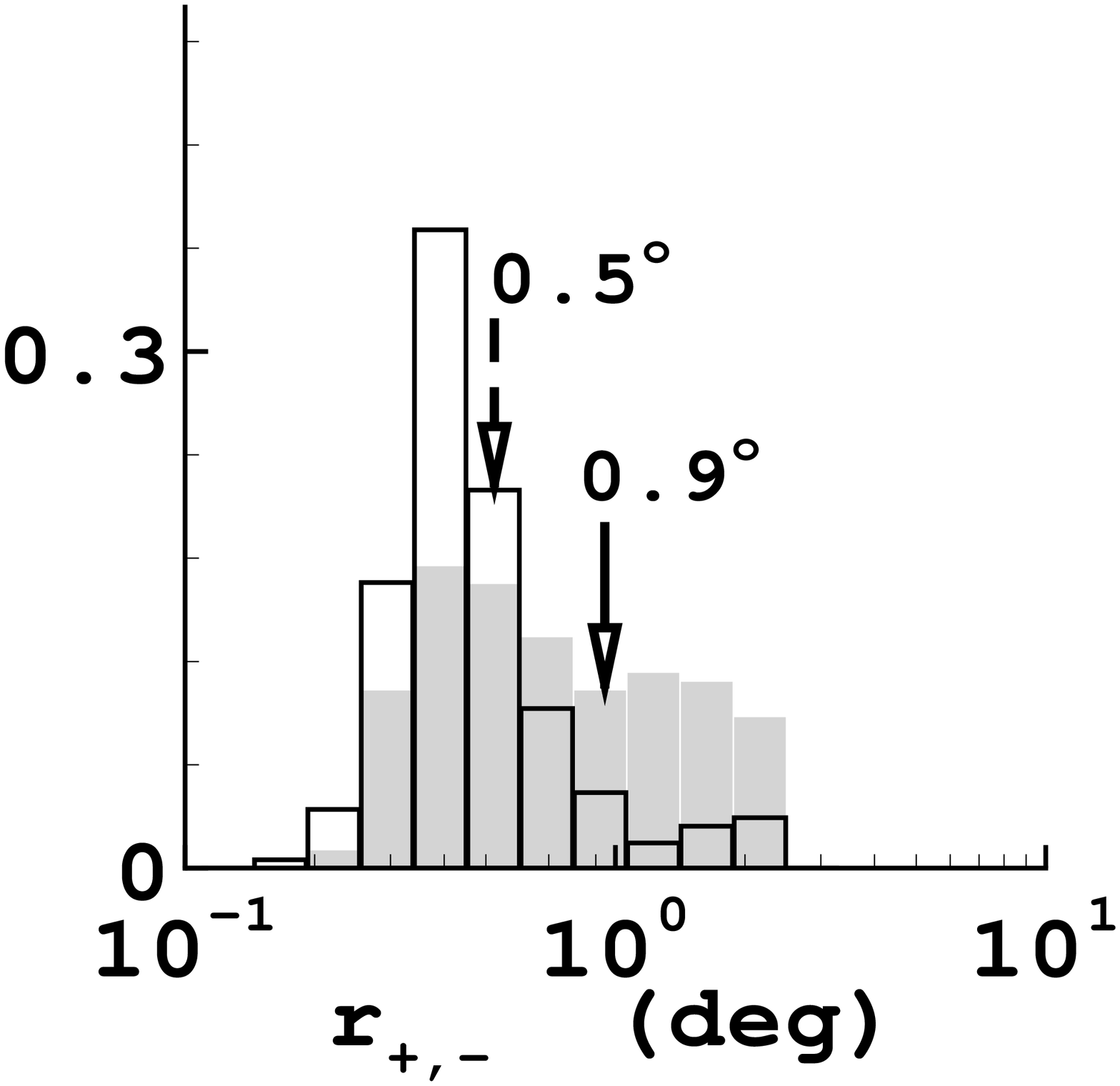}
\includegraphics[height=2.5cm,width=2cm]{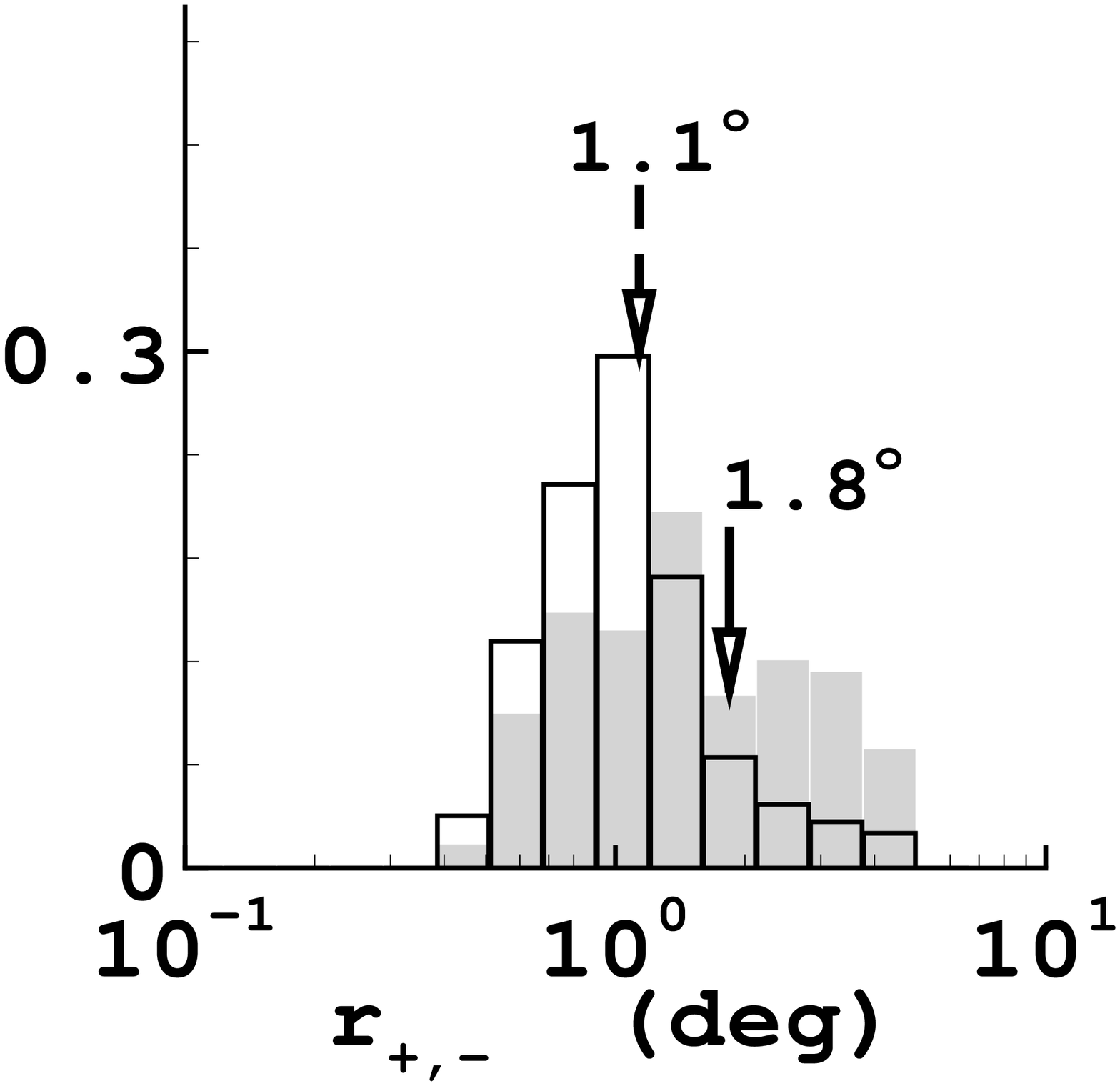}
\includegraphics[height=2.5cm,width=2cm]{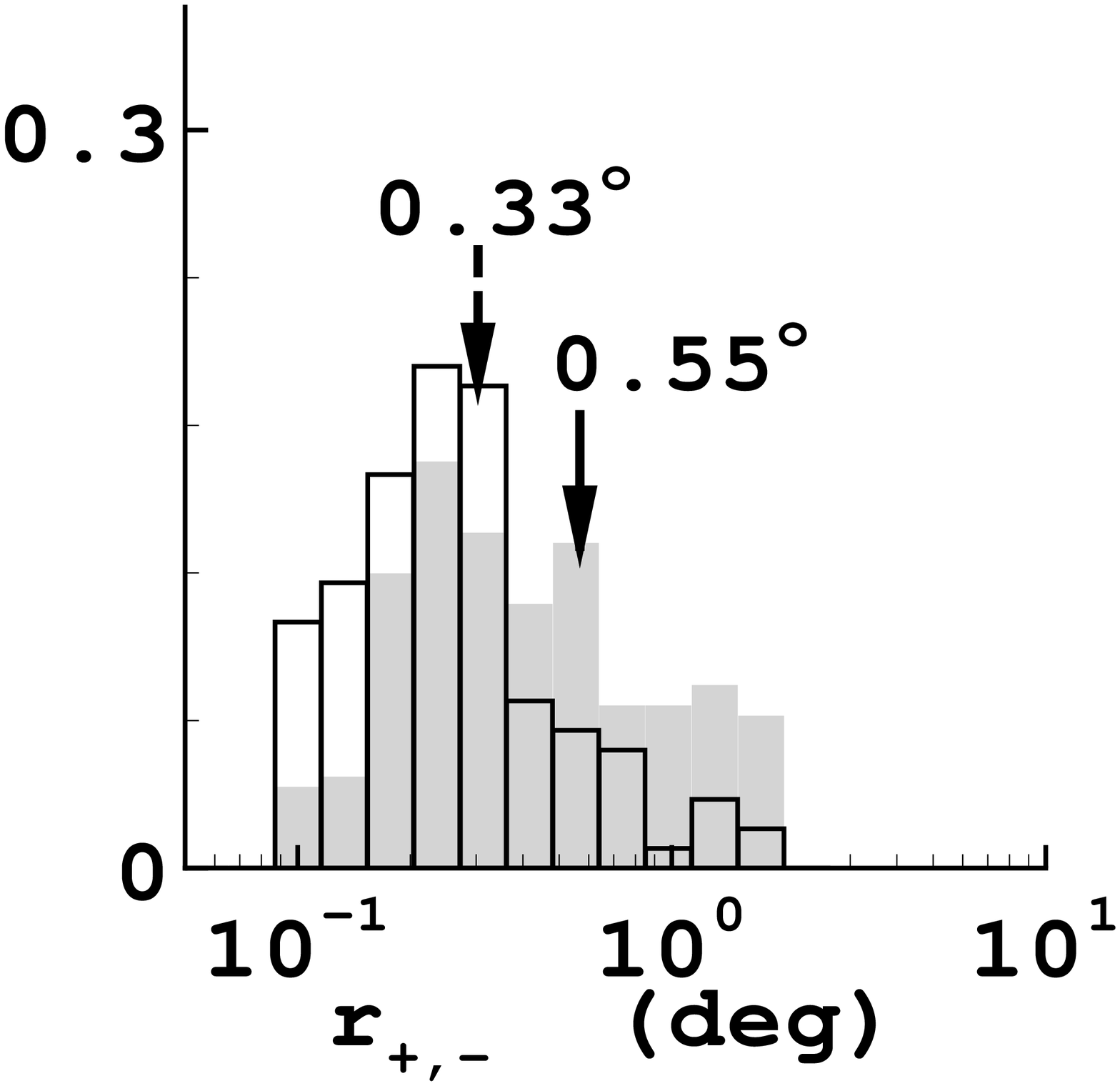}
\includegraphics[height=2.5cm,width=2cm]{fig2_14L.eps}
\includegraphics[height=2.5cm,width=2cm]{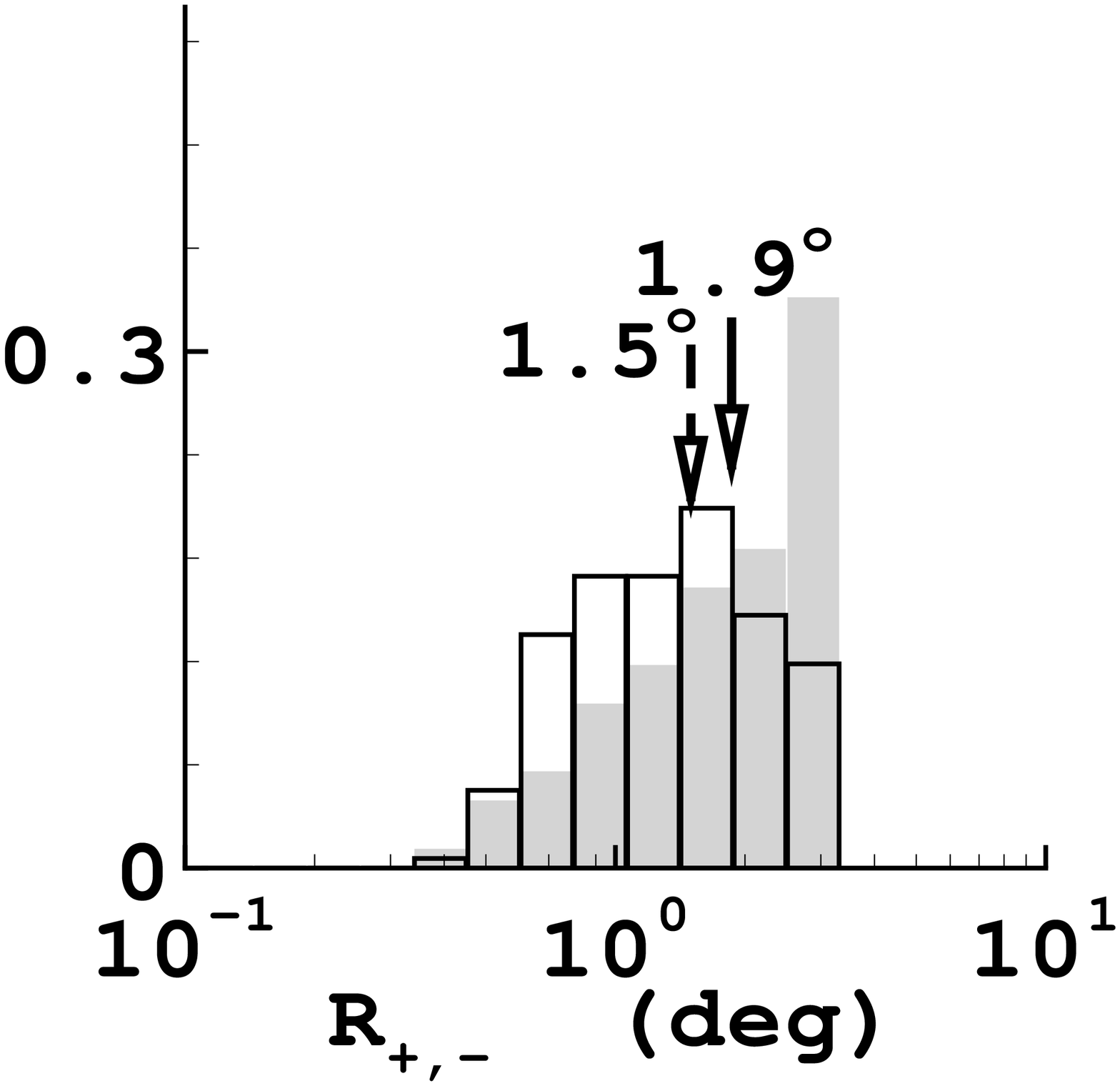}
\includegraphics[height=2.5cm,width=2cm]{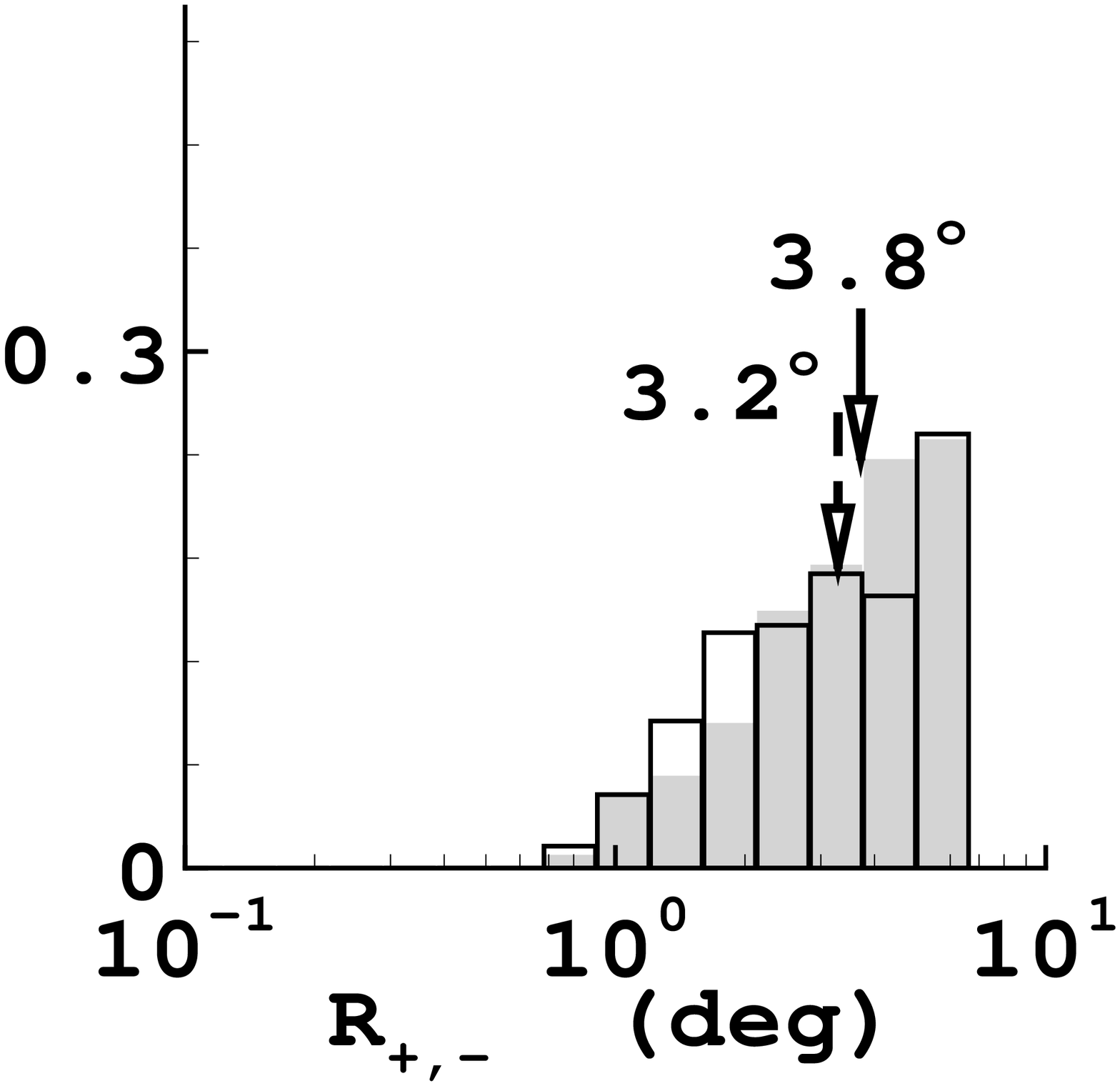}
\includegraphics[height=2.5cm,width=2cm]{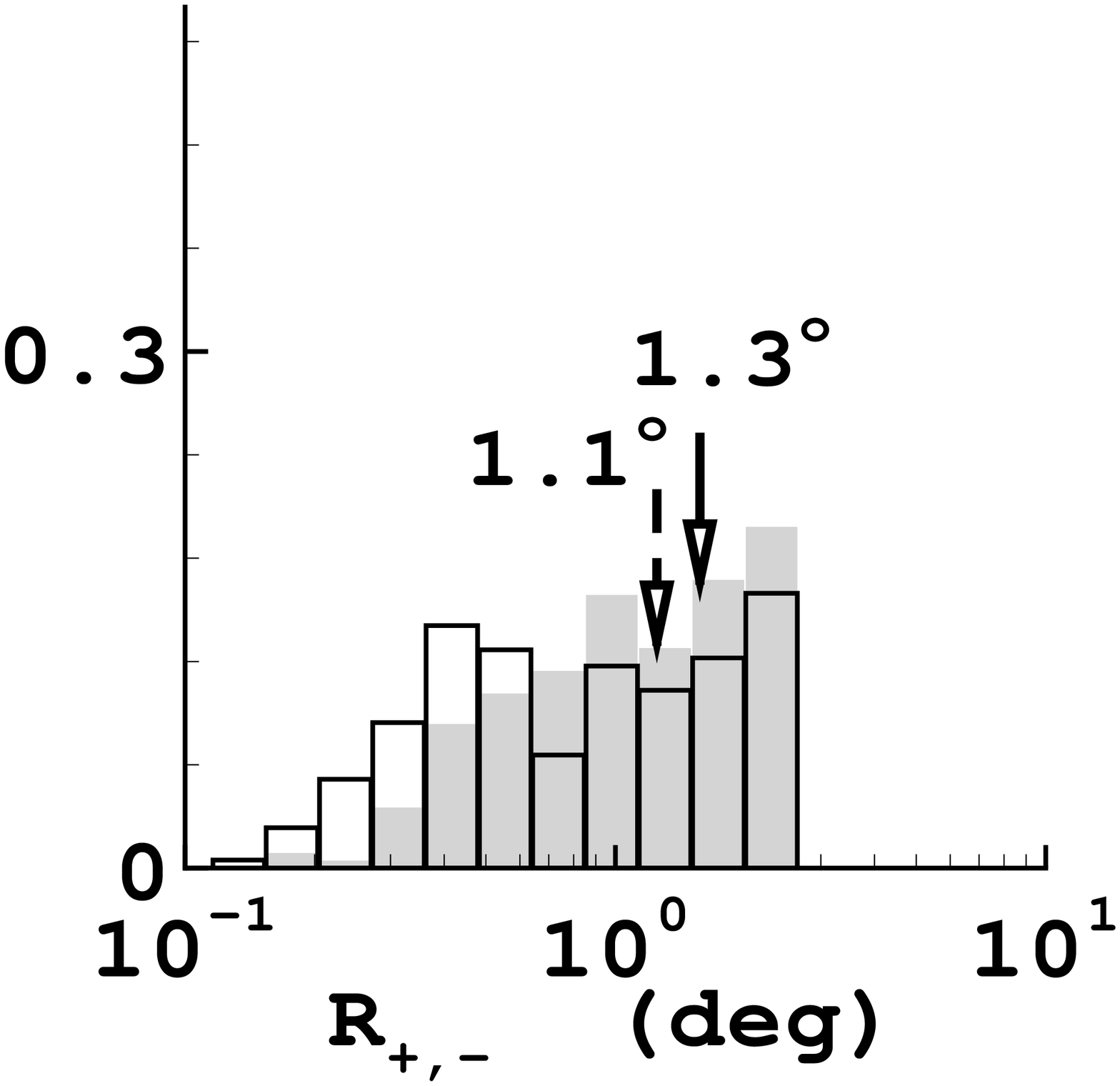}
\includegraphics[height=2.5cm,width=2cm]{fig2_15L.eps}
\includegraphics[height=2.5cm,width=2cm]{fig2_4L.eps}
\includegraphics[height=2.5cm,width=2cm]{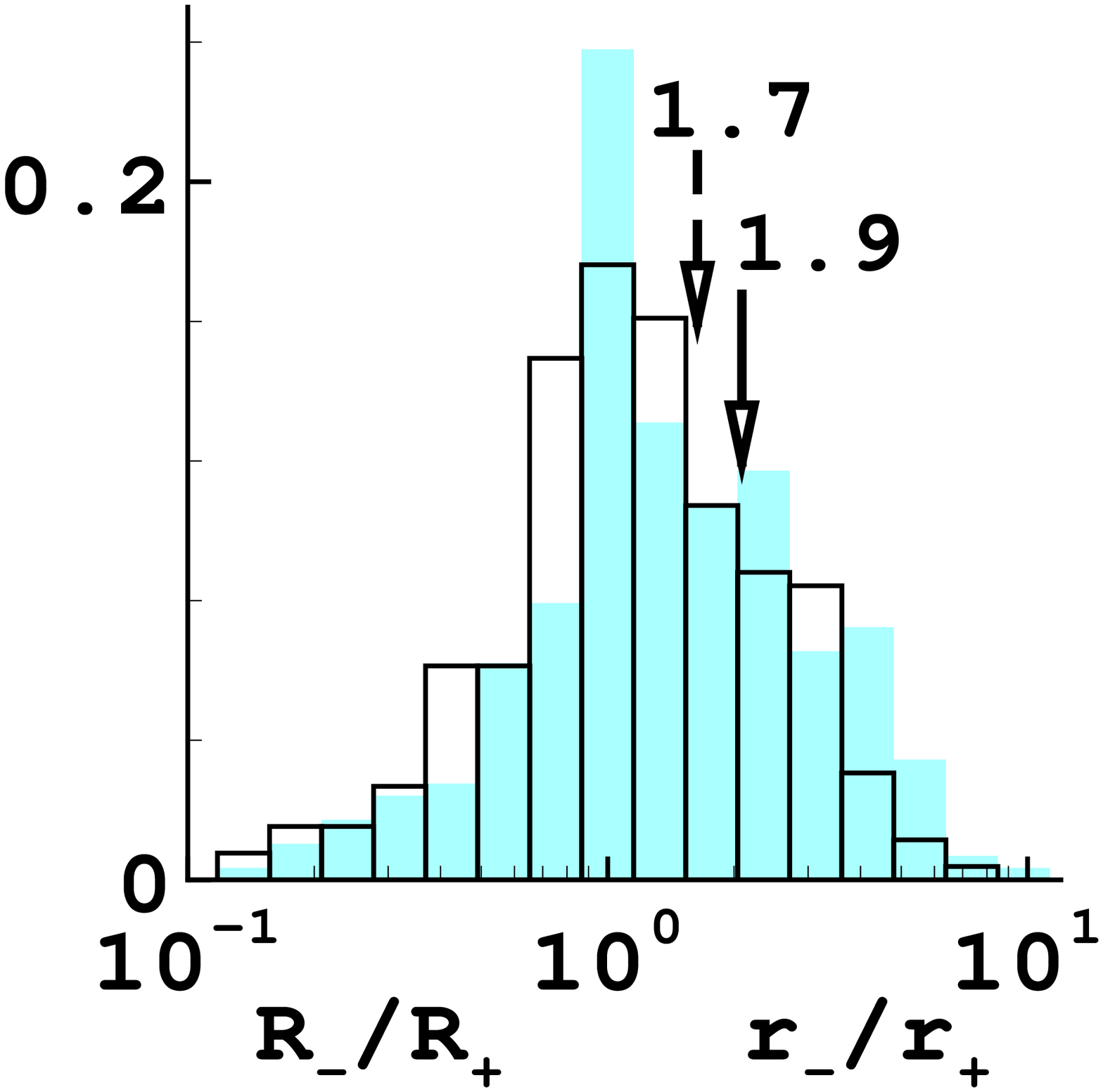}
\includegraphics[height=2.5cm,width=2cm]{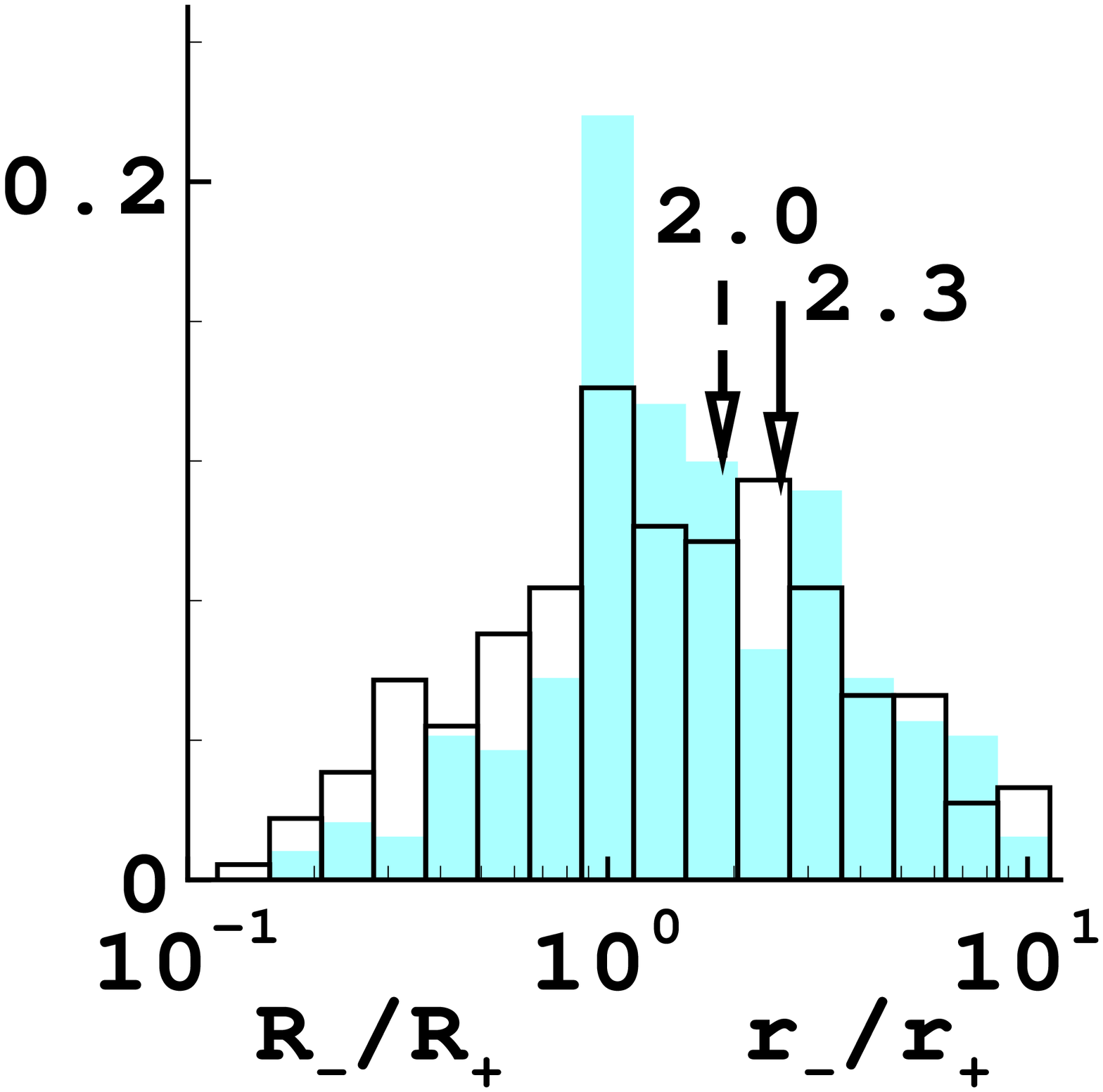}
\includegraphics[height=2.5cm,width=2cm]{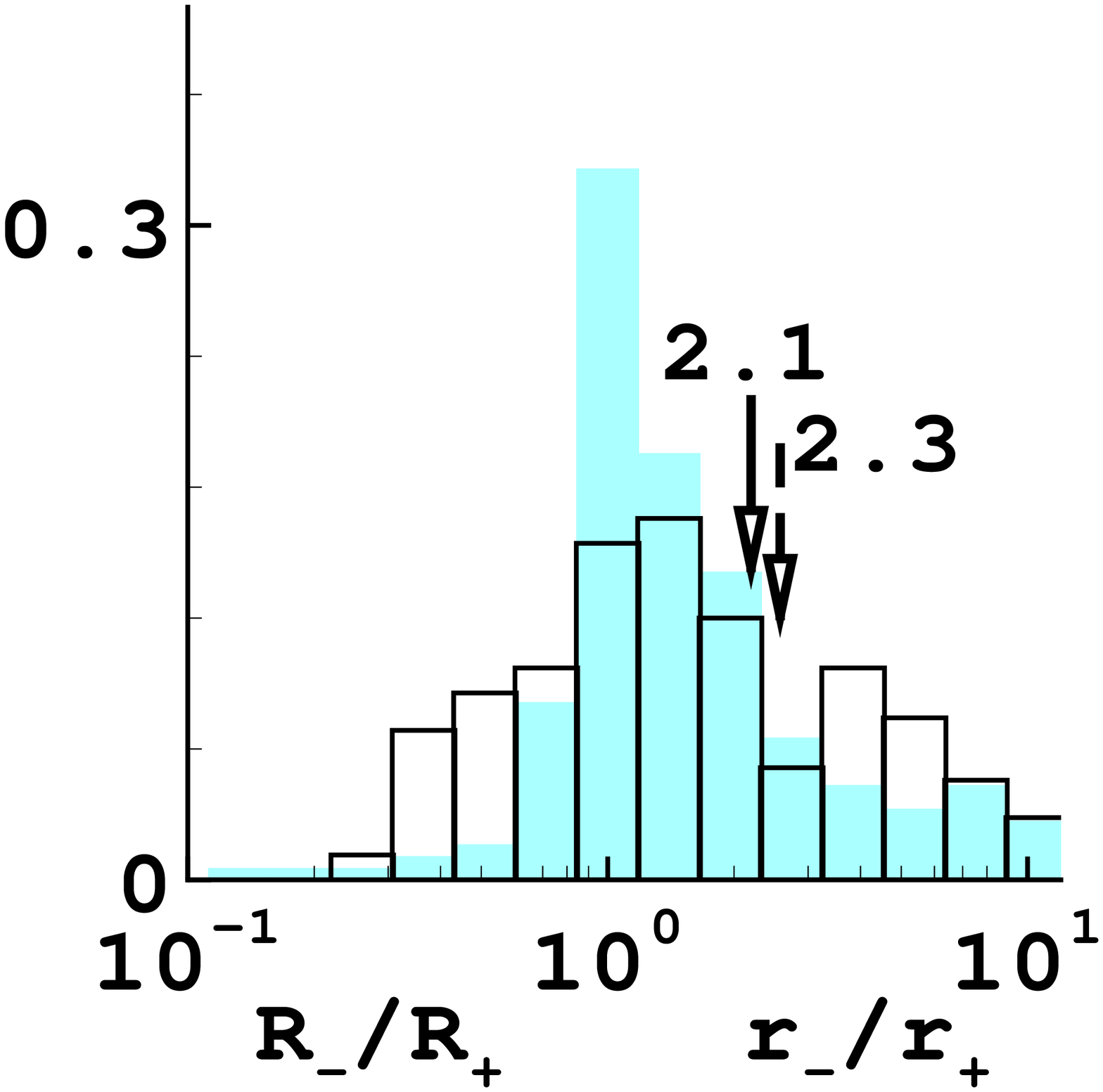}
\caption{\small Summary of extraclassical spatial summation for the four model configuration.
  In rows 1-3 unshaded histograms are for gratings at high contrast
  ($+$), shaded at low contrast ($-$). Throughout the figure, means
  are indicated by arrows, solid arrows refer to shaded histograms,
  dashed arrows to unfilled histograms.  (ROW 1) Distributions of the
  suppression index $SI_1$. 
  All suppression is exclusively due to short-range cortical connectivity. (ROW 2) Receptive 
  field size distributions ($r_{+,-}$). Distributions are shifted to the right for low contrast 
  indicating an increase in receptive field size. Receptive field sizes in the model are in agreement 
  with experimental data \cite{cav02}. (ROW 3) Surround size distributions ($R_{+,-}$) for high and 
  low contrast. Surround sizes are in agreement with experimental data \cite{cav02}. (ROW 4) Histograms of 
  the ratios of the receptive field and surround sizes at low and high contrast, $r_{-}/r_{+}$
  (blue shaded) and $R_{-}/R_{+}$ (unfilled). Growth of surround sizes is about the same as
  for receptive field sizes and each case shows a similar, significant growth of about
  a factor of two (Wilcoxon test on ratio larger than unity: $p<0.001$ for both receptive field 
  and surround growth). 
}
\label{fig:data}
\end{minipage}
\end{figure}

\subsection{Mechanisms of contrast dependent receptive field size}
Consider a situation where both $g_{E}$ and $g_{I}$ have their maxima
at the same aperture size $r_{E}=r_{I}=r_{\star}$ and are
monotonically increasing for $r_{A}<r_{\star}$ and monotonically
decreasing for $r_{A}>r_{\star}$, as depicted in Figure 12A.  We can
distinguish three classes with respect to the relative location of the
maxima in spike responses $r_{S}$ and the conductances $r_{\star}$,
namely \{X: $r_{iS}<r_{\star}$\}, \{Y: $r_{S}=r_{\star}$\} and \{Z:
$r_{S}>r_{\star}$\}. It follows from (\ref{eq:fr}) that if we define the 
parameter $G_{0}(v)=(\left| v_{I}\right| + v)/(v_{E} - v)$ then we can characterize 
the difference between classes X and Z by the way that $G$ crosses $G_{0}(1)$ around
$r_{S}$ as depicted in Figure 12A. For class Y the parameter G is
not of any particular use as it can assume arbitrary behavior around
$r_{S}$. It follows from (\ref{eq:vmem}) that similar observations hold for the maximum 
in the membrane potential $r_{v}$ and we need simply to replace $G_{0}(1)$ with
$G_{0}(v(r_{v}))$.  Obviously, a growth of receptive field can occur without any
change in the spatial summation extent ($r_{\star}$) of the
conductances.  Suppose we wish to remain within the same class X or Z,
then receptive field growth, can be induced, for instance, by an overall increase (X)
or an overall decrease (Z) in relative gain $G(r_{A})$ as shown in
Figure 12A (dashed line). Receptive field growth also can be caused by more
drastic changes in $G$ so that the transitions X $\rightarrow$ Y, X
$\rightarrow$ Z or Y $\rightarrow$ Z occur for a high $\rightarrow$ low contrast
change. The situation is somewhat more involved when we allow for non-suppressed 
responses and conductances, and for different positions of the maxima of $g_{E}$ and
$g_{I}$, however, the essence of our conclusions remains the same.

From a more precise analysis based on the relative gain parameter $G$, we find that for more than 50\% of the cells with significant receptive field growth, a transition takes place from a high contrast RF size less or equal to the spatial summation extent of excitation and inhibition, to a low contrast receptive field size which exceeds both. This analysis is summarized in the remainder of Figure 12.

Cells were classified (Fig. 12B) according to the relative positions of their maxima in spike response
($r_{S}$) and excitatory ($r_{E}$) and inhibitory ($r_{I}$)
conductances, using F0+F1 components. Membrane potential
responses yield similar results.  Comparing this classification at
high and low contrast we observe a striking difference for cells with
significant receptive field growths, i.e. with growth ratios $>$1.5 (Fig. 12B,
bottom), indicative of X $\rightarrow$ Y, X $\rightarrow$ Z and Y
$\rightarrow$ Z transitions (as discussed in the simplified example
above). 

 In this realistic situation there are of course many more
transitions (i.e. $13^{2}$), however, that we indeed observe a prevalence
for qualitatively these transitions can be demonstrated in two ways using
slightly modified definitions of the X,Y,Z classes.  First (Figure 12C,
left), if we redefine the X,Y,Z classes with respect to $r_{S}$ and
$r_{E}$ while ignoring $r_{I}$, i.e. \{X: $r_{S}<r_{E}$\}, \{Y:
$r_{S}=r_{E}$\} and \{Z: $r_{S}>r_{E}$\}, then the transition
distribution for cells with significant receptive field growth shows that in about
60\% of these cells a X $\rightarrow$ Z or Y $\rightarrow$ Z
transition occurs.  Taken together with the fact that roughly 10\% of
the cells with significant receptive field growth (Figure 12B, bottom) have $r_{I}\leq
r_{S}< r_{E}$ at high contrast and $r_{E}<r_{S}\leq r_{I}$ at low
contrast, we can conclude that for more than 50\% of the cells with
significant receptive field growth, a transition takes place from a high contrast RF
size less or equal to the spatial summation extent of excitation and
inhibition, to a low contrast receptive field size which exceeds both (by at least
one aperture).  Note that these transitions occur in addition to
any growth in $r_{E}$ or $r_{I}$.  Secondly (Figure 12C, right), the same
conclusion is reached when we redefine the X,Y,Z classes with respect
to $r_{S}$ and $r_{I}$ while ignoring $r_{E}$ (\{X: $r_{S}<r_{I}$\},
\{Y: $r_{S}=r_{I}$\} and \{Z: $r_{S}>r_{I}$\}), Now a X $\rightarrow$
Z or Y $\rightarrow$ Z transition occurs in about 70\% of the cells
with significant receptive field growth, while about 20\% of the cells with
significant receptive field growth (Figure 12B, bottom) have $r_{E}\leq r_{S}<r_{I}$
at high contrast and $r_{I}<r_{S}\leq r_{E}$ at low contrast.

Figure 12C also demonstrates the presence of a rich diversity in relative
gain changes in our model, since all transitions (for all cells,
unfilled histograms) occur with some reasonable probability.  Finally,
Figure 12C establishes that there is a relationship between the difference
in the prevalence of the surround suppression mechanisms at high and low
contrast and receptive field growth. To see this, first note
that for the redefined Y and Z classes with respect to $r_{S}$ and $r_{I}$, 
the surround suppression, if any, must be caused by mechanisms B or C.  Thus, since 
the total probability (Figure 12C, right) of transitions $\cdot\rightarrow$ Y and 
$\cdot\rightarrow$ Z (not including Y $\rightarrow$ Y, Y $\rightarrow$ Z 
and Z $\rightarrow$ Z) is clearly larger than for transitions 
$\cdot\rightarrow$ X (not including X $\rightarrow$ X), this means that surround suppression 
mechanisms B and C must be more prevalent at low contrast than at high 
contrast, which is confirmed by our data (not shown).

\begin{figure}[here]
\label{moredata}
\centering
\begin{minipage}{1\columnwidth}
\centering
\includegraphics[height=15cm,width=8.5cm]{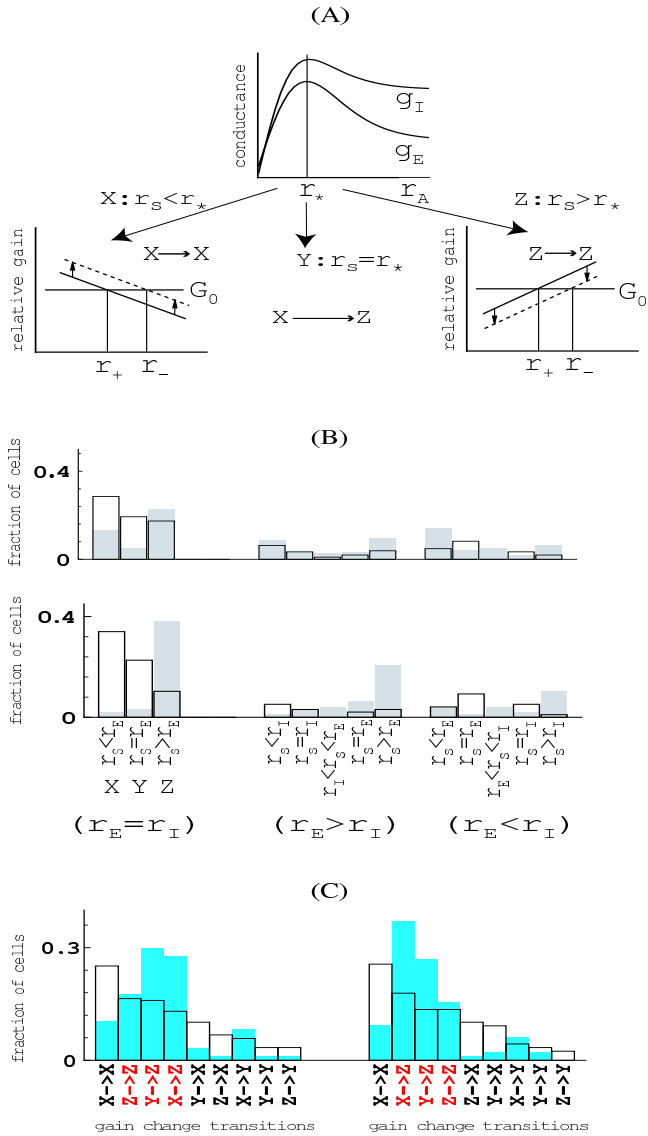}
\end{minipage}
\caption{\small (A) Schematic illustration of mechanisms for receptive field growth under equal and
constant spatial summation extent of the conductances ($r_{E}=r_{I}=r_{\star}$).
(B) Distributions of the relative positions of the maxima (receptive field sizes) of 
spike responses $r_{S}$ and conductances $r_{E}$ and $r_{I}$, for the M0 configuration
(other cases give similar results).  A division is made with respect
to the maxima in the conductances, this corresponds to the left
($r_{E}=r_{I}$), central ($r_{E}>r_{I}$), and right ($r_{E}<r_{I}$)
part of the figure. Each panel is further subdivided with respect to
the maximum in the spike response $r_{S}$.  Upper histograms are for
all cells in the sample, lower histograms are for cells that have receptive field growth 
$r_{-}/r_{+}>1.5$. Unfilled histograms are for high contrast, shaded histograms are for low
contrast. (C) Prevalence of 
transitions between positions of maxima in
spike responses and excitatory conductances (left) 
and in spike responses and inhibitory conductances (right) for a high $\rightarrow$ 
low contrast change. See text for definitions of X, Y, Z classes. Data are evaluated for all cells 
(unfilled histograms) and See text for for cells with a receptive field growth $r_{-}/r_{+}>1.5$ (shaded
histograms).
}
\end{figure}
\end{document}